%% file: main.tex
\documentclass[twocolumn,superscriptaddress,secnumarabic,amssymb, nobibnotes, aps, bm,prl,floatfix,longbibliography]{revtex4-2} 
\usepackage[english]{babel}
\usepackage[utf8]{inputenc}
\usepackage{graphicx}
\usepackage{multirow}
\usepackage[dvipsnames]{xcolor}
\usepackage{bm}
\usepackage{mathtools}
\usepackage{amsmath}
\usepackage{esint}
\usepackage{rotating}
\usepackage{comment}
\definecolor{oceanboatblue}{rgb}{0.0, 0.47, 0.75}
\definecolor{orange}{rgb}{1,0.5,0}
\definecolor{goodgreen}{rgb}{0.1,0.5,0}
\definecolor{goodred}{rgb}{0.7,0,0}
\usepackage[colorlinks,urlcolor=goodgreen,citecolor=blue,linkcolor=goodred]{hyperref}
\usepackage{cleveref}
\usepackage{tikz}
\usepackage{tocvsec2}
\usepackage{scrextend}


\usepackage[normalem]{ulem}

\usepackage{slashbox}
\usepackage{cancel} 

\begin{document}

\title{Chiral edge-state signatures of Berry monopole merging and splitting in a bilayer photonic crystal slab with synthetic momentum}

\author{Ngoc Duc Le}
\email{lnduc@iop.vast.vn}
\affiliation{Donostia International Physics Center, 20018 Donostia-San Sebasti\'{a}n, Spain}
\affiliation{Institute of Physics, Vietnam Academy of Science and Technology, 10 Dao Tan, Giang Vo, Ha Noi, Viet Nam}
\affiliation{Univ. Lille, CNRS, UMR 8523-PhLAM-Physique des Lasers Atomes et Molécules, F-59000 Lille, France}

\author{D.-H.-Minh Nguyen}
\email{d.h.minh.ng@gmail.com}
\affiliation{Donostia International Physics Center, 20018 Donostia-San Sebasti\'{a}n, Spain}
\affiliation{Advanced Polymers and Materials: Physics, Chemistry and Technology, Chemistry Faculty (UPV/EHU), Paseo M. Lardizabal 3, 20018 San Sebastian, Spain}

\author{Dung Xuan Nguyen} 
\email{dungmuop@gmail.com}
\affiliation{Center for Theoretical Physics of Complex Systems, Institute for Basic Science (IBS), Daejeon, 34126, Republic of Korea}
\affiliation{Institute for Interdisciplinary Research in Science and Education, ICISE, Quy Nhon, Vietnam}

\author{Hai Son Nguyen}
\email{hai-son.nguyen@ec-lyon.fr}
\affiliation{Ecole Centrale de Lyon, CNRS, INSA Lyon, Université Claude Bernard Lyon 1, CPE Lyon, CNRS, INL, UMR5270, Ecully 69130, France}
\affiliation{CNRS-International-NTU-Thales Research Alliance (CINTRA), IRL 3288, Singapore 637553, Singapore}
\affiliation{Institut Universitaire de France (IUF), 75231 Paris, France}

\author{Dario Bercioux}
\email{dario.bercioux@dipc.org}
\affiliation{Donostia International Physics Center, 20018 Donostia-San Sebasti\'{a}n, Spain}
\affiliation{IKERBASQUE, Basque Foundation for Science, Euskadi Plaza, 5, 48009 Bilbao, Spain}

\begin{abstract}
Berry monopoles are point degeneracies in parameter space where the Berry curvature emanates radially, analogous to magnetic monopoles. Here, we report the adiabatic merging and splitting of Berry monopoles in a bilayer photonic crystal slab combining one genuine and one synthetic momentum. Two monopoles approach, merge, and split within this hybrid parameter space. The process is described by an effective coupled-mode model and confirmed by full-wave simulations. We further propose a scheme to experimentally probe the Berry monopole merging and splitting by monitoring the evolution of chiral edge states localized at the interface between two topological phases with inverted band structures.
\end{abstract}
\date{\today}
\maketitle

\section{Introduction} 
Dirac famously introduced the concept of a point-like \emph{magnetic monopole} as the magnetic counterpart to electric charge~\cite{Dirac1931,Dirac1948}. While such particles remain unobserved in high-energy physics~\cite{Acharya2022}, monopole-like quasiparticles have emerged in condensed-matter systems such as spin-ice pyrochlores~\cite{Castelnovo2008, Morris2009, Skjrvo_2019} and chiral skyrmion lattices~\cite{Milde2013, Hoffmann_2017}. 
In 1984, Berry showed that adiabatic evolution of a quantum state gives rise to a geometric phase, equal to the flux of a fictitious magnetic field—the Berry curvature—originating from point degeneracies known as \emph{Berry monopoles}~\cite{Berry1984,BernevigHughes2013,Cayssol_2021}. 
Prominent examples are Weyl points in topological semimetals, acting as Berry monopoles in momentum space~\cite{Volovik1987,Armitage2018}; analogous Weyl points have also been realized in photonic~\cite{Chen2016} and phononic crystal systems~\cite{Li2017}. As with electric charges, Berry monopoles carrying the same topological charge are expected to repel. They cannot annihilate, but their merging produces a higher-charge object that is generically unstable.

%
%
\begin{figure}[!h]
    \centering
    \includegraphics[width=0.95\linewidth]{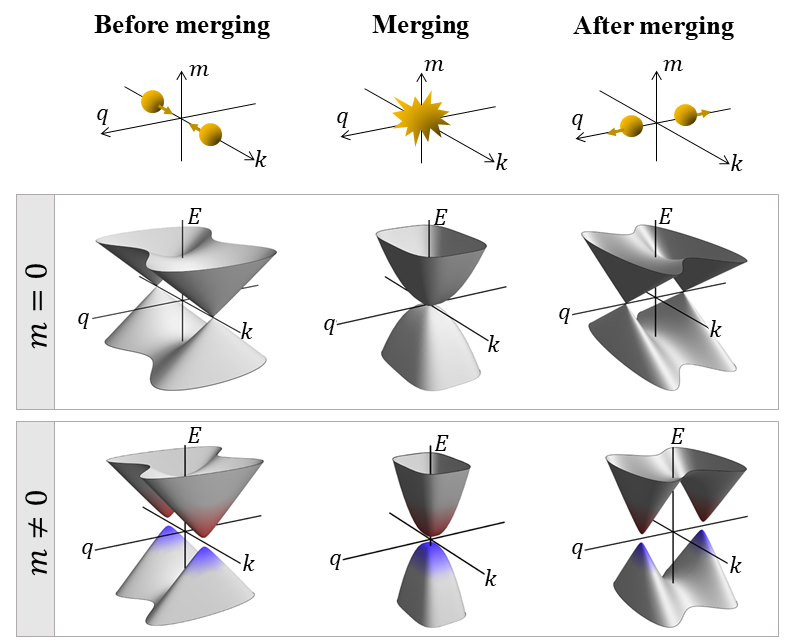}
    \caption{\textbf{Berry monopole merging and splitting in hybrid momentum space.} 
\textit{Top:} In the $(k, q, m)$ space, two Berry monopoles with the same charge approach along the $k$-axis, merge at $(0, 0, 0)$, and split along the $q$-axis. 
\textit{Middle:} At $m = 0$, the system is gapless with band degeneracies (linear and quadratic band touching points) in the $(k, q)$ plane, corresponding to Berry monopoles. 
\textit{Bottom:} For $m \ne 0$, the bands are gapped and exhibit finite Berry curvature (red: positive, blue: negative).}
\label{fig:Concepts}
\end{figure}
%
%
Synthetic quantum matter has emerged as a powerful framework for simulating exotic quantum systems in controllable settings~\cite{Ozawa_2019, Grass2025, Yu2025}, including higher-dimensional phenomena. The four-dimensional (4D) quantum Hall effect~\cite{FrohlichPedrini2000, ZhangHu2001, Price2015,Lohse2018, Zilberberg2018, Bouhiron2024} has been explored through synthetic dimensions~\cite{Boada2012, Yu2025}, topological pumping~\cite{Kraus2013, Lohse2018, Zilberberg2018}, and engineered connectivity~\cite{Jukic2013, Price2020}. While synthetic dimensions expand real space into higher dimensions, the emerging concept of \textit{synthetic momenta} instead enlarges the momentum space by incorporating additional parameters into the crystal momentum. These include relative layer displacements~\cite{Nguyen2021, Lee2022, Nguyen2023}, geometric modulations~\cite{Wang2017, Fonseca2024}, and material-dependent properties~\cite{Ma2021}. Weyl points have been realized in such synthetic momentum spaces~\cite{Wang2017, Nguyen2023, Fonseca2024}, making them a compelling platform for investigating Berry monopoles.

Here, we propose a photonic platform that enables the generation, merging, and splitting of same-charge Berry monopoles along distinct directions of the hybrid momentum space. The system is described by an effective model in a synthetic three-dimensional parameter space $(k, q, m)$, where $k$ is the genuine momentum, $q$ is a synthetic momentum, and $m$ is a tunable gap parameter, the parametric space is illustrated in Fig.~\ref{fig:Concepts}. Berry monopoles appear at band degeneracy points located in the $m = 0$ plane, where they act as quantized sources of Berry curvature. When $m \ne 0$, the system opens a gap and transitions into topological phases characterized by finite Chern numbers, a so-called Chern insulator. By continuously tuning the parameters, the monopoles move toward each other along the $k$-axis, merge at the origin $(k,q) = (0,0)$, and split along the $q$-axis. We also demonstrate that chiral edge states emerge at a band-inverted interface, and that their evolution serves as a direct experimental probe of the Berry monopole merging and splitting. Finally, we discuss the experimental feasibility of this setup, laying the groundwork for realizing Berry monopole dynamics in photonic platforms.

The paper is organized as follows. Section~\ref{sec:Model} describes the bilayer photonic crystal slab by presenting an effective model derived from the coupled-wave theory. This section defines the synthetic momentum $q$ that extends the hybrid momentum space and introduces the deformation parameter $e$ that controls the Berry monopole motion. Section~\ref{sec:Monopole_motion} describes the monopole motion in the parameter space by showing the photonic band structure and the Berry curvature maps in the gapped regime. Section~\ref{sec:edge_state} illustrates the evolution of the chiral edge states, which can serve as experimental proof of the Berry monopole merging and splitting. Finally, Section~\ref {sec:experiments} is dedicated to discussing the experimental feasibility of our setup.

\section{Bilayer photonic crystal slab with synthetic momentum}
\label{sec:Model}
\begin{figure}
    \centering
    \includegraphics[width=1\linewidth]{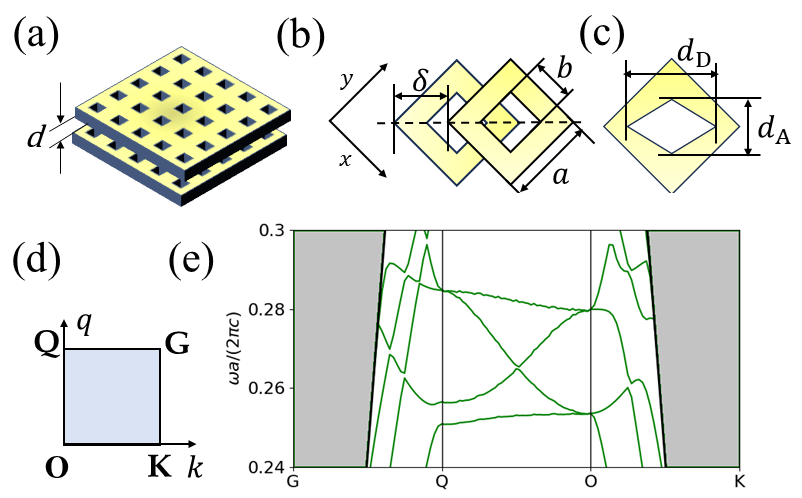}
    \caption{\textbf{Bilayer 2D photonic crystal slab with synthetic momentum.} (a) The design of the bilayer 2D photonic crystal slab with square holes. (b) Top view of the unit cell. (c) Deforming the square hole into a rhombus of diagonals $d_D$ and $d_A$. (d) The hybrid Brillouin zone. (e) The band structure in the hybrid Brillouin zone for the structure with $h/a = 0.35$ and $b/a = 0.38$ shows a quadratic band touching point between the two lowest bands at the point O. The photonic crystal slab is made of silicon (refractive index $n_\text{Si} = 3.54$) and immersed in silica (refractive index $n_\text{SiO$_2$} = 1.46$.)
    }\label{fig:Structures}
\end{figure}
%
%
Our system consists of a bilayer structure formed by two parallel two-dimensional (2D) photonic crystal slabs, separated by an interlayer distance $d$ (Fig.~\ref{fig:Structures}a). Each slab features a square lattice with identical lattice constant $a$, thickness $h$, and a unit cell comprising a square hole of edge length $b$. The two layers are laterally offset along their main diagonals (Fig.~\ref{fig:Structures}b), with a center-to-center displacement $\delta$. The structure is periodic under translations $\delta \to \delta + \sqrt{2}a$, giving rise to a dimensionless synthetic momentum defined as
\begin{equation}
    q = \frac{\delta}{\sqrt{2}a} - \frac{1}{2}
\end{equation}
with $q \in [-1/2, 1/2]$. This construction is analogous to the synthetic dimension used in one-dimensional multilayer grating systems~\cite{Nguyen2021, Nguyen2023}. By choosing $k_x = k_y = k/\sqrt{2} + 1/2$, the system can be reinterpreted as a hybrid (1+1)D configuration, described by a genuine momentum $k$ and a synthetic momentum $q$, rather than as a conventional 2D photonic crystal (Fig.~\ref{fig:Structures}). Interestingly, under the action of the time-reversal operator, the genuine momentum $k$ changes sign ($k \to -k$) as expected when $t \to -t$, whereas the synthetic momentum $q$ remains unchanged ($q \to q$) due to its geometric origin—it arises from the relative spatial shift between the two layers, rather than from a dynamical variable. As a result, when the system is viewed in the hybrid momentum space $(k, q)$, time-reversal symmetry is intrinsically broken. This symmetry breaking gives rise to non-trivial topological properties in the system.

Figure~\ref{fig:Structures}e shows the photonic band structure in the hybrid Brillouin zone for a representative bilayer structure, calculated using the Plane Wave Expansion (PWE) method implemented in the MIT Photonic Bands (MPB) package~\cite{Johnson2001}. The two lowest-energy bands exhibit a degeneracy at the high-symmetry point O ($k = q = 0$), marked by the blue arrow in the figure. This point corresponds to a quadratic band-touching point, with both bands displaying parabolic dispersion along the $k$ and $q$ directions. In the following, we focus on the physical phenomena that emerge near this degeneracy point.

Next, we deform each square hole into a rhombus whose diagonals are defined by 
\begin{equation}
    d_\text{D/A}^\ell = \sqrt{2} b_\ell \frac{1 \pm e}{1 \mp e}
\end{equation}
where $e$ is the \textit{diagonal deformation parameter}, constrained within $-1 < e < 1$ (Fig.~\ref{fig:Structures}c), we allow the hole size to be different for the two layers: $\ell=\pm$, $+$ for the upper layer and $-$ for the lower one. This deformation preserves the hole area—and thus the filling factor—while allowing geometric anisotropy to be continuously tuned via $e$. Here, $d^\ell_\text{D}$ corresponds to the main diagonal along the diagonal $x = y$ direction, and $d^\ell_\text{A}$ to the transverse diagonal along the $x = -y$ direction. To further break the vertical symmetry, we slightly differentiate the hole sizes in the two layers, denoted by $b_+$ and $b_-$. Both layers, however, share the same deformation parameter $e$, which serves as a key control parameter for tuning the system's optical behavior. 
%
%
\begin{figure}[!th]
    \centering
    \includegraphics[width=1\linewidth]{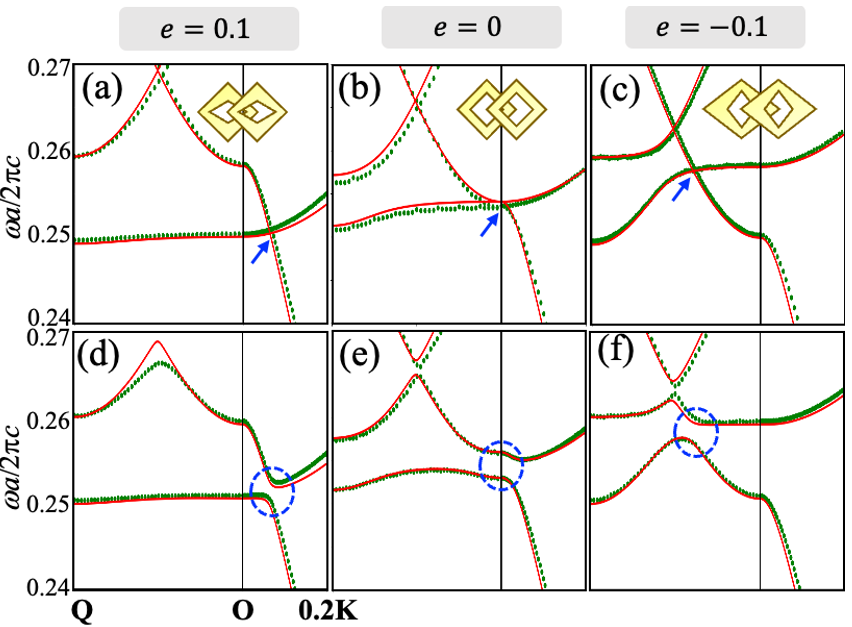}
    \caption{\textbf{Band structure comparison.}  
Photonic band structures computed using the effective model (red lines) and the PWE method (green dots).  
\textit{Top row:} symmetric case with $b_+ = b_- = 0.38a$. \textit{Bottom row:} asymmetric case with $b_+ = 0.46a$ and $b_- = 0.30a$.  
Each column corresponds to a different deformation parameter: $e = 0.1$ (left), $e = 0$ (center), and $e = -0.1$ (right).  
Degeneracy points are marked by blue arrows; gap openings are highlighted by blue circles.}
    \label{fig:Collision}
\end{figure}
%
%
The lowest-energy transverse electric (TE) eigenmodes near the high-symmetry point O can be described by an effective Hamiltonian derived from coupled-wave theory~\cite{Streifer1977, Liang2011, Peng2011, Peng2012}:
\begin{equation}
    \mathcal{H}(k, q, m) = 
    \begin{pmatrix}
        \Delta_+ & \Omega \\ 
        \Omega^{\dagger} & \Delta_- 
    \end{pmatrix},
\end{equation}
where $\Delta_\ell$ is the Hamiltonian of individual layer $\ell$ ($\ell = +$ for the upper layer and $\ell = -$ for the lower layer~\footnote{We use the symbols $\pm$ to refer to both the layer indices $\ell = \pm$ and the diagonal directions $d^\ell_\text{D/A}$. The reader is advised to distinguish between these notations.}):
\begin{equation}
    \begin{aligned}
     \Delta_\ell =& \omega_\ell + \frac{v_\ell}{\sqrt{2}} k^2
     \\
    +&  
    \begin{pmatrix}
        \eta + v_\ell k & W_\ell & W_\ell & U_l(1+\alpha) \\ 
        W_\ell & -\eta & U_\ell(1-\alpha) & W_\ell \\ 
        W_\ell & U_\ell(1-\alpha) & -\eta & W_\ell \\ 
        U_\ell(1+\alpha) & W_\ell & W_\ell & \eta -v_\ell k 
    \end{pmatrix}. 
    \end{aligned}
\label{eq:MonolayerHamiltonian}
\end{equation}
The asymmetry between the layers is captured by the \textit{gap parameter} $m$, which modifies the coupling constants as $\omega_{\pm} = \omega(1 \pm r_\omega m), \quad 
v_{\pm} = v(1 \pm r_v m), \quad 
U_{\pm} = U(1 \pm m), \quad 
W_{\pm} = W(1 \pm r_W m)$. 
The parameters $(k, q, m)$ define a three-dimensional space in which topological features emerge. When $m = 0$, the layers are identical; when $m \ne 0$, they differ in thickness and/or hole size. The parameters $\alpha$ and $\eta$ originate from the deformation of the square hole into a rhombus shape and vary monotonically as functions of the deformation parameter $e$. Both vanish when the holes are square ($e = 0$).

The interlayer coupling is described by the block
\[\Omega = e^{-d/d_0} \cdot
\text{diag} \left\{ 
    V_{00} e^{-i2\pi q},\ V_{0,-1},\ V_{-1,0},\ V_{-1,-1} e^{i2\pi q} 
\right\},\] 
in which the matrix elements are dependent on the genuine momentum $k$: 
\begin{equation}
    \begin{aligned}
        V_{00}(k) =& - \left( V + \beta k + \frac{\beta}{\sqrt{2}} k^2 \right) 
        \\ 
        V_{0-1}(k) =& V_{-10}(k) = V + \frac{\beta}{\sqrt{2}} k^2 
        \\
        V_{-1-1} (k) =& - \left( V - \beta k + \frac{\beta}{\sqrt{2}} k^2 \right)
    \end{aligned}
\end{equation}
The parameters of the effective model fitted with MPB calculations are: 
    $\omega = 0.2992$, $v = 0.3131$, $U = -0.01525$, $W = 0.00176$;
    $r_{\omega} = 0.1353$, $r_v = 0.0892$, $r_W = 2.4215$, $m = 0.00045$;
    $d_0 = 0.35$, $d = 0.1$, $V = 0.039$, $\beta = -0.3$. 
    $(\alpha,\eta) = (-0.062,0.0032)$ for $e = 0.1$; $(\alpha,\eta) = (0.0,0.0)$ for $e = 0.0$; $(\alpha,\eta) = (0.062,-0.0032)$ for $e = -0.1$.
A full derivation of this effective model is given in the Supplemental Materials (SM)~\cite{SM}.

\section{Berry monopole merging and splitting}
\label{sec:Monopole_motion}
Figures~\ref{fig:Collision}a–\ref{fig:Collision}c show the photonic band structures in the hybrid Brillouin zone $(k,q)$ for the symmetric case where the two layers are identical ($b_+ = b_-$), corresponding to $m = 0$. In this configuration, the system remains gapless. For $e > 0$, two Dirac cones appear symmetrically along the $k$-axis with respect to the origin O (Fig.~\ref{fig:Collision}a). As $e$ decreases, the cones move closer and merge when $e = 0$, forming a quadratic band touching point at O~\cite{Chong2008} (Fig.~\ref{fig:Collision}b). When $e < 0$, the cones reappear but now separate along the $q$-axis (Fig.~\ref{fig:Collision}c). Breaking inversion symmetry by making the hole sizes slightly different ($b_+ \ne b_-$, i.e., $m \ne 0$) opens a gap at both the Dirac cones and the quadratic band touching point (Figs.~\ref{fig:Collision}d–\ref{fig:Collision}f). In all cases, the effective model agrees perfectly with full numerical simulations using the MPB package. 

\begin{figure}
    \centering
    \includegraphics[width=1\linewidth]{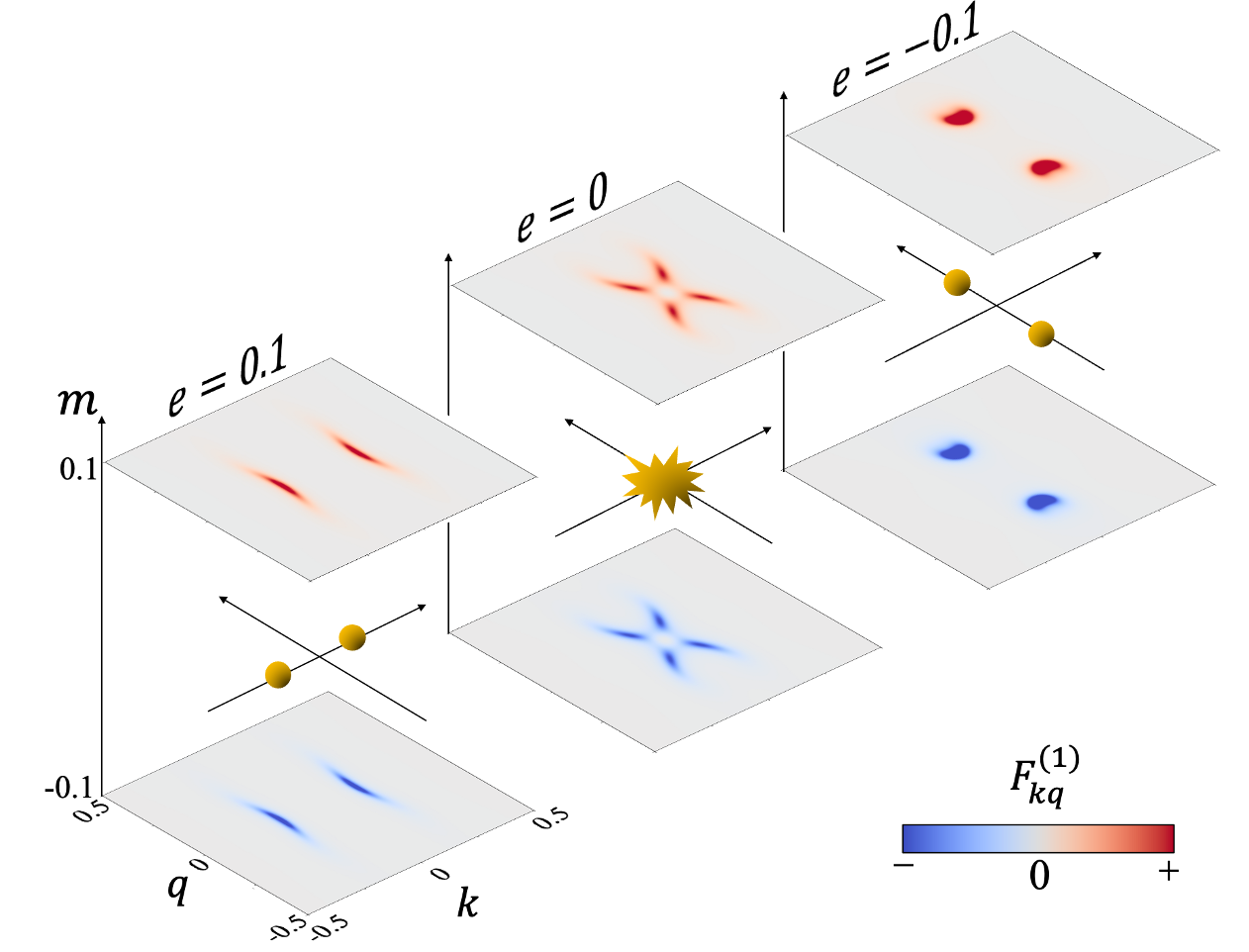}
    \caption{\textbf{The Berry curvature  maps.} The Berry curvature $\mathcal{F}^{(1)}_{kq}$ of band 1 for $m = \pm 0.1$ and $e = 0.1, 0, -0.1$. The results are calculated using the effective model with fitted parameters and MPB band-structure calculations.}
    \label{fig:BerryCurvature}
\end{figure}

Thanks to the intrinsic breaking of time-reversal symmetry—originating from the synthetic momentum $q$—the system becomes a Chern insulator as soon as the degeneracies are lifted. We compute the Berry curvature of band $n$ ($1 \le n \le 8$) using~\cite{Xiao2010} 
\begin{equation}
\mathcal{F}_{kq}^{(n)}(k,q) = -2\, \text{Im} \sum_{n' \ne n} 
\frac{\langle n | \partial_k H | n' \rangle \langle n' | \partial_q H | n \rangle}{(E_n - E_{n'})^2}
\end{equation} 
Figure~\ref{fig:BerryCurvature} presents the Berry curvature of the lowest band ($n = 1$) for $e > 0$, $e = 0$, and $e < 0$, under band inversion with $m = 0.1$ and $m = -0.1$. The Berry curvature changes sign across the transition, confirming the topological nature of the gap. Moreover, the Berry curvature \emph{hotspots}—regions of concentrated curvature—follow the same trajectory as the Dirac cones in the gapless regime: for $e > 0$, they lie on the $k$-axis; as $e$ decreases, they merge at the origin when $e = 0$, forming a four-lobed pattern; for $e < 0$, they split and move along the $q$-axis.

The theory of Berry phase~\cite{Berry1984} provides a unified interpretation of these phenomena through the concept of Berry monopoles. These monopoles are located in the plane $m = 0$ of the parameter space $(k,q,m)$, where the two bands are degenerate and act as sources of the Berry field $\boldsymbol{\mathcal{F}} = (\mathcal{F}_{qm}^{(1)}, \mathcal{F}_{mk}^{(1)}, \mathcal{F}_{kq}^{(1)})$. The sign of $\mathcal{F}_{kq}^{(1)}$ changes across $m = 0$: it is positive for $m > 0$ and negative for $m < 0$, indicating that the Berry field radiates outward from these monopoles. The monopole strength, analogous to magnetic charge, is obtained by integrating the Berry flux over a closed surface enclosing the monopole~\cite{Berry1984, BernevigHughes2013} $ g = \frac{1}{4\pi} \oiint \boldsymbol{\mathcal{F}}(k,q,m) \cdot d\mathbf{S}
\label{eq:MonopoleStrength}$. This yields $g = +1/2$ for each Dirac cone and $g = +1$ for the quadratic band-touching point. As a result, the Chern number of the lowest band is $+1$ for $m > 0$ and $-1$ for $m < 0$, regardless of the deformation parameter~$e$. Overall, \emph{two Berry monopoles approach each other along the genuine momentum ($k$) axis, merge at the origin (O), and split along the synthetic momentum ($q$) axis}.\\

Unlike the familiar scenario in which Berry monopoles with opposite charges merge and annihilate~\cite{Gail2012}, corresponding to the opening of a gap, in our model, same-charge Berry monopoles split apart after having merged into a single charge-2 monopole. In the gapless case ($m = 0$), no gap opens after the transition, and the dispersion remains semimetallic, with two Dirac cones separating along a trajectory set by the Berry monopoles. The merging and splitting of two Berry monopoles of the same charge can be considered as a particular case of ``two-node process" in Ref.~\cite{Naselli2024}. 
The analogous splitting of a charge-2 double-Weyl point into two charge-1 Weyl points was experimentally observed in a 3D chiral woodpile photonic crystal via FTIR spectroscopy~\cite{Jorg2022}. Crucially, the Berry monopoles in Refs.~\cite{Jorg2022, Naselli2024} live in a genuine momentum space, where both the merging of topological charges are defined with respect to ordinary crystal momenta. In contrast, the Berry monopoles in our work live in a hybrid momentum space $(k,q)$, in which the construction breaks the time-reversal symmetry intrinsically, enabling nontrivial Chern phases on either side of the transition.   

%
%
%
%
%
%
\begin{figure*}[!th]
    \centering
\includegraphics[width=\textwidth]{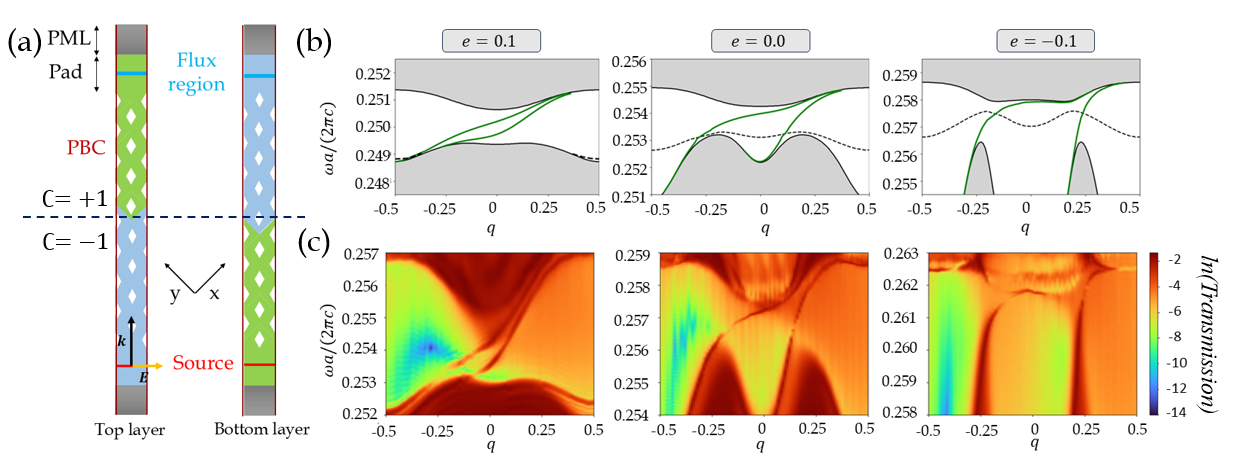}
    \caption{\textbf{Edge states spectra.} (a) Schematic of the FDTD simulation setup. The heterojunction is formed by joining a bilayer, each layer having 12 periods, with thickness $h = 0.35a$ (Chern number $\mathcal{C}=-1$) to its inverted counterpart (with the two layers swapped, $\mathcal{C}=+1$), separated by an interface (dashed line). The hole size is $b = 0.30a$ for the blue regions and $b = 0.46a$ for the green regions. Periodic boundary conditions are applied along the interface, while perfectly matched layers are used perpendicular to it. A linearly polarized TE source (red) is placed at one end of the structure, and the transmitted flux is measured at the opposite end using a detector region (blue).(b) Band structure of the heterojunction interface computed using the effective model, showing chiral edge states (green lines) and bulk bands (grey). A third edge state (dashed black line), which remains within the lower bulk band and does not connect the two bands, is predicted by the model. This state is topologically trivial and does not appear in the FDTD transmission spectra. Parameters used are the same as in Fig.~\ref{fig:Collision}. (c) FDTD transmission spectra, plotted on a logarithmic scale. Both the effective model and the FDTD simulations show the presence of two robust chiral edge states, in agreement with the difference $\Delta \mathcal{C} = +2$ in the Chern number between the two sides of the interface.
}
    \label{fig:FDTD-TransmissionSpectra}
\end{figure*}
%
%


\section{Chiral edge state transformation}
\label{sec:edge_state}
We propose a scheme to probe Berry monopole merging and splitting by tracking the evolution of chiral edge states at the interface between two topological photonic phases with inverted band structures. This can be realized by constructing a heterojunction with a zigzag interface (Fig.~\ref{fig:FDTD-TransmissionSpectra}a) between two bilayer photonic crystal slabs that differ only by the sign of the gap parameter $m$. Although the bulk band structures of both slabs are identical, their Chern numbers are opposite ($+1$ \emph{vs.} $-1$), resulting in the emergence of two topologically protected edge states at the interface.

Figure~\ref{fig:FDTD-TransmissionSpectra}b shows the band dispersion of the heterojunction, calculated using the effective model for three representative cases: $e > 0$, $e = 0$, and $e < 0$. Both bulk bands and chiral edge states are clearly visible. To assess the experimental feasibility of this scheme, we perform Finite Difference Time Domain (FDTD) simulations using the MEEP package~\cite{Oskooi2010} (see Fig.~\ref{fig:FDTD-TransmissionSpectra}a for the setup). The resulting transmission spectra, shown in Fig.~\ref{fig:FDTD-TransmissionSpectra}c, agree well with the effective model and confirm the presence of robust chiral edge states. These results demonstrate that the edge states persist under structural deformation and, more importantly, serve as a direct probe of the process of merging and splitting of the Berry monopole. As the deformation parameter $e$ varies, the edge-state evolution tracks the trajectory of Berry monopoles in synthetic momentum space, as discussed below.

Before collision ($e > 0$): the two edge states are ``entangled"—both originate from the same local valley ($q < 0$) in the lower bulk band and terminate at the same point ($q > 0$) in the upper band. In the bulk, this configuration corresponds to two Berry monopoles separated along the genuine momentum ($k$) axis. However, across the interface, translational symmetry is broken, so $k$ is no longer a good quantum number. As a result, the separation of the monopoles along $k$ becomes ill-defined, and the two associated edge states appear intertwined in synthetic momentum space. At collision ($e = 0$): the two edge states begin to ``disentangle". One originates from a local valley ($q < 0$), while the other starts at $q = 0$ in the lower bulk band. Nevertheless, they still enter the upper bulk band at the same point. This configuration corresponds to the collision of the Berry monopoles at the origin of the $(k, q)$ plane. After collision ($e < 0$): the disentanglement is complete. The two edge states now originate from distinct valleys of the lower bulk band and terminate at different $q$ values in the upper band. This maps directly to the post-collision configuration, where the Berry monopoles have split and are now separated along the synthetic momentum axis $q$. This proposed scenario provides a powerful and experimentally accessible route to visualize Berry monopole merging and splitting, using edge-state evolution as a topological probe in the extended $(k, q, m)$ parameter space.
 
The effective model gives an additional M-shaped edge state shown as dashed line in Fig.~\ref{fig:FDTD-TransmissionSpectra}b. As $e$ decreases from $0.1$ down to $-0.1$, this edge state shifts upward in energy and moves inside the gap from the lower band to the upper band. However, it does not connect the two bands, so at the current stage, we conclude that is topologically trivial. Notably, the FDTD transmission spectra do not reproduce this additional edge state. 
It is possible that this edge state is strongly localized at the interface and does not couple to the other end of the heterojunction to be detectable in the transmission spectra (see the SM~\cite{SM} for more detailed discussion). 

\section{Experimental feasibility}
\label{sec:experiments}
Here, we address the possible experimental demonstration, in a photonic platform, of the process of merging and splitting of Berry monopoles we have introduced so far. A first demonstration does not require continuous actuation of the interlayer displacement: one may fabricate a series of bilayer samples with different prescribed relative shifts $\delta$, or assemble two separately fabricated slabs with controlled lateral offsets, and then reconstruct the monopole merging and splitting sequence from the measured spectra. This static approach is already sufficient to compare the three representative regimes before, at, and after the collision. The bilayer architecture itself is well within current experimental capabilities. Related bilayer photonic structures have already been realized in silicon moir\'e photonic crystals, on-chip optical bilayer photonic crystals, bilayer semiconductor metasurfaces, and WS$_2$ bilayer metasurfaces~\cite{saadi2025, Tang2023, Wang2026, Choi2025}. In addition, shift-dependent optical responses have recently been measured in bilayer metasurfaces, further supporting the practical controllability of the interlayer displacement in coupled photonic platforms~\cite{Tan2025}. For a reconfigurable implementation, the bilayer can be integrated with MEMS actuation. Recent MEMS-integrated moir\'e photonic devices have demonstrated \emph{in situ} tunability together with nanometre-scale positioning of coupled optical layers~\cite{Tang2025}. In our system, a residual displacement uncertainty $\Delta\delta$ translates into an uncertainty in synthetic momentum, $\Delta q=\frac{\Delta\delta}{\sqrt{2}a}$. For a lattice constant $a$ of a few hundred nanometres, even a displacement uncertainty of a few to a few tens of nanometres corresponds to a relatively small uncertainty in $q$. Moreover, the experiment can be performed quasi-statically by stepping the actuator to successive setpoints and recording the optical response only after mechanical settling. In this sense, the proposed measurement does not rely on coherent fast-time adiabatic motion, but rather on reproducible static or quasi-static sampling of $q$. Finally, the robustness observed in our FDTD calculations in the presence of structural perturbations further supports the proposal's feasibility. Altogether, these considerations indicate that Berry monopole merging and splitting should be observable either in a multi-sample static implementation or in a single MEMS-reconfigurable device.

\section{Conclusion \& Outlook}
In conclusion, we theoretically and numerically study the phenomenon of Berry monopole merging and splitting in a topological photonic system with synthetic momenta. We propose a practical scheme to probe this phenomenon via the formation and evolution 
of two chiral edge states at the interface between bilayers with opposite Chern numbers. These edge states provide a direct experimental signature of monopole merging and splitting in momentum space. Our system is experimentally feasible thanks to standard nanofabrication techniques and MEMS technology. Our findings provide a viable topological photonic platform to study Berry monopole physics with both theoretical significance and experimental feasibility. This 2D photonic crystal setup can be exploited to explore the physics in 4D. The two relative displacements along the $x$ and $y$ directions allow us to define two synthetic momenta $(q_x,q_y)$. This permits defining a (2+2)D hybrid momentum space with two genuine momenta $(k_x,k_y)$ and two synthetic momenta. The physics in 4D is potentially rich for both Hermitian and non-Hermitian regimes, promising unusual properties that do not exist in two- or three-dimensions. We aim to explore the 4D analogs of the quantum Hall effect, Weyl semimetals, and tensor monopoles~\cite{Palumbo2018, Palumbo2019, Zhu2020} within synthetic momentum space. Moreover, synthetic momenta would permit going beyond the familiar toroidal momentum space and exploring the properties of 4D non-orientable momentum spaces~\cite{Fonseca2024, Konig2026}. Finally, one can study polaritonic Chern insulator~\cite{He2023} in high dimensions by adding nonlinearity to this photonic platform with synthetic momenta. 

\section{Acknowledgments}

N.D.L., D.H.M.N, and D.B. acknowledge the support from the Transnational Common Laboratory $Quantum-ChemPhys$,  the Department of Education of the Basque Government through the project PIBA\_2023\_1\_0007 (STRAINER), and the financial support received from the IKUR Strategy under the collaboration agreement 
between the Ikerbasque Foundation and DIPC on behalf of the Department of Education of the Basque Government, the Gipuzkoa Provincial Council within the QUAN-000021-01 project, by Grant PID2024-162933NB-I00 (QUILL), funded by MICIU/AEI/10.13039/501100011033, ERDF/EU, and the project ``Artificial Quantum Matter: From 2D Materials to Spin Lattice Systems" - BBVA Foundation Fundamentos Program 2024". D.H.M.N is supported by the Spanish Ministerio de Ciencia, Innovaci\'{o}n y Universidades grant PRE2021-097126. H.S.N is supported by the French National Research Agency (ANR) under the project POLAROID (ANR-24-CE24-7616-01).  D.X.N. is supported
by the Institute for Basic Science in Korea through the Project IBS-R024-D1.
N.D.L is grateful to H. Chau Nguyen and Simone Zanotti for helpful discussions.
The authors acknowledge technical support and computer time allocation provided by the DIPC supercomputing center, the HPC M\'{e}socentre de Calcul Scientifique Intensif de l'Universit\'{e} de Lille, and the cluster of the Vietnam Institute of Physics.


\nocite{Streifer1977,Liang2011,Peng2011,Peng2012,Xiao2010,Girvin_Yang_2019,BernevigHughes2013,Tisseur2000,Tisseur2001,Dedieu2003,Guttel2017,Oskooi2010,Oskooi2011,Oskooi2008,SynMo4Topo,Data_Zenodo} 
\bibliography{bibliography}



\input{Supp/SupplementalMaterials-arXiv}

\end{document}

%% file: Supp/SupplementalMaterials-arXiv.tex
\definecolor{oceanboatblue}{rgb}{0.0, 0.47, 0.75}
\definecolor{orange}{rgb}{1,0.5,0}
\definecolor{goodgreen}{rgb}{0.1,0.5,0}
\definecolor{goodred}{rgb}{0.7,0,0}


\title{Supplementary Materials for \\ Chiral edge-state signatures of Berry monopole merging and splitting in a bilayer photonic crystal slab with synthetic momentum}

\setcounter{section}{0}
\setcounter{equation}{0}
\setcounter{figure}{0}
\setcounter{table}{0}
\setcounter{page}{1}

\renewcommand{\theequation}{S\arabic{equation}}
\renewcommand{\thefigure}{S\arabic{figure}}
\renewcommand{\vec}[1]{\boldsymbol{#1}}

\onecolumngrid
\newpage

\begin{center}
    \textbf{Chiral edge-state signatures of Berry monopole merging and splitting in a bilayer photonic crystal slab with synthetic momentum}
\end{center}

\tableofcontents

\section{Derivation of the effective model: single layer, bilayer and synthetic momenta}

In this section, we derive the effective model for two-dimensional photonic crystal slabs single and bilayers using the couple-wave theory ~\cite{Streifer1977,Liang2011,Peng2011,Peng2012}. In the references, the reader finds the detailed derivations of the coupled-wave theory for the cases of 1D grating and 2D photonic crystal slab single layer at the $\Gamma$ point. Here, we move on to the M point of the 2D slab square lattice and the case of bilayer.

\subsection{Effective Hamiltonian of single layer}
\label{sec:SingleLayerHamiltonian}

\begin{figure}[h]
    \centering
    \includegraphics[width=0.75\linewidth]{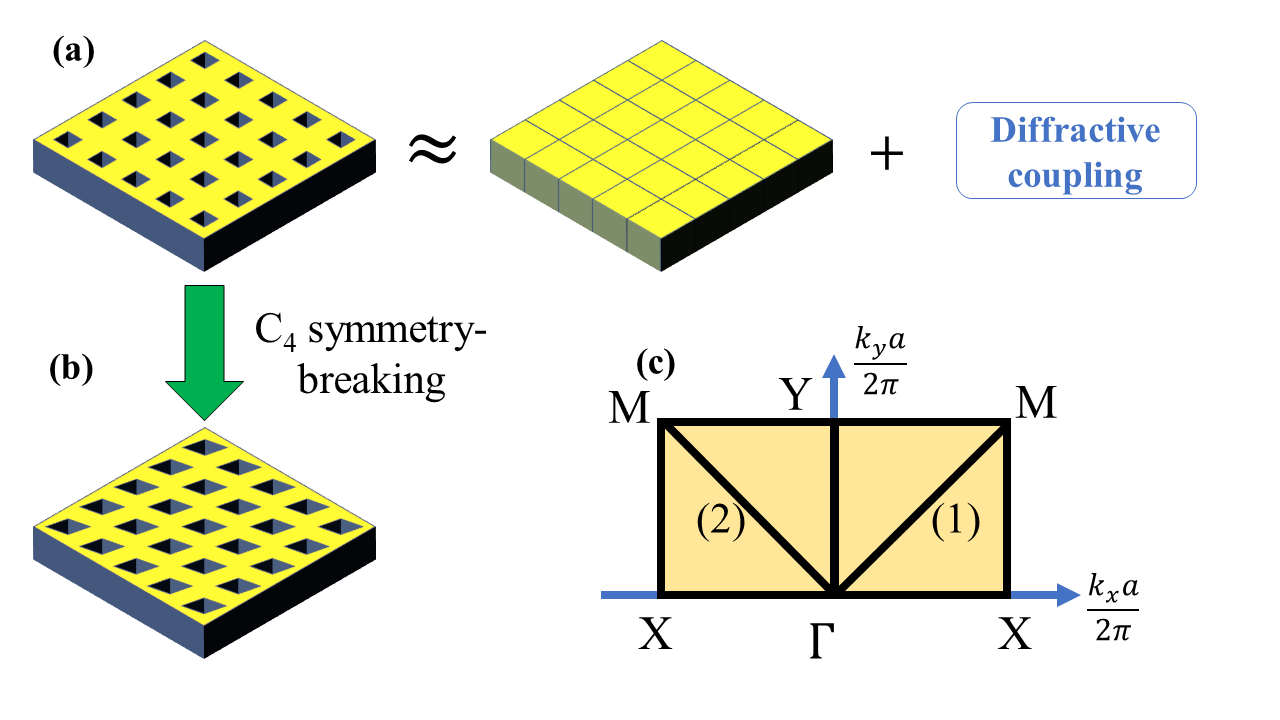}
    \caption{(a) A 2D photonic crystal slab is approximatively equivalent to a non-corrugated slab with effective refractive index. (b) The 2D photonic crystal slab with rhombus hole is obtained by breaking the C\textsubscript{4} symmetry of the 2D photonic crystal slab with square hole. (c) The Brillouin zone of the 2D photonic crystal slab with rhombus hole inside a square unit cell.}
    \label{fig:Monolayer}
\end{figure}

Consider a 2D photonic crystal slab of thickness $h$ and having a square lattice of lattice parameter $a$ (Fig.~\ref{fig:Monolayer}). The unit cell is therefore a square of edge length $a$ and has a hole of edge $b (b<a)$ in the middle. In this work, we consider two types of holes: the square one (C\textsubscript{4} symmetric) and the rhombus one (broken C\textsubscript{4} symmetry). The structure is made of Si of refractive index $n_{\text{Si}} = 3.54$ immerged in SiO\textsubscript{2} (refractive index $n_{\text{SiO\textsubscript{2}}} = 1.46$). The filling factor is therefore $f = (b/a)^2$. The reciprocal lattice is also a square lattice with lattice parameter $K = 2\pi / a$. The high-symmetry points are located at the points $m (K/2) \mathbf{\hat{x}} + n (K/2) \mathbf{\hat{y}} (m,n \in \mathbb{Z})$. In this photonic crystal slab, guided modes are classified according to their polarization: TE modes have electric field vector in the $xy$ plane, and TM modes have magnetic field vector in the $xy$ plane. In this work, we focus on the TE-polarized modes whose electric field vector is given by $\mathbf{E} = (E_x,E_y,0)$. According  to the Bloch theorem, the E-field is written in Bloch form: 
\begin{equation}
    \mathbf{E} (\mathbf{r}) = e^{i\boldsymbol{\kappa}\cdot\boldsymbol{\rho}} \mathbf{u}_{\boldsymbol{\kappa}} (\mathbf{r})   
\end{equation}
where $\boldsymbol{\kappa} = (\kappa_x,\kappa_y)$ belongs to the first Brillouin zone and $\boldsymbol{\rho} = (x,y)$.
That means for $j = x,y$, we write: 
\begin{equation}
    E_j(\mathbf{r}) = e^{i\boldsymbol{\kappa}\cdot \boldsymbol{\rho}} u_{j \boldsymbol{\kappa}} (\mathbf{r}) 
    = e^{i(\kappa_xx+\kappa_yy)} \sum_{mn \in \mathbb{Z}} E_{jmn}(z) e^{i(mKx+nKy)} 
    = \sum_{m,n \in \mathbb{Z}} E_{jmn}(z) e^{i(k_{mx}x + k_{ny}y)}
    \label{eq:BlochWaveFunction}
\end{equation}
with $k_{mx} = mK+\kappa_x$ and $k_{ny}=nK+\kappa_y$ ($0 \le \kappa_x, \kappa_y < K$). From the Maxwell's equations, we deduce the following eigenvalue equation for the electromagnetic field: 
\begin{equation}
    \nabla \times \nabla \times \mathbf{E} (\mathbf{r}) = \left( \frac{\omega}{c} \right)^2 \varepsilon (\mathbf{r}) \mathbf{E} (\mathbf{r})
    \label{eq:MasterEquationE}
\end{equation} 

Here $\varepsilon(\mathbf{r})$ is the periodic dielectric profile of the photonic crystal slab, which can be Fourier transformed as follows: 
\begin{equation}
    \varepsilon (\mathbf{r}) = \bar{\varepsilon} (z) + \sum_{(m,n) \ne (0,0)} \xi_{mn}(z) e^{i(mKx+nKy)}
    \label{eq:DielectricProfile}
\end{equation}
Here we denote $\bar{\varepsilon}$ instead of $\xi_{00}(z)$.
We explicitly develop \eqref{eq:MasterEquationE}:
\begin{equation}
    \begin{aligned}
        \frac{\partial^2 E_x}{\partial y^2} + \frac{\partial^2 E_x}{\partial z^2} - \frac{\partial^2 E_y}{\partial x \partial y} =& - \left( \frac{\omega}{c} \right)^2 \varepsilon E_x 
        \\ 
        \frac{\partial^2 E_y}{\partial x^2} + \frac{\partial^2 E_y}{\partial z^2} - \frac{\partial^2 E_x}{\partial x \partial y} =& - \left( \frac{\omega}{c} \right)^2 \varepsilon E_y 
        \\
        \frac{\partial}{\partial z} \left( \frac{\partial E_x}{\partial x} + \frac{\partial E_y}{\partial y} \right) =& 0 
    \end{aligned}
    \label{eq:MasterEquationE1}
\end{equation}

By substituting \eqref{eq:BlochWaveFunction} and \eqref{eq:DielectricProfile} into \eqref{eq:MasterEquationE1}, we obtain the following system of equations for all $m,n \in \mathbb{Z}$:

\begin{subequations}
    \begin{equation}
        \left[ \frac{d^2}{dz^2} - k_{ny}^2 + \left(\frac{\omega}{c}\right)^2 \bar{\varepsilon} (z) \right] E_{xmn}(z) + k_{mx}k_{ny} E_{ymn}(z) = - \sum_{  \substack{m'n' \\ (m',n') \ne (m,n)}} \left( \frac{\omega}{c} \right)^2 \xi_{\substack{m-m'\\n-n'}} (z) E_{xm'n'}(z) 
        \label{eq:MasterEquationE2A}
    \end{equation}
    \begin{equation}
        \left[ \frac{d^2}{dz^2} - k_{mx}^2 + \left( \frac{\omega}{c} \right)^2 \bar{\varepsilon} (z) \right] E_{ymn}(z) + k_{mx} k_{ny} E_{xmn}(z) = - \sum_{\substack{m'n' \\ (m',n')\ne(m,n)}} \left( \frac{\omega}{c} \right)^2 \xi_{\substack{m-m'\\n-n'}} (z) E_{ym'n'}(z) 
        \label{eq:MasterEquationE2B}
    \end{equation}
    \begin{equation}
         \frac{d}{dz} \left[ k_{mx} E_{xmn}(z) + k_{ny} E_{ymn}(z) \right] = 0
         \label{eq:MasterEquationE2C}
    \end{equation}
\end{subequations}

To solve this system of equations, we define the following linear combinations of solutions:

\begin{subequations}
\begin{equation}
    E_{+mn}(z) = k_{mx}E_{xmn}(z) + k_{ny}E_{ymn}(z)
    \label{eq:Eplusmn}
\end{equation}  
\begin{equation}
    E_{-mn}(z) = k_{ny}E_{xmn}(z) - k_{mx}E_{ymn}(z)
    \label{eq:Eminusmn}
\end{equation}
\end{subequations}

By multiplying both sides of \eqref{eq:MasterEquationE2A} by $k_{mx}$ and both sides of \eqref{eq:MasterEquationE2B} by $k_{ny}$ and adding the two equations sides-by-sides, we arrive to: 
\begin{equation}
    \bar{\varepsilon} (z) E_{+mn}(z) = - \sum_{\substack{m'n'\\(m',n')\ne(m,n)}}  \xi_{\substack{m-m'\\n-n'}}(z) [ k_{mx} E_{xm'n'}(z) + k_{ny} E_{ym'n'}(z) ] 
    \label{eq:EquationEplusmn}
\end{equation}

Here the second-order derivatives with respect to $z$ vanish because of \eqref{eq:MasterEquationE2C}. And by multiplying both sides of \eqref{eq:MasterEquationE2A} by $k_{ny}$ and both sides of \eqref{eq:MasterEquationE2B} by $k_{mx}$, then subtract two equations sides-by-sides, we obtain: 
\begin{equation}
    \left[ \frac{d^2}{dz^2} + \left( \frac{\omega}{c} \right)^2 \bar{\varepsilon} (z) - k_{mx}^2 - k_{ny}^2 \right] E_{-mn} (z) = - \sum_{\substack{m'n' \\ (m',n') \ne (m,n)}} 
    \left( \frac{\omega}{c} \right)^2 \xi_{\substack{m-m'\\n-n'}} (z) [ k_{ny} E_{xm'n'}(z) - k_{mx} E_{ym'n'}(z) ]
    \label{eq:HelmholtzEquationEminusmn}
\end{equation} 

In the scope of our effective theory, we divide the modes into two sets: (i) the set $\mathcal{B}$ of the modes of interest, that we call the \textbf{\textit{basic modes}}, and (ii) the set $\mathcal{O}$ containing all the other modes. Because we are interested in the M-point, the effective Hamiltonian contains the coupling between modes having the same momentum $(\kappa_x,\kappa_y)$ in the vicinity of the four lowest M points: M$(K/2,K/2)$, M$(-K/2,K/2)$, M$(-K/2,-K/2)$ and M$(K/2,-K/2)$. They correspond to $(m,n) = (0,0)$, $(m,n) = (-1,0)$, $(m,n) = (-1,-1)$ and $(m,n) = (0,-1)$, respectively. These modes form the set of basic modes $\mathcal{B}$. The basic modes wavefunctions have the following form:
\begin{equation}
    \begin{aligned}
        E_{xmn}(z) =& \cos \theta_{mn} A_{mn} \Theta_{mn} (z) 
        \\ 
        E_{ymn}(z) =& \sin \theta_{mn} A_{mn} \Theta_{mn} (z)
    \end{aligned}
    \label{eq:BasicModesEfieldComponents}
\end{equation}

Here $A_{mn}$ is the wave amplitude, it can be a complex number. $\Theta_{mn}(z)$ is the envelop wavefunction. $\theta_{mn}$ is the angle form by the electric field $\mathbf{E}_{mn}$ and the $x$ axis. For TE modes, the electric field $\mathbf{E}_{mn}$ is perpendicular to the wave vector $\mathbf{k}_{mn}$, hence we can define the angle $\theta_{mn}$ as follows: 
\begin{equation}
    \begin{aligned}
        \cos \theta_{mn} =& \frac{k_{ny}}{\sqrt{k_{mx}^2 + k_{ny}^2}}
        \\
        \sin \theta_{mn} =& - \frac{k_{mx}}{\sqrt{k_{mx}^2+k_{ny}^2}}
    \end{aligned}
    \label{eq:Definitionthetamn}
\end{equation}

We remark that the definition of the angle $\theta_{mn}$ in \eqref{eq:Definitionthetamn}  applies for both basic modes and higher-order modes, that means for all $(m,n)$. The definition \eqref{eq:Definitionthetamn} and the basic mode wavefunctions \eqref{eq:BasicModesEfieldComponents} straightforwardly imply that the wavefunctions $E_{+mn}(z)$ of the basic modes vanishes:
\begin{equation}
    E_{+mn}(z) = 0 
    \text{ for } (m,n) \in \mathcal{B}
    \label{eq:BasicModesEplusmn}
\end{equation}
and the wavefunction $E_{-mn}(z)$ has the following form:
\begin{equation}
    E_{-mn}(z) = \sqrt{k_{mx}^2 + k_{ny}^2} A_{mn} \Theta_{mn}(z)
    = \bar{A}_{mn} \Theta_{mn}(z) 
    \text{ for } (m,n) \in \mathcal{B}
\end{equation}
where $\bar{A}_{mn} = \sqrt{k_{mx}^2+k_{ny}^2} A_{mn}$.

The wave vectors $\mathbf{k}_{mn}$ of the four basic modes have different amplitudes, so the phase matching condition of the four basic modes are slightly different, implying that they do not have identical envelop function - the wave profile along the $z$ direction. However, because we are interested on the waves in the vicinity of the M point, we can assume that these four basic waves have the same envelop function $\Theta_0(z)$ as the photonic at the M point in a non-corrugated homogeneous dielectric slab with dielectric constant $\bar{\varepsilon}(z)$:  
\begin{equation}
 \Theta_{mn} (z) \approx \Theta_0(z) \text{ for all } (m,n) \in \mathcal{B}
\end{equation}

The envelop function $\Theta_0(z)$ can be nonzero outside the photonic crystal slab (evanescent wave) and satisfies the normalization condition:
\begin{equation}
    \int_{-\infty}^{+\infty} | \Theta_0(z)|^2 dz = 1
\end{equation}

We denote the coordinate of the M point as the vector $\mathbf{g} = (g_x,g_y) = (K/2,K/2)$, and $\omega_M$ the angular frequency of the photonic mode at the M point of a uncorrugated homogeneous slab. The envelop function $\Theta_0(z)$ satisfies the Helmholtz equation at the point M: 
\begin{equation}
    \left[ \frac{d^2}{dz^2} + \bar{\varepsilon}(z) \left( \frac{\omega_M}{c} \right)^2 - g_x^2 - g_y^2 \right] \Theta_0 (z) = 0
    \label{eq:HelmholtzEquationEnvelopFunction}
\end{equation}

For the basic mode $(m,n)$,  the envelop function $\Theta_{mn}(z) \approx \Theta_0(z)$ 
and we rewrite \eqref{eq:HelmholtzEquationEminusmn} as follows:
\begin{equation}
\begin{aligned}
    \left[ \frac{d^2}{dz^2} + \left(\frac{\omega}{c} \right)^2 \bar{\varepsilon}(z) - k_{mx}^2 - k_{ny}^2 \right] \bar{A}_{mn} \Theta_0 (z) 
    =& - \sum_{\substack{(m',n') \in \mathcal{B} \\ (m',n') \ne (m,n)}} \left( \frac{\omega}{c} \right)^2 \xi_{\substack{m-m' \\ n-n'}} (z) 
    [ k_{ny} E_{xm'n'}(z) - k_{mx} E_{ym'n'}(z) ]
    \\
    & - \sum_{(m',n') \in \mathcal{O}} \left( \frac{\omega}{c} \right)^2 \xi_{\substack{m-m'\\n-n'}} (z) [ k_{ny} E_{xm'n'} (z) - k_{mx} E_{ym'n'}(z) ]
\end{aligned}
\label{eq:EquationBasicModes}
\end{equation}

We eliminate the second-order derivative by multiplying \eqref{eq:HelmholtzEquationEnvelopFunction} by $\bar{A}_{mn}$ and then subtracting \eqref{eq:EquationBasicModes} to the resulting equation side-by-side, giving: 
\begin{equation}
    \begin{aligned}
        \left[ \left( \frac{\omega^2-\omega_M^2}{c^2} \right) \bar{\varepsilon}(z) - \mathbf{k}_{mn}^2 + \mathbf{g}^2 \right] \bar{A}_{mn} \Theta_0 (z) 
        =& - \sum_{\substack{(m',n') \in \mathcal{B} \\ (m',n') \ne (m,n)}} \left( \frac{\omega}{c} \right)^2 \xi_{\substack{m-m' \\ n-n'}} (z) 
    [ k_{ny} E_{xm'n'}(z) - k_{mx} E_{ym'n'}(z) ]
    \\
    & - \sum_{(m',n') \in \mathcal{O}} \left( \frac{\omega}{c} \right)^2 \xi_{\substack{m-m'\\n-n'}} (z) [ k_{ny} E_{xm'n'} (z) - k_{mx} E_{ym'n'}(z) ]
    \end{aligned}
    \label{eq:EquationBasicModes1}
\end{equation}

We denote $k_{mn} = \sqrt{k_{mx}^2 + k_{ny}^2}$.
For $(m',n') \in \mathcal{B}$, we have: 
\begin{equation}
    \begin{aligned}
        k_{ny} E_{xm'n'}(z) - k_{mx} E_{ym'n'}(z) =& \sqrt{k_{mx}^2 + k_{ny}^2} ( \cos \theta_{mn} E_{xm'n'}(z) + \sin \theta_{mn} E_{ym'n'} (z) )  
        \\
        =& k_{mn} (cos \theta_{mn} \cos \theta_{m'n'} + \sin \theta_{mn} \sin \theta_{m'n'} ) A_{m'n'} \Theta_0 (z) 
        \\
        =& \frac{k_{mn}}{k_{m'n'}} \cos (\theta_{mn} - \theta_{m'n'}) \bar{A}_{m'n'} \Theta_0 (z)
    \end{aligned}
    \label{eq:Eminusmnmprimenprime}
\end{equation}

By substituting \eqref{eq:Eminusmnmprimenprime} into \eqref{eq:EquationBasicModes1}, we have: 
\begin{equation}
    \begin{aligned}
        \left[ ( \omega^2 - \omega_M^2 ) \bar{\varepsilon} (z) - c^2 (\mathbf{k}_{mn}^2 - \mathbf{g}^2) \right] \bar{A}_{mn} \Theta_0 (z) =& 
        - \omega^2 \sum_{\substack{(m',n') \in \mathcal{B} \\ (m',n') \ne (m,n)}}  \xi_{\substack{m-m' \\ n-n'}} (z) 
        \frac{k_{mn}}{k_{m'n'}} \cos (\theta_{mn} - \theta_{m'n'}) \bar{A}_{m'n'} \Theta_0 (z)
        \\
        & - \omega^2 \sum_{(m',n') \in \mathcal{O}}  \xi_{\substack{m-m'\\n-n'}} (z) [ k_{ny} E_{xm'n'} (z) - k_{mx} E_{ym'n'}(z) ]
    \end{aligned}
    \label{eq:EquationBasicModes2}
\end{equation}

Here, the factor $1/c^2$ cancel in both sides. For our photonic crystal slab, we define the function $f_{\varepsilon}$:
\begin{equation}
    f_{\varepsilon} (z) = 
    \begin{cases}
        1 & \text{ if } z \in \text{slab} \\
        0 & \text{ if } z \notin \text{slab}
    \end{cases}
\end{equation}
so we can write for all $m,n \in \mathbb{Z}$, $(m,n) \ne (0,0)$:
\begin{equation}
    \xi_{mn} (z) =
    \begin{cases}
        \xi_{mn} & \text{ if } z \in \text{slab} \\ 
        0 & \text{ if } z \notin \text{slab} 
    \end{cases} 
    = f_{\varepsilon} (z) \xi_{mn}
\end{equation}

In particular, with $(m,n) = (0,0)$, the average dielectric constant is:
\begin{equation}
    \bar{\varepsilon} (z) = 
    \begin{cases}
        & \xi_{00} = \bar{\varepsilon} \text{ if } z \in \text{ slab }
        \\
        & \varepsilon_{env} \text{ if } z \notin \text{ slab }
    \end{cases}
\end{equation}
where $\varepsilon_{env}$ is the dielectric constant of the envirnonment, here is SiO\textsubscript{2}.

By multiplying both sides of \eqref{eq:EquationBasicModes2} by $\Theta_0(z)^*$ and then take the integral over $dz$ from $-\infty$ to $+\infty$, we obtain: 
\begin{equation}
    \begin{aligned}
        & (\omega^2 - \omega_M^2) \left( \int_{-\infty}^{+\infty} \bar{\varepsilon}(z) | \Theta_0(z)|^2 dz \right) \bar{A}_{mn} - c^2(\mathbf{k}_{mn}^2-\mathbf{g}^2) \bar{A}_{mn} 
        \\
        =& - \omega^2 \sum_{ \substack{(m',n') \in \mathcal{B} \\ (m',n') \ne (m,n)} } \xi_{\substack{m-m' \\ n-n'}} \frac{k_{mn}}{k_{m'n'}} \cos (\theta_{mn} - \theta_{m'n'}) \bar{A}_{m'n'} \left( \int_{-\infty}^{+\infty} f_{\varepsilon} (z) |\Theta_0(z)|^2 dz \right)
        \\ 
        & - \omega^2 \sum_{\substack{(m',n') \in \mathcal{O}}} \xi_{\substack{m-m' \\ n-n'}} 
        \left( \int_{-\infty}^{+\infty} f_{\varepsilon}(z) \Theta_0(z)^* [ k_{ny} E_{xm'n'}(z) - k_{mx} E_{ym'n'}(z) ] dz \right) 
    \end{aligned}
    \label{eq:EquationBasicModes3}
\end{equation}

We define: 
\begin{equation}
    \alpha_{\varepsilon} = \int_{-\infty}^{+\infty} f_{\epsilon} (z) |\Theta_0(z)|^2 dz
    = \int_{\text{slab}} |\Theta_0(z)|^2 dz
\end{equation}
and:
\begin{equation}
        \zeta_{\varepsilon} = \int_{-\infty}^{+\infty} \bar{\varepsilon}(z) | \Theta_0(z)|^2 dz 
\end{equation}

Remark that $\zeta_{\varepsilon} \ne \alpha_{\varepsilon} \bar{\varepsilon}$ (to be precise: $\zeta_{\epsilon} > \alpha_{\varepsilon} \bar{\varepsilon}$). We rewrite \eqref{eq:EquationBasicModes3} as follows:
\begin{equation}
\begin{aligned}
    [(\omega^2 - \omega_M^2) \zeta_{\varepsilon} - c^2 (\mathbf{k}_{mn}^2 - \mathbf{g}^2)] \bar{A}_{mn} =& 
    - \omega^2 \alpha_{\varepsilon} \sum_{\substack{(m',n') \in \mathcal{B} \\ (m',n') \ne (m,n) }} 
    \xi_{\substack{m-m' \\ n-n'}} \frac{k_{mn}}{k_{m'n'}} \cos (\theta_{mn} - \theta_{m'n'}) \bar{A}_{m'n'}
    \\
    & - \omega^2 \sum_{\substack{(m',n') \in \mathcal{O}}} \xi_{\substack{m-m' \\ n-n'}} 
        \left( \int_{-\infty}^{+\infty} f_{\varepsilon}(z) \Theta_0(z)^* [ k_{ny} E_{xm'n'}(z) - k_{mx} E_{ym'n'}(z) ] dz \right) 
\end{aligned}
\label{eq:EquationBasicModes4}
\end{equation}

By dividing both sides by $k_{mn}$, \eqref{eq:EquationBasicModes4} becomes:
\begin{equation}
    \begin{aligned}
        [(\omega^2 - \omega_M^2) \zeta_{\epsilon}  - c^2 (\mathbf{k}_{mn}^2 - \mathbf{g}^2) ] A_{mn} 
        =&
        - \omega^2 \alpha_{\varepsilon} \sum_{\substack{(m',n')\in \mathcal{B} \\ (m',n') \ne (m,n)}} \xi_{\substack{m-m' \\ n-n'}} \cos (\theta_{mn} - \theta_{m'n'}) A_{m'n'} 
        \\
        & - \omega^2 \sum_{\substack{(m',n') \in \mathcal{O}}} \xi_{\substack{m-m' \\ n-n'}} 
        \left( \int_{-\infty}^{+\infty} f_{\varepsilon} (z) \Theta_0^*(z) [ \cos \theta_{mn} E_{xm'n'} (z) + \sin \theta_{mn} E_{ym'n'} (z) ] dz \right) 
    \end{aligned}
    \label{eq:EquationBasicModes5}
\end{equation}

In the right-hand side of \eqref{eq:EquationBasicModes5}, the first term represents the direct coupling between basic modes, and the second term represents the coupling between basic modes via the remaining modes. To solve \eqref{eq:EquationBasicModes5}, we need to find the expressions of $E_{xm'n'}(z)$ and $E_{ym'n'}(z)$ as functions of $A_{mn}$.  We recall that \eqref{eq:HelmholtzEquationEminusmn} is valid for both basic modes and higher-order modes, so for the mode $(m',n')$, we construct the Green's function $G_{m'n'}(z,z')$ that satisfies:
\begin{equation}
    \left[ \frac{d^2}{dz^2} + \left( \frac{\omega}{c} \right)^2 \bar{\varepsilon} (z) - k_{m'x}^2 - k_{n'y}^2 \right] G_{m'n'} (z,z') = - \frac{1}{c^2} \delta (z - z')  
\end{equation}

The Green's function $G_{m'n'}(z,z')$ is approximated as:
\begin{equation}
    G_{m'n'} (z,z') \approx \frac{1}{2 \beta_{m'n'}(z) c^2} e^{ - \beta_{m'n'}(z) | z - z'| } 
\end{equation}
where $\beta_{m'n'}(z) = \sqrt{k_{m'x}^2 + k_{n'y}^2 - \left(\frac{\omega}{c}\right)^2 \bar{\varepsilon} (z)}$.
Therefore, for all modes $(m',n')$ including basic modes and higher-order modes, the wavefunction $E_{-m'n'}(z)$ is given by:
\begin{equation}
    E_{-m'n'}(z) = \int_{-\infty}^{+\infty} 
    G_{m'n'}(z,z') \sum_{\substack{(m'',n'') \\ (m'',n'') \ne (m',n')}} \omega^2 \xi_{\substack{m'-m'' \\ n'-n''}} (z') [ k_{n'y} E_{xm''n''}(z') - k_{m'x} E_{ym''n''}(z') ] dz'
    \label{eq:EminusmnGreenFunction}
\end{equation}


From the definition of $E_{-mn}(z)$ in \eqref{eq:Eminusmn} and the definition of $\theta_{mn}$ in \eqref{eq:Definitionthetamn}, we rewrite $E_{-m'n'}(z) = k_{m'n'} [ \cos \theta_{m'n'} E_{xm'n'}(z) + \sin \theta_{m'n'} E_{ym'n'} (z) ]$ so that \eqref{eq:EminusmnGreenFunction} gives us the following equation for $E_{xm'n'}(z)$ and $E_{ym'n'}(z)$:
\begin{equation}
\begin{aligned}
    & \cos \theta_{m'n'} E_{xm'n'}(z) + \sin \theta_{m'n'} E_{ym'n'} (z) 
    \\
    & = \omega^2 \sum_{\substack{(m'',n'') \\ (m'',n'')\ne (m',n')}} \xi_{\substack{m'-m'' \\ n'-n''}}
    \int_{-\infty}^{+\infty} G_{m'n'}(z,z')  f_{\varepsilon} (z') 
    [ \cos \theta_{m'n'} E_{xm''n''}(z') + \sin \theta_{m'n'} E_{ym''n''}(z') ] dz'
\end{aligned}
\label{eq:LinearSystemExmn}
\end{equation}

From the definition of $E_{+mn}(z)$ in \eqref{eq:Eplusmn} and the definition of $\theta_{mn}$ in \eqref{eq:Definitionthetamn}, we rewrite $E_{+m'n'}(z) = k_{m'n'}[-\sin \theta_{m'n'} E_{xm'n'}(z) + \cos \theta_{m'n'} E_{ym'n'}(z)]$. By substituting into \eqref{eq:EquationEplusmn}, we obtain: 
\begin{equation}
\begin{aligned}
    & - \sin \theta_{m'n'} E_{xm'n'} (z) + \cos \theta_{m'n'} E_{ym'n'} (z) \\
    & = - \sum_{\substack{(m'',n'') \\ (m'',n'') \ne (m',n')}} \frac{\xi_{\substack{m'-m''\\n'-n''}}}{\bar{\varepsilon} (z)} 
    f_{\varepsilon} (z)
    [ -\sin \theta_{m'n'} E_{xm''n''} (z) + \cos \theta_{m'n'} E_{ym''n''} (z) ] 
\end{aligned}
\label{eq:LinearSystemEymn}
\end{equation}

Equations \eqref{eq:LinearSystemExmn} and \eqref{eq:LinearSystemEymn} form a system of linear equations for $E_{xm'n'}(z)$ and $E_{ym'n'}(z)$. For a high-order mode $(m',n') \in \mathcal{O}$, we solve this system of linear equations to extract the values of $E_{xm'n'}(z)$ and $E_{ym'n'}(z)$. Next, we substitute $E_{xm'n'}(z)$ and $E_{ym'n'}(z)$ back into \eqref{eq:EquationBasicModes5}, resulting in \footnote{One can set the right-hand sides of Eqs.~\eqref{eq:LinearSystemExmn} and \eqref{eq:LinearSystemEymn} as $A$ and $B$, respectively. Therefore, we obtain $E_{xm'n'}(z) = A \cos \theta_{m'n'} - B \sin \theta_{m'n'}$ and $E_{ym'n'}(z) = B \cos \theta_{m'n'} + A \sin \theta_{m'n'}$. Consequently, $\cos \theta_{mn} E_{xm'n'}(z) + \sin \theta_{mn} E_{ym'n'}(z) = \cos (\theta_{mn} - \theta_{m'n'}) A + \sin (\theta_{mn} - \theta_{m'n'}) B$}:
\begin{equation}
    \begin{aligned}
        & \cos \theta_{mn} E_{xm'n'} (z) + \sin \theta_{mn} E_{ym'n'} (z)
        \\
        =& \cos (\theta_{mn} - \theta_{m'n'}) \omega^2 
        \sum_{\substack{(m'',n'') \\ (m'',n'') \ne (m',n')}} \xi_{\substack{m'-m'' \\ n'-n''}}
        \int_{-\infty}^{+\infty} G_{m'n'}(z,z') f_{\varepsilon} (z') 
        [ \cos \theta_{m'n'} E_{xm''n''}(z') + \sin \theta_{m'n'} E_{ym''n''}(z') ] dz'
        \\
        & - \sin (\theta_{mn} - \theta_{m'n'}) 
        \sum_{\substack{(m'',n'') \\ (m'',n'') \ne (m',n')}} \xi_{\substack{m'-m''\\n'-n''}} 
        \frac{f_{\varepsilon}(z)}{\bar{\varepsilon}(z)} 
        [ -\sin \theta_{m'n'} E_{xm''n''}(z) + \cos \theta_{m'n'} E_{ym''n''}(z) ]
    \end{aligned}
    \label{eq:EquationBasicModes6}
\end{equation}

Before substituting \eqref{eq:EquationBasicModes6} into \eqref{eq:EquationBasicModes5}, we remark that the coupling between basic modes and high-order modes is more important than the coupling between high-order modes. Therefore, we can omit the contributions from high-order modes, that means $(m'',n'') \in \mathcal{O}$ in the two sums in the right-hand side of \eqref{eq:EquationBasicModes6}, resulting in the approximation: 
\begin{equation}
    \begin{aligned}
        & \cos \theta_{mn} E_{xm'n'} (z) + \sin \theta_{mn} E_{ym'n'} (z)
        \\
        \approx & \cos (\theta_{mn} - \theta_{m'n'}) \omega^2 
        \sum_{\substack{(m'',n'') \in \mathcal{B} \\ (m'',n'') \ne (m',n')}} \xi_{\substack{m'-m'' \\ n'-n''}}
        \int_{-\infty}^{+\infty} G_{m'n'}(z,z') f_{\varepsilon} (z') 
        [ \cos \theta_{m'n'} E_{xm''n''}(z') + \sin \theta_{m'n'} E_{ym''n''}(z') ] dz'
        \\
        & - \sin (\theta_{mn} - \theta_{m'n'}) 
        \sum_{\substack{(m'',n'') \in \mathcal{B} \\ (m'',n'') \ne (m',n')}} \xi_{\substack{m'-m''\\n'-n''}} 
        \frac{f_{\varepsilon}(z)}{\bar{\varepsilon}(z)} 
        [ -\sin \theta_{m'n'} E_{xm''n''}(z) + \cos \theta_{m'n'} E_{ym''n''}(z) ]
    \end{aligned}
    \label{eq:EquationBasicModes7}
\end{equation}

For $(m'',n'') \in \mathcal{B}$, by substituting \eqref{eq:BasicModesEfieldComponents} into \eqref{eq:EquationBasicModes7}, we have:
\begin{equation}
    \begin{aligned}
        & \cos \theta_{mn} E_{xm'n'} (z) + \sin \theta_{mn} E_{ym'n'} (z)
        \\
        \\
        \approx & \omega^2 \cos (\theta_{mn} - \theta_{m'n'})  
        \sum_{\substack{(m'',n'') \in \mathcal{B} \\ (m'',n'') \ne (m',n')}} 
        \xi_{\substack{m'-m'' \\ n'-n''}} 
        \cos (\theta_{m'n'} - \theta_{m''n''}) 
        \left( \int_{-\infty}^{+\infty} G_{m'n'}(z,z') f_{\varepsilon} (z') \Theta_0 (z') dz' \right)
        A_{m'' n''} 
        \\ 
        & + \sin (\theta_{mn} - \theta_{m'n'}) \sum_{\substack{(m'',n'') \in \mathcal{B} \\ {m'',n'') \ne (m',n')}}} 
        \xi_{\substack{m'-m'' \\ n' - n''}} \sin (\theta_{m'n'} - \theta_{m''n''}) \frac{f_{\varepsilon}(z)}{\bar{\varepsilon}(z)} \Theta_0(z) A_{m''n''} 
    \end{aligned}
    \label{eq:EquationBasicModes8}
\end{equation}

By substituting \eqref{eq:EquationBasicModes8} into \eqref{eq:EquationBasicModes5}, we obtain:
\begin{equation}
    \begin{aligned}
        & [ (\omega^2 - \omega_M^2) \zeta_{\varepsilon} - c^2 (\mathbf{k}_{mn}^2 - \mathbf{g}^2) ] A_{mn} 
        \\
        =& - \omega^2 \alpha_{\varepsilon} \sum_{\substack{(m',n') \in \mathcal{B} \\ (m',n') \ne (m,n)}} \xi_{\substack{m-m' \\ n-n'}} \cos (\theta_{mn} - \theta_{m'n'}) A_{m'n'} 
        \\
        & - \omega^4 \sum_{(m',n') \in \mathcal{O}} \sum_{(m'',n'') \in \mathcal{B}} 
        \xi_{\substack{m-m'\\n-n'}} \xi_{\substack{m'-m''\\n'-n''}} \times 
        \\ 
        & \times 
        \cos (\theta_{mn} - \theta_{m'n'}) \cos (\theta_{m'n'} - \theta_{m''n''}) 
        \left( \int_{-\infty}^{+\infty} \int_{-\infty}^{+\infty} f_{\varepsilon} (z) \Theta_0^*(z) G_{m'n'}(z,z') f_{\varepsilon} (z') \Theta_0(z') dz dz' \right) 
        A_{m''n''} 
        \\
        & -\omega^2 \sum_{(m',n') \in \mathcal{O}} \sum_{(m'',n'') \in \mathcal{B}} 
        \xi_{\substack{m-m'\\n-n'}} \xi_{\substack{m'-m''\\n'-n''}} 
        \sin (\theta_{mn} - \theta_{m'n'}) \sin (\theta_{m'n'} - \theta_{m''n''}) 
        \left( \int_{-\infty}^{+\infty} \frac{f_{\varepsilon}^2 (z)}{\bar{\varepsilon}(z)} | \Theta_0 (z) |^2  dz  \right) A_{m''n''}
    \end{aligned}
\end{equation}

We define: 
\begin{equation}
    \mu_{mn} = \int_{-\infty}^{+\infty} \int_{-\infty}^{+\infty} f_{\varepsilon} (z) \Theta_0^*(z) G_{mn}(z,z') f_{\varepsilon}(z') \Theta_0(z') dz dz'
\end{equation}

We remark that:
\begin{equation}
    \int_{-\infty}^{+\infty} \frac{f_{\varepsilon}^2 (z)}{\bar{\varepsilon} (z)} |\Theta_0(z)|^2 dz  
    = \int_{slab} \frac{1}{\bar{\varepsilon}} |\Theta_0(z)|^2 dz 
    = \frac{\alpha_{\varepsilon}}{\bar{\varepsilon}}
\end{equation}

By interchanging the symbols $(m',n')$ and $(m'',n'')$, we obtain: 
\begin{equation}
    \begin{aligned}
        & [(\omega^2 - \omega_M^2) \zeta_{\varepsilon} - c^2 (\mathbf{k}_{mn}^2 - \mathbf{g}^2) ] A_{mn} 
        \\
        =& - \omega^2 \alpha_{\varepsilon} \sum_{\substack{(m',n') \in \mathcal{B} \\ (m',n') \ne (m,n) }} \xi_{\substack{m-m'\\n-n'}} \cos (\theta_{mn} - \theta_{m'n'}) A_{m'n'} 
        \\
        & - \omega^4 \sum_{(m',n') \in \mathcal{B}} 
        \left[ \sum_{\substack{(m'',n'')\in \mathcal{O}}} \xi_{\substack{m-m''\\n-n''}} \xi_{\substack{m''-m'\\n''-n'}} \cos (\theta_{mn}-\theta_{m''n''})  \cos (\theta_{m''n''}-\theta_{m'n'}) \mu_{m'' n''} \right] A_{m'n'} 
        \\ 
        & - \omega^2 \frac{\alpha_{\varepsilon}}{\bar{\varepsilon}} \sum_{(m',n') \in \mathcal{B}} 
        \left[ \sum_{\substack{(m'',n'')\in \mathcal{O}}} \xi_{\substack{m-m''\\n-n''}} \xi_{\substack{m''-m'\\n''-n'}} \sin (\theta_{mn}-\theta_{m''n''})  \sin (\theta_{m''n''}-\theta_{m'n'}) \right] A_{m'n'} 
    \end{aligned}
    \label{eq:EquationBasicModes9}
\end{equation}

In the vicinity of the M point, $\omega \approx \omega_M$, we can approximate $\omega^2 - \omega_M^2 \approx 2 \omega_M (\omega - \omega_M)$, $\omega^2 \approx \omega_M^2$ and $\omega^4 \approx \omega_M^4$. By applying this approximation and then dividing both sides by $2\omega_M \zeta_{\varepsilon}$, we arrive to the following equation: 
\begin{equation}
    \begin{aligned}
        & \left[ (\omega - \omega_M) - \frac{c^2 (\mathbf{k}_{mn}^2 - \mathbf{g}^2 )}{2\omega_M \zeta_{\varepsilon}} \right] A_{mn} 
        \\
        =& - \frac{\alpha_{\varepsilon} \omega_M}{2 \zeta_{\epsilon}} \sum_{\substack{(m',n') \in \mathcal{B} \\ (m',n') \ne (m,n)}} \xi_{\substack{m-m'\\n-n'}} \cos (\theta_{mn} - \theta_{m'n'}) A_{m'n'}
        \\
        & - \frac{\omega_M^3}{2\zeta_{\varepsilon}} \sum_{\substack{(m',n') \in \mathcal{B}}} 
        \left[ \sum_{(m'',n'') \in \mathcal{O}} \mu_{m'' n''} \xi_{\substack{m-m'' \\ n-n''}} \xi_{\substack{m''-m' \\ n'' -n'}} \cos (\theta_{mn} - \theta_{m'' n''}) \cos (\theta_{m'' n''} - \theta_{m'n'}) \right] A_{m'n'} 
        \\ 
        & -  \frac{\alpha_{\varepsilon} \omega_M}{2 \zeta_{\varepsilon} \bar{\varepsilon}} 
        \sum_{(m',n') \in \mathcal{B}} 
        \left[ \sum_{\substack{(m'',n'')\in \mathcal{O}}} \xi_{\substack{m-m''\\n-n''}} \xi_{\substack{m''-m'\\n''-n'}} \sin (\theta_{mn}-\theta_{m''n''})  \sin (\theta_{m''n''}-\theta_{m'n'}) \right] A_{m'n'} 
    \end{aligned}
    \label{eq:EquationBasicModes10}
\end{equation}

We remark that $\cos (\theta_{mn} - \theta_{m'n'})$ is the scalar product of the two unit vectors $\mathbf{\hat{e}}_{mn}$ and $\mathbf{\hat{e}}_{m'n'}$ along the direction of the electric fields $\mathbf{E}_{mn}$ and $\mathbf{E}_{m'n'}$:
\begin{equation}
    \cos (\theta_{mn} - \theta_{m'n'}) = \mathbf{\hat{e}}_{mn} \cdot \mathbf{\hat{e}}_{m'n'}
\end{equation}

We arrive to an eigenvalue problem in terms of the 4 basic modes.
The element of the Hamiltonian corresponding to the row $(m,n)$ and the column $(m',n')$ is given by: 

\begin{equation}
    \mathcal{H}_{mn,m'n'}^{\text{1layer}}(\mathbf{k}) = \mathcal{H}_{mn,m'n'}^{(A)} (\mathbf{k}) + \mathcal{H}_{mn,m'n'}^{(B)} (\mathbf{k}) + \mathcal{H}_{mn,m'n'}^{(C)} (\mathbf{k})
    \label{eq:Hamiltonian1Layer}
\end{equation}
where:
\begin{subequations}
    \begin{equation}
        \mathcal{H}_{mn,m'n'}^{(A)}(\mathbf{k}) = \omega_M + \frac{c^2 ( \mathbf{k}_{mn}^2 - \mathbf{g}^2 )}{2 \omega_M \zeta_{\epsilon}} \delta_{mm'} \delta_{nn'} 
        \label{eq:Hamiltonian1LayerA}
    \end{equation}
    \begin{equation}
        \mathcal{H}_{mn,m'n'}^{(B)}(\mathbf{k}) = - \frac{\alpha_{\varepsilon} \omega_M}{2\zeta_{\varepsilon}} \xi_{\substack{m-m' \\ n-n'}} (\mathbf{\hat{e}}_{mn} \cdot \mathbf{\hat{e}}_{m'n'}) (1-\delta_{mm'} \delta_{nn'}) 
        \label{eq:Hamiltonian1LayerB}
    \end{equation} 
    \begin{equation}
    \begin{aligned}
        \mathcal{H}_{mn,m'n'}^{(C)}(\mathbf{k}) =& - \frac{\omega_M^3}{2\zeta_{\varepsilon}} \sum_{(m'',n'') \in \mathcal{O}} \mu_{m''n''} \xi_{\substack{m-m''\\n-n''}} \xi_{\substack{m''-m'\\n''-n'}} (\mathbf{\hat{e}}_{mn} \cdot \mathbf{\hat{e}}_{m''n''}) (\mathbf{\hat{e}}_{m''n''} \cdot \mathbf{\hat{e}}_{m'n'}) 
        \\ 
        & -  \frac{\alpha_{\varepsilon} \omega_M}{2 \zeta_{\varepsilon} \bar{\varepsilon}} 
        \sum_{\substack{(m'',n'')\in \mathcal{O}}} \xi_{\substack{m-m''\\n-n''}} \xi_{\substack{m''-m'\\n''-n'}} \sin (\theta_{mn}-\theta_{m''n''})  \sin (\theta_{m''n''}-\theta_{m'n'}) 
    \end{aligned}
        \label{eq:Hamiltonian1LayerC}
    \end{equation}
\end{subequations}

In the right-hand side of \eqref{eq:Hamiltonian1Layer}, the contribution $\mathcal{H}^{(A)}(\mathbf{k})$ is the energy of each of the basic modes, the contribution $\mathcal{H}^{(B)}(\mathbf{k})$ is the direct coupling between pairs of different basic modes, and the contribution $\mathcal{H}^{(C)}$ is the coupling between basic modes via high-order modes.
In the following, we examine each of these three contributions. 

\paragraph{\textbf{Energy of the basic modes:}}
This contribution is related to the diagonal elements of the Hamitltonian.
We shift the momentum so that $\mathbf{k} = \boldsymbol{0}$ at the M point, that means:
\begin{equation}
    \begin{aligned}
        \kappa_x =& \frac{K}{2} + k_x \\
        \kappa_y =& \frac{K}{2} + k_y 
    \end{aligned}
\end{equation}

The difference between the squares of the momenta is expressed in terms of $k_x$ and $k_y$ as: 
\begin{equation}
    \begin{aligned}
        \mathbf{k}_{mn}^2 - \mathbf{g}^2 =& \left[ k_x + \left( m + \frac{1}{2} \right) K \right]^2
        + \left[ k_y + \left( n + \frac{1}{2} \right) K \right]^2 - \frac{K^2}{2} 
        \\
        =& k_x^2 + k_y^2 + (2m+1) K k_x + (2n+1) K k_y + (m^2 + n^2 + m + n) K^2 
    \end{aligned}
\end{equation} 

For the basic modes $(m,n) \in \lbrace (0,0), (0,-1), (-1,0), (-1,-1) \rbrace$, we obtain: 
\begin{equation}
    \begin{aligned}
        \mathbf{k}_{00}^2 - \mathbf{g}^2 =& k_x^2 + k_y^2 + K(k_x + k_y)
        \\
        \mathbf{k}_{0-1}^2 - \mathbf{g}^2 =& k_x^2 + k_y^2 + K(k_x - k_y) 
        \\
        \mathbf{k}_{-10}^2 - \mathbf{g}^2 =& k_x^2 + k_y^2 + K(-k_x + k_y)
        \\
        \mathbf{k}_{-1-1}^2 - \mathbf{g}^2 =& k_x^2 + k_y^2 + K(-k_x - k_y) 
    \end{aligned}
\end{equation}


We define the \textbf{\textit{group velocity}} of the basic modes: 
\begin{equation}
    v = \frac{\sqrt{2} K c^2}{2 \omega_M \zeta_{\varepsilon}}
\end{equation}

The matrix elements of the Hamiltonian $\mathcal{H}^{(A)}$ are: 
\begin{equation}
    \begin{aligned}
        \mathcal{H}_{00,00}^{(A)} (\mathbf{k}) =& \omega_M + \frac{v}{\sqrt{2}} (k_x + k_y) + \frac{v}{\sqrt{2}K} (k_x^2 + k_y^2) 
        \\
        \mathcal{H}_{0-1,0-1}^{(A)} (\mathbf{k}) =& \omega_M + \frac{v}{\sqrt{2}} (k_x - k_y) + \frac{v}{\sqrt{2}K} (k_x^2 + k_y^2) 
        \\
        \mathcal{H}_{-10,-10}^{(A)} (\mathbf{k}) =& \omega_M + \frac{v}{\sqrt{2}} (-k_x + k_y) + \frac{v}{\sqrt{2}K} (k_x^2 + k_y^2) 
        \\
        \mathcal{H}_{-1-1,-1-1}^{(A)} (\mathbf{k}) =& \omega_M + \frac{v}{\sqrt{2}} (-k_x - k_y) + \frac{v}{\sqrt{2}K} (k_x^2 + k_y^2) 
    \end{aligned}
\end{equation}

All the off-diagonal elements of $\mathcal{H}^{(A)}$ vanish.

\paragraph{\textbf{Direct coupling between pairs of basic modes:}} 
This contribution is related to the off-diagonal elements of the Hamiltonian.
Because our slab is invariant under the mirror symmetries about the two planes $x = y$ and $x = -y$:
\begin{equation}
    \begin{aligned}
        (x,y) &\rightarrow (y,x) \\
        (x,y) &\rightarrow (-y,-x) 
    \end{aligned}
\end{equation}
we deduce that $\xi_{mn} = \xi_{nm}$, $\xi_{mn} = \xi_{-n-m}$.
Because the dielectric profile $\varepsilon(\mathbf{r})$ is a real number, $\xi_{mn} = \xi_{-m-n}^*$.
Therefore all the coefficients $\xi_{mn}$ are real and:
\begin{equation}
    \xi_{mn} = \xi_{nm} = \xi_{-m-n} = \xi_{-n-m}    
\end{equation}

For this reason, $\mathcal{H}^{(B)}$ is a real symmetric matrix.
The matrix elements of this contribution is given in Table.~\ref{tab:MatrixElementsDirectCoupling}.
We remark that exactly at the M point, the vectors $\mathbf{\hat{e}}_{mn}$ and $\mathbf{\hat{e}}_{m'n'}$ of modes propagating perpendicularly are orthogonal and have vanishing scalar product:
\begin{equation}
    \mathbf{\hat{e}}_{00} \cdot \mathbf{\hat{e}}_{0-1} 
    = \mathbf{\hat{e}}_{00} \cdot \mathbf{\hat{e}}_{-10} 
    = \mathbf{\hat{e}}_{0-1} \cdot \mathbf{\hat{e}}_{-1-1}
    =\mathbf{\hat{e}}_{0-1} \cdot \mathbf{\hat{e}}_{-1-1}
    = 0 
\end{equation}

\begin{table}[]
    \centering
    \begin{tabular}{|c|c|c|c|c|}
    \hline 
    \backslashbox{$(m,n)$}{$(m',n')$} & \makebox[3em]{(0,0)} & \makebox[3em]{(0,-1)} & \makebox[3em]{(-1,0)} & \makebox[3em]{(-1,-1)} \\
    \hline 
    (0,0) &  0   &  $\xi_{01} \mathbf{\hat{e}}_{00} \cdot \mathbf{\hat{e}}_{0-1}$  &  $\xi_{10} \mathbf{\hat{e}}_{00} \cdot \mathbf{\hat{e}}_{-10}$  & $\xi_{11} \mathbf{\hat{e}}_{00} \cdot \mathbf{\hat{e}}_{-1-1}$   \\
    \hline 
    (0,-1) & $\xi_{0-1} \mathbf{\hat{e}}_{0-1} \cdot \mathbf{\hat{e}}_{00}$   & 0  & $\xi_{1-1} \mathbf{\hat{e}}_{0-1} \cdot \mathbf{\hat{e}}_{-10}$    &  $\xi_{10} \mathbf{\hat{e}}_{0-1} \cdot \mathbf{\hat{e}}_{-1-1}$ \\
    \hline 
    (-1,0) & $\xi_{-10} \mathbf{\hat{e}}_{-10} \cdot \mathbf{\hat{e}}_{00}$ & $\xi_{-11} \mathbf{\hat{e}}_{-10} \cdot \mathbf{\hat{e}}_{0-1}$ &  0  &  $\xi_{01} \mathbf{\hat{e}}_{-10} \cdot \mathbf{\hat{e}}_{-1-1}$    \\ 
    \hline 
    (-1,-1) & $\xi_{-1-1} \mathbf{\hat{e}}_{-1-1} \cdot \mathbf{\hat{e}}_{00}$ & $\xi_{-10} \mathbf{\hat{e}}_{-1-1} \cdot \mathbf{\hat{e}}_{0-1}$ & $\xi_{0-1} \mathbf{\hat{e}}_{-1-1} \cdot \mathbf{\hat{e}}_{-10}$      & 0 \\ 
    \hline 
    \end{tabular}
    \caption{Matrix elements of the direct coupling between pairs of basic modes (in unit of $-\alpha_{\varepsilon} \omega_M / (2 \zeta_{\varepsilon})$).}
    \label{tab:MatrixElementsDirectCoupling}
\end{table}

Therefore, the matrix elements for modes propagating perpendicularly vanish \textit{at the } M \textit{point}.
However, because of the $k$-dependence of the unit vectors $\mathbf{\hat{e}}_{mn}$ and $\mathbf{\hat{e}}_{m'n'}$, when we go out of the M point, the modes no longer propagate orthogonally, this scalar product becomes non-vanishing and the matrix elements become non-zero.
We also remark that the four matrix elements between counter-propagating modes have the same value because:
\begin{equation}
    \mathbf{\hat{e}}_{00} \cdot \mathbf{\hat{e}}_{-1-1}
    = \mathbf{\hat{e}}_{0-1} \cdot \mathbf{\hat{e}}_{-10}
    = -1 
\end{equation}

Because $\xi_{11} = \xi_{1-1} = \xi_{-11} = \xi_{-1-1}$, we define: 
\begin{equation}
    U^{(B)} = \frac{\alpha_{\varepsilon} \omega_M \xi_{11}}{2 \zeta_{\varepsilon}} 
\end{equation}

Although the scalar products $\mathbf{\hat{e}}_{mn} \cdot \mathbf{\hat{e}}_{m'n'}$ are $k$-dependent, we consider this dependency as negligible perturbations and write the matrix $\mathcal{H}^{(B)}$ with matrix elements at the M point: 
\begin{equation}
    \mathcal{H}^{(B)} = 
    \begin{pmatrix}
        0 & 0 & 0 & U^{(B)} \\ 
        0 & 0 & U^{(B)} & 0 \\ 
        0 & U^{(B)} & 0 & 0 \\ 
        U^{(B)} & 0 & 0 & 0 
    \end{pmatrix}
\end{equation}

The above formula is written in the basis $\lbrace |00\rangle, |0-1\rangle, |-10\rangle, |-1-1\rangle \rbrace$.

\paragraph{\textbf{Coupling between basic modes via high-order modes:}} 

The matrix elements of $\mathcal{H}^{(C)}$ are dependent on the momentum $(k_x,k_y)$ due to the $k$-dependency of the unit vector $\mathbf{\hat{e}}_{mn}$, $\mathbf{\hat{e}}_{m'n'}$, $\mathbf{\hat{e}}_{m''n''}$ and the Green's function inside $\mu_{m''n''}$.
For the sake of simplicity, we omit these $k$-dependencies because of the small value of $|k|$ compared to $K$.
That means we assume that $\mathbf{\hat{e}}_{mn}$, $\mathbf{\hat{e}}_{m'n'}$, $\mathbf{\hat{e}}_{m''n''}$ and $\mu_{m''n''}$ take their values at the M point ($k_x = k_y = 0$).
First of all, from \eqref{eq:Hamiltonian1LayerC} it is easy to see that $\mathcal{H}^{(C)}$ is a symmetric matrix:
\begin{equation}
    \mathcal{H}_{mn,m'n'}^{(C)} = \mathcal{H}_{m'n',mn}^{(C)}
\end{equation}

Next, we evaluate the diagonal matrix elements of $\mathcal{H}^{(C)}$:
\begin{equation}
\begin{aligned}
    \mathcal{H}_{mn,mn}^{(C)} =& - \frac{\omega_M^3}{2\zeta_{\varepsilon}} \sum_{(m'',n'') \in \mathcal{O}} 
    \mu_{m''n''} \xi_{\substack{m-m'' \\ n-n''}}^2 
    \cos^2 (\theta_{mn}-\theta_{m''n''})  
    + \frac{\alpha_{\varepsilon} \omega_M}{2 \zeta_{\varepsilon} \bar{\varepsilon}} 
    \sum_{\substack{(m'',n'') \in \mathcal{O}}} \xi_{\substack{m-m'' \\ n-n''}}^2 \sin^2 (\theta_{mn}-\theta_{m''n''})
    \\
    =& - \frac{\omega_M^3}{4\zeta_{\varepsilon}} \sum_{(m'',n'') \in \mathcal{O}} 
    \mu_{m''n''} \xi_{\substack{m-m'' \\ n-n''}}^2 
    [ 1 + \cos 2 (\theta_{mn}-\theta_{m''n''}) ]
    + \frac{\alpha_{\varepsilon} \omega_M}{4 \zeta_{\varepsilon} \bar{\varepsilon}} 
    \sum_{\substack{(m'',n'') \in \mathcal{O}}} \xi_{\substack{m-m'' \\ n-n''}}^2 [1 - \cos 2(\theta_{mn} - \theta_{m''n''})]
\end{aligned}
\end{equation}

Here we use the facts that $\xi_{\substack{m-m''\\n-n''}} = \xi_{\substack{m''-m\\n''-n}}$. 
For the 4 basic modes $(m,n) \in \lbrace (0,0), (0,-1), (-1,0), (-1,-1) \rbrace$, $\theta_{00} = -\pi/4$, $\theta_{0-1}=-3\pi/4$, $\theta_{-10}=\pi/4$ and $\theta_{-1-1}=3\pi/4$.
We have: 
\begin{equation}
    \begin{aligned}
        \mathcal{H}_{00,00}^{(C)} =& - \frac{\omega_M^3}{4\zeta_{\varepsilon}} \sum_{(m'',n'') \in \mathcal{O}} \mu_{m''n''} \xi_{\substack{-m''\\-n''}}^2 [1 - \sin (2 \theta_{m''n''})] 
        + \frac{\alpha_{\varepsilon} \omega_M}{4 \zeta_{\varepsilon} \bar{\varepsilon}} \sum_{(m'',n'') \in \mathcal{O}}  \xi_{\substack{-m''\\-n''}}^2 [1 + \sin (2 \theta_{m''n''})] 
        \\
        \mathcal{H}_{0-1,0-1}^{(C)} =& - \frac{\omega_M^3}{4\zeta_{\varepsilon}} \sum_{(m'',n'') \in \mathcal{O}} \mu_{m''n''} \xi_{\substack{-m''\\-1-n''}}^2 [1 + \sin (2 \theta_{m''n''})]  
        + \frac{\alpha_{\varepsilon} \omega_M}{4 \zeta_{\varepsilon} \bar{\varepsilon}} \sum_{(m'',n'') \in \mathcal{O}}  \xi_{\substack{-m''\\-1-n''}}^2 [1 - \sin (2 \theta_{m''n''})] 
        \\
        \mathcal{H}_{-10,-10}^{(C)} =& - \frac{\omega_M^3}{4\zeta_{\varepsilon}} \sum_{(m'',n'') \in \mathcal{O}} \mu_{m''n''} \xi_{\substack{-1-m''\\-n''}}^2 [1 + \sin (2 \theta_{m''n''})] 
        + \frac{\alpha_{\varepsilon} \omega_M}{4 \zeta_{\varepsilon} \bar{\varepsilon}} \sum_{(m'',n'') \in \mathcal{O}}  \xi_{\substack{-1-m''\\-n''}}^2 [1 - \sin (2 \theta_{m''n''})] 
        \\
        \mathcal{H}_{-1-1,-1-1}^{(C)} =& - \frac{\omega_M^3}{4\zeta_{\varepsilon}} \sum_{(m'',n'') \in \mathcal{O}} \mu_{m''n''} \xi_{\substack{-1-m''\\-1-n''}}^2 [1 - \sin (2 \theta_{m''n''})]  
        + \frac{\alpha_{\varepsilon} \omega_M}{4 \zeta_{\varepsilon} \bar{\varepsilon}} \sum_{(m'',n'') \in \mathcal{O}}  \xi_{\substack{-1-m''\\-1-n''}}^2 [1 + \sin (2 \theta_{m''n''})] 
    \end{aligned}
    \label{eq:HCDiagonal}
\end{equation}




The coupling is strongest between basic modes and the higher-order modes having the lowest energy levels.
Therefore, we consider $(m'',n'') \in \lbrace (1,0),(0,1),(-1,1),(-2,0),(-2,-1),(-1,-2),(0,-2),(1,-1) \rbrace$ and neglect contribution from the other high-order modes.
For those modes: $\mathbf{k}_{mn}^2 - \mathbf{g}^2 = 2K^2$.
The values of $\xi_{-m'',-n''}$, $\xi_{-m'',-1-n''}$, $\xi_{-1-m'',-n''}$ and $\xi_{-1-m'',-1-n''}$ for those eight modes are given in Table.~\ref{tab:xiForModes2}.
By definition, the value $k_{m'x}^2 + k_{n'y}^2$ is the same for the eight high-order modes we are considering, so they have the same Green function $G_{m'n'}(z,z')$ and the same value $\mu_{m'n'}$:
\begin{equation}
    \mu_{m'n'} = \mu
\end{equation}


Let $\sin 2\theta_{10} = s$, we remark that $\theta_{10} + \theta_{01} = -\pi/2$, 
$\theta_{-11}=-\theta_{01}$, $\theta_{-20}=-\theta_{10}$,
$\theta_{-2-1}=\theta_{10}+\pi$,
$\theta_{-1-2}=\theta_{01}+\pi$,
$\theta_{0-2}=\theta_{-11}+\pi$,
$\theta_{1-1}=\theta_{-20}+\pi$,
so:
\begin{equation}
    \begin{aligned}
        \sin 2\theta_{10} =& \sin 2 \theta_{01} =& s \\
        \sin 2\theta_{-11} =& \sin 2\theta_{-20} =& -s \\
        \sin 2 \theta_{-2-1} =& \sin 2 \theta_{-1-2} =& s 
        \\
        \sin 2 \theta_{0-2} =& \sin 2 \theta_{1-1} =& -s 
    \end{aligned}
\end{equation}

Therefore, by explicitly developing \eqref{eq:HCDiagonal} over the eight lowest energy modes in $\mathcal{O}$ and by recalling that $\xi_{mn} = \xi_{nm} = \xi_{-m-n} = \xi_{-n-m}$, we obtain:
\begin{equation}
    \begin{aligned}
        \mathcal{H}_{00,00}^{(C)} =& - \frac{\omega_M^3}{2\zeta_{\varepsilon}} \mu [ (\xi_{0-1}^2 + \xi_{12}^2)(1-s) + (\xi_{1-1}^2+\xi_{02}^2)(1+s) ]
         + \frac{\alpha_{\varepsilon} \omega_M}{2\zeta_{\varepsilon} \bar{\varepsilon}} [ (\xi_{0-1}^2 + \xi_{12}^2)(1+s) + (\xi_{1-1}^2+\xi_{02}^2)(1-s) ]
        \\
        \mathcal{H}_{-10,-10}^{(C)} =& - \frac{\omega_M^3}{2\zeta_{\varepsilon}} \mu [ (\xi_{0-1}^2 + \xi_{-12}^2)(1-s) + (\xi_{11}^2+\xi_{02}^2)(1+s) ]
         + \frac{\alpha_{\varepsilon} \omega_M}{2\zeta_{\varepsilon} \bar{\varepsilon}} [ (\xi_{0-1}^2 + \xi_{-12}^2)(1+s) + (\xi_{11}^2+\xi_{02}^2)(1-s) ]
        \\
        \mathcal{H}_{0-1,0-1}^{(C)} =& - \frac{\omega_M^3}{2\zeta_{\varepsilon}} \mu [ (\xi_{01}^2 + \xi_{1-2}^2)(1-s) + (\xi_{11}^2+\xi_{0-2}^2)(1+s) ]
         + \frac{\alpha_{\varepsilon} \omega_M}{2\zeta_{\varepsilon} \bar{\varepsilon}} [ (\xi_{01}^2 + \xi_{1-2}^2)(1+s) + (\xi_{11}^2+\xi_{0-2}^2)(1-s) ]
        \\
        \mathcal{H}_{-1-1,-1-1}^{(C)} =& - \frac{\omega_M^3}{2\zeta_{\varepsilon}} \mu [ (\xi_{01}^2 + \xi_{-1-2}^2)(1-s) + (\xi_{1-1}^2+\xi_{0-2}^2)(1+s) ]
         + \frac{\alpha_{\varepsilon} \omega_M}{2\zeta_{\varepsilon} \bar{\varepsilon}} [ (\xi_{01}^2 + \xi_{-1-2}^2)(1+s) + (\xi_{1-1}^2+\xi_{0-2}^2)(1-s) ]
    \end{aligned}
\end{equation}

Again, the property $\xi_{mn} = \xi_{nm} = \xi_{-m-n} = \xi_{-n-m}$ leads to:
\begin{equation}
    \mathcal{H}_{00,00}^{(C)} = \mathcal{H}_{-1-1,-1-1}^{(C)} = \omega_+^{(C)} 
    \text{ and }
    \mathcal{H}_{0-1,0-1}^{(C)} = \mathcal{H}_{-10,-10}^{(C)} = \omega_-^{(C)}
\end{equation}
when $\mathbf{k} = \boldsymbol{0}$.
Here: 
\begin{equation}
    \begin{aligned}
        \omega_+^{(C)} =& - \frac{\omega_M^3}{2\zeta_{\varepsilon}} \mu [ (\xi_{0-1}^2 + \xi_{12}^2)(1-s) + (\xi_{1-1}^2+\xi_{02}^2)(1+s) ]
         + \frac{\alpha_{\varepsilon} \omega_M}{2\zeta_{\varepsilon} \bar{\varepsilon}} [ (\xi_{0-1}^2 + \xi_{12}^2)(1-s) + (\xi_{1-1}^2+\xi_{02}^2)(1-s) ]
        \\
        \omega_-^{(C)} =& - \frac{\omega_M^3}{2\zeta_{\varepsilon}} \mu [ (\xi_{0-1}^2 + \xi_{-12}^2)(1-s) + (\xi_{11}^2+\xi_{02}^2)(1+s) ]
         + \frac{\alpha_{\varepsilon} \omega_M}{2\zeta_{\varepsilon} \bar{\varepsilon}} [ (\xi_{0-1}^2 + \xi_{-12}^2)(1+s) + (\xi_{11}^2+\xi_{02}^2)(1-s) ]
    \end{aligned}
\end{equation}

Generally, $\xi_{12}$ is different to $\xi_{-12}$ and $\xi_{11}$ is different to $\xi_{1-1}$, so $\omega_+^{(C)}$ and $\omega_-^{(C)}$ are not constrained to be equal to each other.

\begin{table}[t]
    \centering
    \begin{tabular}{|c|c|c|c|c|}
    \hline 
       $(m'',n'')$  & $\xi_{\substack{-m''\\-n''}}$ & $\xi_{\substack{-m''\\-1-n''}}$ & $\xi_{\substack{-1-m''\\-n''}}$ & $\xi_{\substack{-1-m''\\-1-n''}}$ \\
       \hline 
       (1,0)  & $\xi_{\substack{-1\\0}}$ & $\xi_{\substack{-1\\-1}}$ & $\xi_{\substack{-2\\0}}$ & $\xi_{\substack{-2\\-1}}$ \\
       \hline 
       (0,1)  & $\xi_{\substack{0\\-1}}$ & $\xi_{\substack{0\\-2}}$ & $\xi_{\substack{-1\\-1}}$ & $\xi_{\substack{-1\\-2}}$ \\
       \hline 
       (-1,1)  & $\xi_{\substack{1\\-1}}$ & $\xi_{\substack{1\\-2}}$ & $\xi_{\substack{0\\-1}}$ & $\xi_{\substack{0\\-2}}$ \\
       \hline
       (-2,0)  & $\xi_{\substack{2\\0}}$ & $\xi_{\substack{2\\-1}}$ & $\xi_{\substack{1\\0}}$ & $\xi_{\substack{1\\-1}}$ \\
       \hline
       (-2,-1)  & $\xi_{\substack{2\\1}}$ & $\xi_{\substack{2\\0}}$ & $\xi_{\substack{1\\1}}$ & $\xi_{\substack{1\\0}}$ \\
       \hline 
       (-1,-2)  & $\xi_{\substack{1\\2}}$ & $\xi_{\substack{1\\1}}$ & $\xi_{\substack{0\\2}}$ & $\xi_{\substack{0\\1}}$ \\
       \hline 
       (0,-2)  & $\xi_{\substack{0\\2}}$ & $\xi_{\substack{0\\1}}$ & $\xi_{\substack{-1\\2}}$ & $\xi_{\substack{-1\\1}}$ \\
       \hline 
       (1,-1)  & $\xi_{\substack{-1\\1}}$ & $\xi_{\substack{-1\\0}}$ & $\xi_{\substack{-2\\1}}$ & $\xi_{\substack{-2\\0}}$ \\
       \hline 
    \end{tabular}
    \caption{The values of $\xi_{-m'',-n''}$, $\xi_{-m'',-1-n''}$, $\xi_{-1-m'',-n''}$ and $\xi_{-1-m'',-1-n''}$ for the eight high-order modes $(m'',n'') \in \lbrace (1,0),(0,1),(-1,1),(-2,0),(-2,-1),(-1,-2),(0,-2),(1,-1) \rbrace$.}
    \label{tab:xiForModes2}
\end{table}

Next, we consider the coupling between counter-propagating modes.
Because $\mathcal{H}^{(C)}$ is a symmetric matrix, we immediately obtain their values: 
\begin{equation}
    \begin{aligned}
        \mathcal{H}_{00,-1-1}^{(C)} =& \mathcal{H}_{-1-1,00}^{(C)} =& - \frac{\omega_M^3}{2\zeta_{\varepsilon}} 
        \sum_{(m'',n'')\in \mathcal{O}} 
        \mu_{m''n''}
        \xi_{\substack{-m''\\-n''}} \xi_{\substack{m''+1\\n''+1}} 
        (\mathbf{\hat{e}}_{00} \cdot \mathbf{\hat{e}}_{m''n''} )
        (\mathbf{\hat{e}}_{m''n''} \cdot \mathbf{\hat{e}}_{-1-1} )
        \\
        & & - \frac{\alpha_{\varepsilon} \omega_M}{2 \zeta_{\varepsilon} \bar{\varepsilon}} 
        \sum_{(m'',n'')\in \mathcal{O}} 
        \xi_{\substack{-m''\\-n''}} \xi_{\substack{m''+1\\n''+1}} \sin (\theta_{00}-\theta_{m''n''}) \sin(\theta_{m''n''}-\theta_{-1-1})
        \\ 
        &= U_+^{(C)} &
        \\
        \mathcal{H}_{0-1,-10}^{(C)} =& \mathcal{H}_{-10,0-1}^{(C)} =& - \frac{\omega_M^3}{2\zeta_{\varepsilon}} 
        \sum_{(m'',n'')\in \mathcal{O}} 
        \mu_{m''n''}
        \xi_{\substack{-m''\\-1-n''}} \xi_{\substack{m''+1\\n''}} 
        (\mathbf{\hat{e}}_{0-1} \cdot \mathbf{\hat{e}}_{m''n''} )
        (\mathbf{\hat{e}}_{m''n''} \cdot \mathbf{\hat{e}}_{-10} )
        \\
        && - \frac{\alpha_{\varepsilon} \omega_M}{2\zeta_{\varepsilon}\bar{\varepsilon}} 
        \sum_{(m'',n'')\in\mathcal{O}} 
        \xi_{\substack{-m''\\-1-n''}} \xi_{\substack{m''+1\\n''}} 
        \sin (\theta_{0-1}-\theta_{m''n''}) \sin (\theta_{m''n''}-\theta_{-10})
        \\
        &= U_-^{(C)} &
    \end{aligned}
\end{equation}

The next part is to prove that:
\begin{equation}
    \mathcal{H}_{00,0-1}^{(C)} 
    = \mathcal{H}_{00,-10}^{(C)}
    = \mathcal{H}_{0-1,-1-1}^{(C)}
    = \mathcal{H}_{-10,-1-1}^{(C)}
\end{equation} 

Because $\mathcal{H}^{(C)}$ is a symmetric matrix, the values of the other four coupling strength between modes propagating perpendicularly are equal to these four matrix elements.
We have:
\begin{equation}
    \begin{aligned}
        \mathcal{H}_{00,0-1}^{(C)} =& - \frac{\omega_M^3}{2\zeta_{\varepsilon}} \sum_{(m'',n'') \in \mathcal{O}} \mu_{m''n''} \xi_{\substack{-m'' \\ -n''}} \xi_{\substack{m'' \\ n''+1}} 
        ( \mathbf{\hat{e}}_{00} \cdot \mathbf{\hat{e}}_{m''n''} ) ( \mathbf{\hat{e}}_{m''n''} \cdot \mathbf{\hat{e}}_{0-1} ) \\ 
        & - \frac{\alpha_{\varepsilon} \omega_M}{2 \zeta_{\varepsilon} \bar{\varepsilon}} \sum_{(m'',n'')\in \mathcal{O}} 
        \xi_{\substack{-m''\\-n''}} \xi_{\substack{m''\\n''+1}} \sin (\theta_{00}-\theta_{m''n''}) \sin(\theta_{m''n''}-\theta_{0-1})
        \\
        \mathcal{H}_{00,-10}^{(C)} =& - \frac{\omega_M^3}{2\zeta_{\varepsilon}} \sum_{(m'',n'') \in \mathcal{O}} \mu_{m''n''} \xi_{\substack{-m'' \\ -n''}} \xi_{\substack{m''+1 \\ n''}} 
        ( \mathbf{\hat{e}}_{00} \cdot \mathbf{\hat{e}}_{m''n''} ) ( \mathbf{\hat{e}}_{m''n''} \cdot \mathbf{\hat{e}}_{-10} ) \\     
        & - \frac{\alpha_{\varepsilon} \omega_M}{2 \zeta_{\varepsilon} \bar{\varepsilon}} \sum_{(m'',n'')\in \mathcal{O}} 
        \xi_{\substack{-m''\\-n''}} \xi_{\substack{m''+1\\n''}} \sin (\theta_{00}-\theta_{m''n''}) \sin(\theta_{m''n''}-\theta_{-10})
        \\
        \mathcal{H}_{0-1,-1-1}^{(C)} =& - \frac{\omega_M^3}{2\zeta_{\varepsilon}} \sum_{(m'',n'') \in \mathcal{O}} \mu_{m''n''} \xi_{\substack{-m'' \\ -1-n''}} \xi_{\substack{m''+1 \\ n''+1}} 
        ( \mathbf{\hat{e}}_{0-1} \cdot \mathbf{\hat{e}}_{m''n''} ) ( \mathbf{\hat{e}}_{m''n''} \cdot \mathbf{\hat{e}}_{-1-1} ) \\ 
        & - \frac{\alpha_{\varepsilon} \omega_M}{2 \zeta_{\varepsilon} \bar{\varepsilon}} \sum_{(m'',n'')\in \mathcal{O}} 
        \xi_{\substack{-m''\\-1-n''}} \xi_{\substack{m''+1\\n''+1}} \sin (\theta_{0-1}-\theta_{m''n''}) \sin(\theta_{m''n''}-\theta_{-1-1})
        \\
        \mathcal{H}_{-10,-1-1}^{(C)} =& - \frac{\omega_M^3}{2\zeta_{\varepsilon}} \sum_{(m'',n'') \in \mathcal{O}} \mu_{m''n''} \xi_{\substack{-1-m'' \\ -n''}} \xi_{\substack{m''+1 \\ n''+1}} 
        ( \mathbf{\hat{e}}_{-10} \cdot \mathbf{\hat{e}}_{m''n''} ) ( \mathbf{\hat{e}}_{m''n''} \cdot \mathbf{\hat{e}}_{-1-1} )  
        \\
        & - \frac{\alpha_{\varepsilon} \omega_M}{2 \zeta_{\varepsilon} \bar{\varepsilon}} \sum_{(m'',n'')\in \mathcal{O}} 
        \xi_{\substack{-1-m''\\-n''}} \xi_{\substack{m''+1\\n''+1}} \sin (\theta_{-10}-\theta_{m''n''}) \sin(\theta_{m''n''}-\theta_{-1-1})
    \end{aligned}
\end{equation}

Recall that at the M point: $\theta_{00} = -\pi/4$, $\theta_{0-1} = -3\pi/4$, $\theta_{-10} = \pi/4$ and $\theta_{-1-1} = 3\pi/4$.
For the four matrix elements, we have:
\begin{equation}
\begin{aligned}
        \cos (\theta_{00}-\theta_{m''n''}) \cos (\theta_{m''n''}-\theta_{0-1}) 
        =& \sin (\theta_{00}-\theta_{m''n''}) \sin (\theta_{m''n''}-\theta_{0-1}) 
        =& -\frac{1}{2} \cos 2\theta_{m''n''} 
        \\
        \cos (\theta_{00}-\theta_{m''n''}) \cos (\theta_{m''n''}-\theta_{-10}) 
        =& \sin (\theta_{00}-\theta_{m''n''}) \sin (\theta_{m''n''}-\theta_{-10}) 
        =& \frac{1}{2} \cos 2\theta_{m''n''} 
        \\
        \cos (\theta_{0-1}-\theta_{m''n''}) \cos (\theta_{m''n''}-\theta_{-1-1}) 
        =& \sin (\theta_{0-1}-\theta_{m''n''}) \sin (\theta_{m''n''}-\theta_{-1-1}) 
        =& \frac{1}{2} \cos 2\theta_{m''n''} 
        \\
        \cos (\theta_{-10}-\theta_{m''n''}) \cos (\theta_{m''n''}-\theta_{-1-1}) 
        =& \sin (\theta_{-10}-\theta_{m''n''}) \sin (\theta_{m''n''}-\theta_{-1-1}) 
        =& -\frac{1}{2} \cos 2\theta_{m''n''} 
        \\
\end{aligned}
\end{equation}

By defining: 
\begin{equation}
    \Xi = \frac{\omega_M^3}{4\zeta_{\varepsilon}} \mu + \frac{\alpha_{\varepsilon}\omega_M}{4\zeta_{\varepsilon}\bar{\varepsilon}}
\end{equation}
we can write the four matrix elements as:
\begin{equation}
    \begin{aligned}
        \mathcal{H}_{00,0-1}^{(C)} =& \Xi  \sum_{(m'',n'')\in\mathcal{O}} 
        \xi_{\substack{-m''\\-n''}} \xi_{\substack{m''\\n''+1}} \cos 2 \theta_{m''n''}
        \\
        \mathcal{H}_{00,-10}^{(C)} =& -\Xi  \sum_{(m'',n'')\in\mathcal{O}} 
        \xi_{\substack{-m''\\-n''}} \xi_{\substack{m''+1\\n''}} \cos 2 \theta_{m''n''}
        \\
        \mathcal{H}_{0-1,-1-1}^{(C)} =& -\Xi  \sum_{(m'',n'')\in\mathcal{O}} 
        \xi_{\substack{-m''\\-1-n''}} \xi_{\substack{m''+1\\n''+1}} \cos 2 \theta_{m''n''}
        \\
        \mathcal{H}_{-10,-1-1}^{(C)} =& \Xi  \sum_{(m'',n'')\in\mathcal{O}} 
        \xi_{\substack{-1-m''\\-n''}} \xi_{\substack{m''+1\\n''+1}} \cos 2 \theta_{m''n''}
    \end{aligned}
    \label{eq:HighOrderCouplingPerpendicularPropagatingModes}
\end{equation}

Because $\theta_{10} + \theta_{01} = -\pi/2$, $\theta_{-11}=-\theta_{01}$, 
$\theta_{-20}=-\theta_{10}$, $\theta_{-2-1}=\theta_{10}+\pi$, $\theta_{-1-2}=\theta_{01}+\pi$, $\theta_{0-2}=\theta_{-11}+\pi$, $\theta_{1-1}=\theta_{-20}+\pi$.
Therefore, define $c = \cos 2 \theta_{10}$, we have:
\begin{equation}
    \begin{aligned}
        \cos 2\theta_{10} = \cos 2\theta_{1-1} = \cos 2 \theta_{-20} = \cos 2 \theta_{-2-1} =& c \\
        \cos 2\theta_{01} = \cos 2\theta_{-11} = \cos 2\theta_{-1-2} = \cos 2 \theta_{0-2} =& -c
    \end{aligned}
\end{equation} 

We explicitly expand \eqref{eq:HighOrderCouplingPerpendicularPropagatingModes} and obtain:
\begin{equation}
    \begin{aligned}
    \mathcal{H}_{00,0-1}^{(C)} =& \Xi c ( \xi_{\substack{-1\\0}} \xi_{\substack{1\\1}} - \xi_{\substack{0\\-1}} \xi_{\substack{0\\2}} 
    - \xi_{\substack{1 \\ -1}} \xi_{\substack{-1 \\ 2}} 
    + \xi_{\substack{2 \\ 0}} \xi_{\substack{-2 \\ 1}} 
    + \xi_{\substack{2 \\ 1}} \xi_{\substack{-2 \\ 0}}
    - \xi_{\substack{1 \\ 2}} \xi_{\substack{-1 \\ -1}}
    - \xi_{\substack{0 \\ 2}} \xi_{\substack{0 \\ -1}}
    + \xi_{\substack{-1 \\ 1}} \xi_{\substack{1 \\ 0}})
    \\ 
    \mathcal{H}_{00,-10}^{(C)} =& \Xi c ( -\xi_{\substack{-1\\0}} \xi_{\substack{2\\0}} + \xi_{\substack{0\\-1}} \xi_{\substack{1\\1}} 
    + \xi_{\substack{1 \\ -1}} \xi_{\substack{0 \\ 1}} 
    - \xi_{\substack{2 \\ 0}} \xi_{\substack{-1 \\ 0}} 
    - \xi_{\substack{2 \\ 1}} \xi_{\substack{-1 \\ -1}}
    + \xi_{\substack{1 \\ 2}} \xi_{\substack{0 \\ -2}}
    + \xi_{\substack{0 \\ 2}} \xi_{\substack{1 \\ -2}}
    - \xi_{\substack{-1 \\ 1}} \xi_{\substack{2 \\ -1}})
    \\
    \mathcal{H}_{0-1,-1-1}^{(C)} =& \Xi c ( -\xi_{\substack{-1\\-1}} \xi_{\substack{2\\1}} 
    + \xi_{\substack{0\\-2}} \xi_{\substack{1\\2}} 
    + \xi_{\substack{1 \\ -2}} \xi_{\substack{0 \\ 2}} 
    - \xi_{\substack{2 \\ -1}} \xi_{\substack{-1 \\ 1}} 
    - \xi_{\substack{2 \\ 0}} \xi_{\substack{-1 \\ 0}}
    + \xi_{\substack{1 \\ 1}} \xi_{\substack{0 \\ -1}}
    + \xi_{\substack{0 \\ 1}} \xi_{\substack{1 \\ -1}}
    - \xi_{\substack{-1 \\ 0}} \xi_{\substack{2 \\ 0}})
    \\
    \mathcal{H}_{-10,-1-1}^{(C)} =& \Xi c ( \xi_{\substack{-2\\0}} \xi_{\substack{2\\1}} 
    - \xi_{\substack{-1\\-1}} \xi_{\substack{1\\2}} 
    - \xi_{\substack{0 \\ -1}} \xi_{\substack{0 \\ 2}} 
    + \xi_{\substack{1 \\ 0}} \xi_{\substack{-1 \\ 1}} 
    + \xi_{\substack{1 \\ 1}} \xi_{\substack{-1 \\ 0}}
    - \xi_{\substack{0 \\ 2}} \xi_{\substack{0 \\ -1}}
    - \xi_{\substack{-1 \\ 2}} \xi_{\substack{1 \\ -1}}
    + \xi_{\substack{-2 \\ 1}} \xi_{\substack{2 \\ 0}})
    \end{aligned}
\end{equation}

Now we define:
\begin{equation}
\boxed{ 
    W = \Xi c ( \xi_{\substack{-1\\0}} \xi_{\substack{1\\1}} - \xi_{\substack{0\\-1}} \xi_{\substack{0\\2}} 
    - \xi_{\substack{1 \\ -1}} \xi_{\substack{-1 \\ 2}} 
    + \xi_{\substack{2 \\ 0}} \xi_{\substack{-2 \\ 1}} 
    + \xi_{\substack{2 \\ 1}} \xi_{\substack{-2 \\ 0}}
    - \xi_{\substack{1 \\ 2}} \xi_{\substack{-1 \\ -1}}
    - \xi_{\substack{0 \\ 2}} \xi_{\substack{0 \\ -1}}
    + \xi_{\substack{-1 \\ 1}} \xi_{\substack{1 \\ 0}})
}
\label{eq:DefinitionW}
\end{equation}

We recall that $\xi_{mn} = \xi_{nm}$ due to the mirror symmetry with respect to the mirror plane $x = y$.
We obtain:
\begin{equation}
    \mathcal{H}_{00,0-1}^{(C)} = \mathcal{H}_{00,-10}^{(C)} = \mathcal{H}_{0-1,-1-1}^{(C)} = \mathcal{H}_{-10,-1-1}^{(C)} = W
\end{equation}

The corresponding transpose components also take value $W$.

\paragraph{\textbf{Effective model for single layer:}} 
Knowing the matrix elements of $\mathcal{H}^{(A)}$, $\mathcal{H}^{(B)}$, $\mathcal{H}^{(C)}$, we establish the matrix elements for $\mathcal{H}_{\text{1layer}}$.

The diagonal elements are given by: 
\begin{equation}
    \begin{aligned}
        \mathcal{H}_{00,00}^{\text{1layer}} =& \omega_M + \frac{v}{\sqrt{2}} (k_x+k_y) + \frac{v}{\sqrt{2}K} (k_x^2 + k_y^2) + \omega_+^{(C)} \\ 
        \mathcal{H}_{0-1,0-1}^{\text{1layer}} =& \omega_M + \frac{v}{\sqrt{2}} (k_x-k_y) + \frac{v}{\sqrt{2}K} (k_x^2 + k_y^2) + \omega_-^{(C)} \\ 
        \mathcal{H}_{-10,-10}^{\text{1layer}} =& \omega_M + \frac{v}{\sqrt{2}} (-k_x+k_y) + \frac{v}{\sqrt{2}K} (k_x^2 + k_y^2) + \omega_-^{(C)} \\ 
        \mathcal{H}_{-1-1,-1-1}^{\text{1layer}} =& \omega_M + \frac{v}{\sqrt{2}} (-k_x-k_y) + \frac{v}{\sqrt{2}K} (k_x^2 + k_y^2) + \omega_+^{(C)} 
    \end{aligned}
\end{equation}

We define two parameters:
\begin{equation}
    \boxed{
    \begin{aligned}
        \omega_0 =& \omega_M + \frac{\omega_+^{(C)} + \omega_-^{(C)}}{2} 
        \\ 
        \eta =& \frac{\omega_+^{(C)} - \omega_-^{(C)}}{2}
    \end{aligned}
    }
\end{equation}

The diagonal elements are rewritten as:
\begin{equation}
    \begin{aligned}
        \mathcal{H}_{00,00}^{\text{1layer}} =& \omega_0 + \eta + \frac{v}{\sqrt{2}} (k_x+k_y) + \frac{v}{\sqrt{2}K} (k_x^2 + k_y^2) \\ 
        \mathcal{H}_{0-1,0-1}^{\text{1layer}} =& \omega_0 - \eta + \frac{v}{\sqrt{2}} (k_x-k_y) + \frac{v}{\sqrt{2}K} (k_x^2 + k_y^2) \\ 
        \mathcal{H}_{-10,-10}^{\text{1layer}} =& \omega_0 - \eta + \frac{v}{\sqrt{2}} (-k_x+k_y) + \frac{v}{\sqrt{2}K} (k_x^2 + k_y^2) \\ 
        \mathcal{H}_{-1-1,-1-1}^{\text{1layer}} =& \omega_0 + \eta + \frac{v}{\sqrt{2}} (-k_x-k_y) + \frac{v}{\sqrt{2}K} (k_x^2 + k_y^2) 
    \end{aligned}
\end{equation}

Next, we consider the coupling parameters between mode propagating in opposite directions:
\begin{equation}
    \begin{aligned}
        \mathcal{H}_{00,-1-1}^{\text{1layer}} = \mathcal{H}_{-1-1,00}^{\text{1layer}} =& U^{(B)} + U_+^{(C)} \\
        \mathcal{H}_{0-1,-10}^{\text{1layer}} = \mathcal{H}_{-10,0-1}^{\text{1layer}} =& U^{(B)} + U_-^{(C)}
    \end{aligned}
\end{equation}

We define two parameters:
\begin{equation}
    \boxed{
    \begin{aligned}
        U =& U^{(B)} + \frac{U_+^{(C)}+U_-^{(C)}}{2} \\
        \alpha =& \frac{U_+^{(C)}-U_-^{(C)}}{2U}
    \end{aligned}
    }
\end{equation}

With this notation, we rewrite the matrix elements as follows:
\begin{equation}
    \begin{aligned}
        \mathcal{H}_{00,-1-1}^{\text{1layer}} = \mathcal{H}_{-1-1,00}^{\text{1layer}} =& U(1+\alpha) \\
        \mathcal{H}_{0-1,-10}^{\text{1layer}} = \mathcal{H}_{-10,0-1}^{\text{1layer}} =& U(1-\alpha)
    \end{aligned}
\end{equation} 

The remaining eight matrix elements of $\mathcal{H}^{\text{1layer}}$ are coupling strength between modes propagating in opposite directions.
All of them have the same value equal to $W$ where $W$ is defined in \eqref{eq:DefinitionW}.

Overall, the matrix of the Hamiltonian of the single-layer is given by:
\begin{equation}
    \mathcal{H}^{\text{1layer}} = \omega_0 + \frac{v}{\sqrt{2}K} (k_x^2+k_y^2) +
    \begin{pmatrix}
        \eta + \frac{v}{\sqrt{2}} (k_x+k_y) & W & W & U(1+\alpha) \\
        W & -\eta + \frac{v}{\sqrt{2}} (k_x-k_y) & U(1-\alpha) & W \\ 
        W & U(1-\alpha) & -\eta + \frac{v}{\sqrt{2}} (-k_x+k_y) & W \\ 
        U(1+\alpha) & W & W & \eta + \frac{v}{\sqrt{2}} (-k_x-k_y) 
    \end{pmatrix}
\end{equation}

This Hamiltonian acts on a spinor $(A_{00}, A_{-10}, A_{0-1}, A_{-1-1})$ that contains the amplitude of the basic modes, as defined in \eqref{eq:BasicModesEfieldComponents}.
By solving for the eigenvalue problem of $\mathcal{H}^{\text{1layer}}$, we can recover the fields $E_x(\mathbf{r})$ and $E_y(\mathbf{r})$ of the eigenstates.

\paragraph{\textbf{The C\textsubscript{4} limit with square holes:}}
We end the discussion on the effective model of a single layer by investigating the C\textsubscript{4} limit where the rhombus hole becomes a square ($e = 0$).
In this case, apart from the two mirror symmetries about the two diagonals, the structure is invariant with respect to the two mirror symmetries about the $Oxz$ and $Oyz$ planes, that means the mirror symmetries:
\begin{equation}
    \begin{aligned}
        (x,y) & \rightarrow (-x,y) \\
        (x,y) & \rightarrow (x,-y) 
    \end{aligned}
\end{equation}

That leads to the following equality:
\begin{equation}
    \xi_{mn} = \xi_{-mn} = \xi_{m-n} = \xi_{-m-n}
\end{equation}

For the C\textsubscript{4} case, the eight Fourier components have the same values:
\begin{equation}
    \xi_{mn} = \xi_{-mn} = \xi_{m-n} = \xi_{-m-n} 
    = \xi_{nm} = \xi_{-nm} = \xi_{n-m} = \xi_{-n-m}
    \label{eq:C4limit}
\end{equation}
for all $(m,n) \in \mathbb{Z}^2$.
Consequently, $\xi_{12} = \xi_{-12}$ and $\xi_{1-1} = \xi_{11}$, implying that $\omega_+^{(C)} = \omega_-^{(C)}$.
This yields:
\begin{equation}
\boxed{ 
    \eta = 0 
} 
\end{equation}

Next, we explicitly develop $U_+^{(C)}$ and $U_-^{(C)}$ and obtain:
\begin{equation}
    \begin{aligned}
        U_+^{(C)} =& \frac{\omega_M^3}{4\zeta_{\varepsilon}} \mu \sum_{(m'',n'') \in \mathcal{O}} 
        \xi_{\substack{-m''\\-n''}} \xi_{\substack{m''+1\\n''+1}} [1 - \sin (2 \theta_{m''n''})] 
        - \frac{\alpha_{\varepsilon} \omega_M}{4\zeta_{\varepsilon} \bar{\varepsilon}} 
        \sum_{(m'',n'')\in\mathcal{O}} 
        \xi_{\substack{-m''\\-n''}} \xi_{\substack{m''+1\\n''+1}} [1 + \sin (2 \theta_{m''n''})] 
        \\
        =& \frac{\omega_M^3}{4\zeta_{\varepsilon}} \mu 
        [ \xi_{\substack{-1\\0}} \xi_{\substack{2\\1}} (1-s)
        + \xi_{\substack{0\\-1}} \xi_{\substack{1\\2}} (1-s)
        + \xi_{\substack{1\\-1}} \xi_{\substack{0\\2}} (1+s)
        + \xi_{\substack{2\\0}} \xi_{\substack{-1\\1}} (1+s)
        \\
        & + \xi_{\substack{2\\1}} \xi_{\substack{-1\\0}} (1-s)
        + \xi_{\substack{1\\2}} \xi_{\substack{0\\-1}} (1+s)
        + \xi_{\substack{0\\2}} \xi_{\substack{1\\-1}} (1+s)
        + \xi_{\substack{-1\\1}} \xi_{\substack{2\\0}} (1+s)
        ]
        \\
        & + \left( \frac{\omega_M^3}{4 \zeta_{\varepsilon}} \mu \leftrightarrow - \frac{\alpha_{\varepsilon} \omega_M}{4 \zeta_{\varepsilon} \bar{\varepsilon}} , s \leftrightarrow -s \right) 
    \end{aligned}
\end{equation}

\begin{equation}
    \begin{aligned}
        U_-^{(C)} =& \frac{\omega_M^3}{4\zeta_{\varepsilon}} \mu \sum_{(m'',n'') \in \mathcal{O}} 
        \xi_{\substack{-m''\\-1-n''}} \xi_{\substack{m''+1\\n''}} [1 + \sin (2 \theta_{m''n''})] 
        + \frac{\alpha_{\varepsilon} \omega_M}{4\zeta_{\varepsilon} \bar{\varepsilon}} 
        \sum_{(m'',n'')\in\mathcal{O}} 
        \xi_{\substack{-m''\\-1-n''}} \xi_{\substack{m''+1\\n''}} [1 - \sin (2 \theta_{m''n''})]
        \\
        =& \frac{\omega_M^3}{4\zeta_{\varepsilon}} \mu 
        [ \xi_{\substack{-1\\-1}} \xi_{\substack{2\\0}} (1+s)
        + \xi_{\substack{0\\-2}} \xi_{\substack{1\\1}} (1+s)
        + \xi_{\substack{1\\-2}} \xi_{\substack{0\\1}} (1-s)
        + \xi_{\substack{2\\-1}} \xi_{\substack{-1\\0}} (1-s)
        \\
        & + \xi_{\substack{2\\0}} \xi_{\substack{-1\\-1}} (1+s)
        + \xi_{\substack{1\\1}} \xi_{\substack{0\\-2}} (1+s)
        + \xi_{\substack{0\\1}} \xi_{\substack{1\\-2}} (1-s)
        + \xi_{\substack{-1\\0}} \xi_{\substack{2\\-1}} (1-s)
        ]
        \\
        & + \left( \frac{\omega_M^3}{4 \zeta_{\varepsilon}} \mu \leftrightarrow - \frac{\alpha_{\varepsilon} \omega_M}{4 \zeta_{\varepsilon} \bar{\varepsilon}} , s \leftrightarrow -s \right)
    \end{aligned}
\end{equation}

According to \eqref{eq:C4limit}, we see that:
\begin{equation}
    U_+^{(C)} = U_-^{(C)}
\end{equation}

This implies that:
\begin{equation}
\boxed{ 
    \alpha = 0 
}
\end{equation}

The previous equalities are only valid if \eqref{eq:C4limit} is satisfied, that means the rhombus hole becomes a square.
The Hamiltonian of a 2D slab with square hole is given by: 
\begin{equation}
    \mathcal{H}^{1layer} = \omega_0 + \frac{v}{\sqrt{2} K} (k_x^2 + k_y^2) + 
    \begin{pmatrix}
        \frac{v}{\sqrt{2}} (k_x+k_y) & W & W & U \\
        W & \frac{v}{\sqrt{2}} (k_x-k_y) & U & W \\ 
        W & U & \frac{v}{\sqrt{2}} (-k_x+k_y) & W \\ 
        U & W & W & \frac{v}{\sqrt{2}} (-k_x-k_y) 
    \end{pmatrix}
\end{equation}

\paragraph{\textbf{Summary}}
\begin{itemize}
    \item The state spinor has 4 components: $( A_{00}, A_{0-1}, A_{-10}, A_{-1-1})$. They correspond to the strengths of the fields $E_{mn}$. 
    This spinor is gauge-independent.
    \item In the C\textsubscript{4} limit, the anisotropy vanishes: $\eta = 0$ and $\alpha = 0$, the four basic modes are equivalent to each other
    \item The coupling strength $W$ between modes propagating perpendicularly and the difference in coupling constant $\eta$, $\alpha U$ between pairs of counter-propagating modes arise from coupling between basic modes and higher-order modes 
\end{itemize} 

\subsection{Effective Hamiltonian of bilayer}

\paragraph{\textbf{Geometry and dielectric profile:}}
In this section, we construct the effective theory of a bilayer of photonic crystal slab.
The two component slabs may \textit{not} be identical and are separated apart by a distance $d$ that we call the \textit{interlayer distance}.
To be precise, we consider the following geometry:
The two slabs have the same square lattice with lattice parameter $a$, so that the bilayer also has square lattice with the same lattice parameter $a$.
Consequently, both slabs and the bilayer share the same momentum space and the same reciprocal lattice vectors.
The upper layer has thickness $h_1$, hole size $b_1$, hole deformation parameter $e_1$, refractive index $n_1$ and its $z$-mirror plane is located at $z = z_1 = (d+h_1)/2$.
The lower layer has thickness $h_2$, hole size $b_2$, hole deformation parameter $e_2$, refractive index $n_2$ with the $z$-mirror plane located at $z = z_2 = -(d+h_2)/2$.
The lower slab is shifted by a vector $\boldsymbol{\delta} = (\delta_x, \delta_y)$ with respect to the upper slab.
The dielectric profiles of the two slabs are given by the following Fourier series: 
\begin{equation}
    \varepsilon_l (\mathbf{r}) = \bar{\varepsilon}_l(z-z_l) + \sum_{(m,n)\ne(0,0)} \xi_{mn}^l f_l(z-z_l) e^{iK(mx+ny)}
\end{equation}

Here $l = 1$ means the \textit{upper layer} and $l = 2$ means the \textit{lower layer}. 
The functions $\bar{\varepsilon}_l(z)$ are given by:
\begin{equation}
    \bar{\varepsilon}_l(z) = 
    \begin{cases}
        \bar{\varepsilon}_l &\text{ if } |z|<\frac{h_l}{2} \\
        \frac{1}{2} \varepsilon_{env} & \text{ otherwise }
    \end{cases}
\end{equation}
and the function $f_l(z)$ is defined as: 
\begin{equation}
    f_l(z) = 
    \begin{cases}
        1 &\text{ if } |z| < \frac{h_l}{2} \\
        0 &\text{ otherwise }
    \end{cases}
\end{equation}

It allows to write the total dielectric profile as the sum of the two Fourier series of the two slabs:
\begin{equation}
    \varepsilon (\mathbf{r}) = \varepsilon_1(x,y,z) + \varepsilon_2(x-\delta_x,y-\delta_y,z) 
    = \bar{\varepsilon}(z) + \sum_{(m,n)\ne(0,0)} \xi_{mn}(z) e^{iK(mx+ny)}
\end{equation}
where 
\begin{equation}
    \bar{\varepsilon}(z) = \bar{\varepsilon}_1(z-z_1) + \bar{\varepsilon}_2(z-z_2) = 
    \begin{cases}
        \frac{1}{2}\varepsilon_{env} + \bar{\varepsilon}_1 & \text{ if } |z-z_1| < \frac{h_1}{2} \\
        \frac{1}{2}\varepsilon_{env} + \bar{\varepsilon}_2 & \text{ if } |z-z_2| < \frac{h_2}{2} \\
        \varepsilon_{env} & \text{ otherwise }
    \end{cases}
    \label{eq:barepsilonbilayer}
\end{equation}
and 
\begin{equation}
    \xi_{mn}(z) = \xi_{mn}^1 f_1(z-z_1) + \xi_{mn}^2 f_2(z-z_2) e^{-iK(m\delta_x+n\delta_y)}
    \label{eq:xibilayer}
\end{equation}

The exponential in the second term is due to the shift of the lower slab with respect to the upper slab.
This choice of the dielectric profile modifies the Fourier coefficients $\xi_{m,n}^l$ compared to the ones in section \ref{sec:SingleLayerHamiltonian}.
However, it keeps the symmetry of the coefficients $\xi_{mn}$ via exchange of index and the symmetry of the envelope wavefunctions $\Theta_l$ with respect to the mirror planes of the slabs.
Consequently, the monolayer blocks in the bilayer Hamiltonian keep the same form as the one derived in section \ref{sec:SingleLayerHamiltonian}.
\footnote{One can also define the dielectric functions to be:
\begin{equation}
    \bar{\varepsilon}_1(z) = 
    \begin{cases}
        \bar{\varepsilon}_{1/2} &\text{ if } |z-z_{1/2}|<\frac{h_{1/2}}{2} \\
        0 &\text{ if } |z-z_{2/1}|<\frac{h_{2/1}}{2} \\ 
        \frac{\varepsilon_{env}}{2} &\text{ otherwise }
    \end{cases}
\end{equation}
so that the presence of an additional slab does not influence the dielectric Fourier series of the first one.
Although this definition gives the same total dielectric profile of the bilayer, it modifies the envelop wavefunction to be asymmetric with respect to the mirror plane of the monolayer slabs. 
}

Although our multilayer structure is placed inside the environment and has a layer made of the environment between the two slabs, we restrict to consider the TE guided modes in both slabs. 
Their wave profiles are not strictly confined inside the slabs, but decay outside.
This allows \textbf{\textit{the guided modes from different slabs to couple via evanescent waves}}.

\paragraph{\textbf{Electric field profile:}}
The component $j$ ($j = x/y$) of the electric field in the upper slab is given by: 
\begin{equation}
    E_{j \boldsymbol{\kappa}}^1 (\mathbf{r}) = e^{i(\kappa_x x + \kappa_y y)} \sum_{m,n\in\mathbb{Z}} E_{jmn}^1(z-z_1) e^{iK(mx+ny)}
\end{equation}

Here we use the same symbols as in the section on the single layer.
That means $E_{jmn}^1(z)$ is the electric field component if the mirror plane of the slab is at $z = 0$.
Here the center of the slab is shifted from $z$ to $z-z_1$.
The component $j$ ($j = x/y$) of the electric field in the lower slab is given by:
\begin{equation}
\begin{aligned}
    E_{j\boldsymbol{\kappa}}^2(\mathbf{r}) =& e^{i(\kappa_x (x-\delta_x) + \kappa_y (y-\delta_y))} \sum_{m,n\in\mathbb{Z}} E_{jmn}^2(z-z_2) e^{iK[m(x-\delta_x)+n(y-\delta_y)]} 
    \\
    \approx & e^{i(\kappa_xx+\kappa_yy)} e^{-iK(\delta_x+\delta_y)/2} \sum_{m,n\in\mathbb{Z}} E_{jmn}^2(z-z_2) e^{iK[m(x-\delta_x)+n(y-\delta_y)]} 
\end{aligned}
    \label{eq:Ejbilayer}
\end{equation}

Here because we are in the vicinity of the M-point, $\kappa_x \approx K/2$ and $\kappa_y \approx K/2$.
The total electric field in all space is the superposition of $\mathbf{E}_{\boldsymbol{\kappa}}^1$ and $\mathbf{E}_{\boldsymbol{\kappa}}^2$ and has Bloch form.
Its $j$ component is given by: 
\begin{equation}
\begin{aligned}
    E_{j\boldsymbol{\kappa}}(\mathbf{r}) =& e^{i(\kappa_xx+\kappa_yy)} \sum_{m,n\in\mathbb{Z}} E_{jmn}(z) e^{iK(mx+ny)} 
    \\
    =& e^{i(\kappa_xx+\kappa_yy)} 
    \sum_{m,n\in\mathbb{Z}} \left[ E_{jmn}^1(z-z_1) + E_{jmn}^2(z-z_2) e^{-iK[(m+\frac{1}{2})\delta_x+(n+\frac{1}{2})\delta_y]} \right] e^{iK(mx+ny)}
\end{aligned}
\end{equation}

The electric field profile $\mathbf{E}_{\boldsymbol{\kappa}}(\mathbf{r})$ satisfies the Maxwell equation in \eqref{eq:MasterEquationE1}.
Consequently, the system of equations \eqref{eq:MasterEquationE2A}, \eqref{eq:MasterEquationE2B} and \eqref{eq:MasterEquationE2C} are satisfied by the fields $E_{jmn}(\mathbf{r})$ and the Fourier components $\bar{\varepsilon} (z)$, $\xi_{\substack{m-m'\\n-n'}}(z)$, which are given by \eqref{eq:Ejbilayer}, \eqref{eq:barepsilonbilayer} and \eqref{eq:xibilayer}, respectively.

For the slab $l$ ($l = 1,2$), we define the intralayer electric fields $E_{+mn}^l$ and $E_{-mn}^l$ as follows:
\begin{equation}
    \begin{aligned}
        E_{+mn}^l(z) =& k_{mx} E_{xmn}^l(z) + k_{ny} E_{ymn}^l(z) \\ 
        E_{-mn}^l(z) =& k_{ny} E_{xmn}^l(z) - k_{mx} E_{ymn}^l(z)
    \end{aligned}
\end{equation}
and the total electric fields $E_{+mn}$ and $E_{-mn}$ of the bilayer:
\begin{equation}
    \begin{aligned}
        E_{+mn}(z) =& E_{+mn}^1(z-z_1) + E_{+mn}^2(z-z_2) e^{-iK[(m+\frac{1}{2})\delta_x+(n+\frac{1}{2})\delta_y]} \\
        E_{-mn}(z) =& E_{-mn}^1(z-z_1) + E_{-mn}^2(z-z_2) e^{-iK[(m+\frac{1}{2})\delta_x+(n+\frac{1}{2})\delta_y]} 
    \end{aligned}
\end{equation}

One can check that the bilayer electric fields satisfy the same relation as the single layer electric fields:
\begin{equation}
    \begin{aligned}
        E_{+mn}(z) =& k_{mx} E_{xmn}(z) + k_{ny} E_{ymn}(z) \\ 
        E_{-mn}(z) =& k_{ny} E_{xmn}(z) - k_{mx} E_{ymn}(z)
    \end{aligned}
\end{equation} 

\paragraph{\textbf{Basic and higher-order modes:}}
Similar to the single layer slab, the basic modes are the ones whose $(m,n) \in \lbrace (0,0), (-1,0), (-1,-1), (0,-1)\rbrace$.
We have in total eight basic modes, four come from the upper slab and four come from the lower slab.
We also denote the set of basic modes as $\mathcal{B}$ and the set of higher-order modes as $\mathcal{O}$.
The modes with momentum $\mathbf{k} = (k_{mx},k_{ny})$ in both layers also have the same angle $\theta_{mn}$ but may have different amplitudes and envelope wavefunctions.
To be precise, the electric field of the basic modes $(m,n)$ in the upper layer is given by: 
\begin{equation}
    \begin{aligned}
        E_{xmn}^1(z) =& \cos \theta_{mn} A_{mn}^1 \Theta_1(z)
        \\
        E_{ymn}^1(z) =& \sin \theta_{mn} A_{mn}^1 \Theta_1(z)
    \end{aligned}
\end{equation}
and the electric field of the basic modes $(m,n)$ in the lower layer is given by:
\begin{equation}
    \begin{aligned}
        E_{xmn}^2(z) =& \cos \theta_{mn} A_{mn}^2 \Theta_2(z)
        \\
        E_{ymn}^2(z) =& \sin \theta_{mn} A_{mn}^2 \Theta_2(z)
    \end{aligned}
\end{equation}

Here the expressions are given for the case where the mirror planes of the slabs are located in the plane $z = 0$.
The envelope wavefunctions $\Theta_1(z)$ and $\Theta_2(z)$ may not be identical, unless the two slabs are identical, that means $h_1 = h_2$, $b_1 = b_2$ and $e_1 = e_2$.

The envelope wavefunction of the slab $l$ ($l = 1,2$) satisfies the Helmholtz equation:
\begin{equation}
    \left[ \frac{d^2}{dz^2} + \bar{\varepsilon}_l(z) \left(\frac{\omega_l}{c} \right)^2 - \mathbf{g}^2 \right] \Theta_l(z) = 0
    \label{eq:HelmholtzEquationBilayer}
\end{equation}

For the single-layer slab $l$ ($l = 1,2$), we have:
\begin{equation}
    E_{-mn}^l(z) = k_{mn} A_{mn}^l \Theta_l(z) = \bar{A}_{mn}^l \Theta_l(z)
\end{equation}

For an arbitrary mode $(m,n)$ (can be either basic mode or higher-order mode) and a basic mode $(m',n')$ in the layer $l$ ($l = 1,2$), we also have:
\begin{equation}
    k_{ny} E_{xm'n'}^l(z) - k_{mx} E_{ym'n'}^l(z) = k_{mn} \cos (\theta_{mn}-\theta_{m'n'}) A_{m'n'}^l \Theta_l(z)
    = k_{mn} (\mathbf{\hat{e}}_{mn} \cdot \mathbf{\hat{e}}_{m'n'}) A_{m'n'}^l \Theta_l(z)
\end{equation}

\paragraph{\textbf{Solve the Maxwell equation:}}
The equations \eqref{eq:EquationEplusmn} and \eqref{eq:HelmholtzEquationEminusmn} also apply for the \textbf{\textit{bilayer}} electric fields $E_{+mn}$ and $E_{-mn}$.
We extract $\frac{d^2}{dz^2} \Theta_l(z)$ from the Helmholtz equation \eqref{eq:HelmholtzEquationBilayer} and substitute into \eqref{eq:HelmholtzEquationEminusmn}, we obtain:
\begin{equation}
    \begin{aligned}
        & [ (\omega^2 - \omega_1^2) \bar{\varepsilon}_1(z-z_1) + \omega^2 \bar{\varepsilon}_2(z-z_2) - c^2(\mathbf{k}_{mn}^2 - \mathbf{g}^2) ] \bar{A}_{mn}^1 \Theta_1(z-z_1)  
        \\
        &+ [ (\omega^2 - \omega_2^2) \bar{\varepsilon}_2(z-z_2) + \omega^2 \bar{\varepsilon}_1(z-z_1) - c^2(\mathbf{k}_{mn}^2 - \mathbf{g}^2) ] \bar{A}_{mn}^2 \Theta_2(z-z_2) e^{-iK[(m+\frac{1}{2})\delta_x+(n+\frac{1}{2})\delta_y]}  
        \\
        =& - \sum_{\substack{m'n' \\ (m',n')\ne(m,n)}} \omega^2 
        [ \xi_{\substack{m-m'\\n-n'}}^1 f_1(z-z_1) + \xi_{\substack{m-m'\\n-n'}}^2 f_2(z-z_2) e^{-iK[(m-m')\delta_x+(n-n')\delta_y]} ]
        [k_{ny}E_{xm'n'}(z)-k_{mx}
        E_{ym'n'}(z)]
    \end{aligned}
    \label{eq:BilayerMasterEquation1}
\end{equation}

Our objective is to construct an $8\times 8$ Hamiltonian for the bilayer whose diagonal blocks represent the \textit{intralayer couplings} and whose off-diagonal blocks represent the \textit{interlayer coupling}.
For the intralayer blocks, we omit coupling between basic modes via modes in the other slab, so these diagonal blocks have the same form as the single layer Hamiltonian we derived in section \ref{sec:SingleLayerHamiltonian}.
For the interlayer coupling blocks, because the two slabs are separated apart, we do not consider second-order coupling between basic modes via higher-order modes. 
For this reason, we omit the coupling between basic modes and higher-order modes, and the coupling between basic modes themselves via higher-order modes.
The equation \eqref{eq:BilayerMasterEquation1} becomes:
\begin{equation}
    \begin{aligned}
        & [ (\omega^2 - \omega_1^2) \bar{\varepsilon}_1(z-z_1) + \omega^2 \bar{\varepsilon}_2(z-z_2) - c^2 (\mathbf{k}_{mn}^2 - \mathbf{g}^2) ] A_{mn}^1 \Theta_1(z-z_1)  
        \\
        &+ [ (\omega^2 - \omega_2^2) \bar{\varepsilon}_2(z-z_2) + \omega^2 \bar{\varepsilon}_1(z-z_1) - c^2 (\mathbf{k}_{mn}^2 - \mathbf{g}^2) ] A_{mn}^2 \Theta_2(z-z_2) e^{-iK[(m+\frac{1}{2})\delta_x+(n+\frac{1}{2})\delta_y]}  
        \\
        =& - \sum_{\substack{(m',n') \in \mathcal{B} \\ (m',n')\ne(m,n)}} \omega^2 
        [ \xi_{\substack{m-m'\\n-n'}}^1 f_1(z-z_1) + \xi_{\substack{m-m'\\n-n'}}^2 f_2(z-z_2) e^{-iK[(m-m')\delta_x+(n-n')\delta_y]} ]
        \cos (\theta_{mn} - \theta_{m'n'}) \times 
        \\ 
        & \times [A_{m'n'}^1 \Theta_1(z-z_1)+A_{m'n'}^2\Theta_2(z-z_2) e^{-iK[(m'+\frac{1}{2})\delta_x+(n'+\frac{1}{2})\delta_y]}] 
    \end{aligned}
    \label{eq:BilayerMasterEquation2}
\end{equation} 

Here the factor $k_{mn}$ on both sides cancel each other.
By multiplying both sides of \eqref{eq:BilayerMasterEquation2} by $\Theta_1(z-z_1)^*$ and then take integral over $dz$ from $-\infty$ to $+\infty$, we obtain: 
\begin{equation}
    \begin{aligned}
        & [(\omega^2-\omega_1^2) \zeta_{111} + \omega^2 \zeta_{121} - c^2 (\mathbf{k}_{mn}^2 - \mathbf{g}^2) ] A_{mn}^1 
        + [(\omega^2-\omega_2^2) \zeta_{122} + \omega^2 \zeta_{112} - c^2(\mathbf{k}_{mn}^2 - \mathbf{g}^2) \gamma_{12}] A_{mn}^2 e^{-iK[(m+\frac{1}{2})\delta_x+(n+\frac{1}{2})\delta_y]}
        \\ 
        =& - \sum_{\substack{(m',n')\in\mathcal{B} \\ (m',n')\ne(m,n)}} \omega^2 \cos (\theta_{mn}-\theta_{m'n'}) 
        \bigg\lbrace [ \xi_{\substack{m-m'\\n-n'}}^1 \alpha_{111} + \xi_{\substack{m-m'\\n-n'}}^2 \alpha_{121} e^{-iK[(m-m')\delta_x+(n-n')\delta_y]}  ] A_{m'n'}^1 \\
        & + [ \xi_{\substack{m-m'\\n-n'}}^1 \alpha_{112} e^{-iK[(m'+\frac{1}{2})\delta_x+(n'+\frac{1}{2})\delta_y]} + \xi_{\substack{m-m'\\n-n'}}^2 \alpha_{122} e^{-iK[(m+\frac{1}{2})\delta_x+(n+\frac{1}{2})\delta_y]} ] A_{m'n'}^2  \bigg\rbrace 
    \end{aligned}
\end{equation}

Here the integrals are denoted by the symbols:
\begin{equation}
    \begin{aligned}
        \zeta_{abc} =& \int_{-\infty}^{+\infty} \Theta_a^*(z-z_a) \bar{\varepsilon}_b(z-z_b) \Theta_c(z-z_c) dz
        \\ 
        \alpha_{abc} =& \int_{-\infty}^{+\infty} \Theta_a^*(z-z_a) f_b(z-z_b) \Theta_c(z-z_c) dz
        \\
        \gamma_{ab} =& \int_{-\infty}^{+\infty} \Theta_a^*(z-z_a) \Theta_b(z-z_b) dz
    \end{aligned}
    \label{eq:BilayerMasterEquation3}
\end{equation}
for $a,b,c = 1,2$.
Similarly, by multiplying both sides of \eqref{eq:BilayerMasterEquation2} by $\Theta_2^*(z-z_2)$ and then take integration over $dz$ from $-\infty$ to $+\infty$, we obtain:
\begin{equation}
    \begin{aligned}
        & [(\omega^2 - \omega_1^2)\zeta_{211} + \omega^2 \zeta_{221} - c^2(\mathbf{k}_{mn}^2 - \mathbf{g}^2) \gamma_{21}] A_{mn}^1 e^{iK[(m+\frac{1}{2})\delta_x+(n+\frac{1}{2})\delta_y]} 
        + [(\omega^2-\omega_2^2)\zeta_{222}+\omega^2\zeta_{212} - c^2(\mathbf{k}_{mn}^2-\mathbf{g}^2)] A_{mn}^2 
        \\
        =& - \sum_{\substack{(m',n')\in\mathcal{B} \\ (m',n')\ne(m,n)}} \omega^2 \cos (\theta_{mn}-\theta_{m'n'}) 
        \bigg\lbrace \bigg[ \xi_{\substack{m-m'\\n-n'}}^1 \alpha_{211} e^{iK[(m+\frac{1}{2})\delta_x+(n+\frac{1}{2})\delta_y]} + \xi_{\substack{m-m'\\n-n'}}^2 \alpha_{221} e^{-iK[(m'+\frac{1}{2})\delta_x+(n'+\frac{1}{2})\delta_y]} \bigg] A_{m'n'}^1 
        \\ 
        &+ \bigg[ \xi_{\substack{m-m'\\n-n'}}^1 \alpha_{212} e^{iK[(m-m')\delta_x + (n-n')\delta_y]} + \xi_{\substack{m-m'\\n-n'}}^2 \alpha_{222} \bigg] A_{m'n'}^2 
        \bigg\rbrace 
    \end{aligned}
    \label{eq:BilayerMasterEquation4}
\end{equation}

We consider the regime where the two layers do not differ very significantly $\omega_1 \approx \omega_2$ so that we can examine the frequency $\omega$ in the vicinity of both $\omega_1$ and $\omega_2$. 
Because $\omega \approx \omega_1$, we can approximate $\omega^2 - \omega_1^2 \approx 2 \omega_1 (\omega - \omega_1)$. 
Equation \eqref{eq:BilayerMasterEquation3} becomes:
\begin{equation}
    \begin{aligned}
        & \bigg[ (\omega-\omega_1) + \frac{\omega_1}{2} \frac{\zeta_{121}}{\zeta_{111}} - \frac{c^2 (\mathbf{k}_{mn}^2-\mathbf{g}^2)}{2\omega_1 \zeta_{111}} \bigg] A_{mn}^1 
        + \bigg[ \frac{\omega_1^2-\omega_2^2}{2\omega_1} \frac{\zeta_{122}}{\zeta_{111}} + \frac{\omega_1}{2} \frac{\zeta_{112}}{\zeta_{111}} - \frac{c^2 (\mathbf{k}_{mn}^2-\mathbf{g}^2)}{2\omega_1} \frac{\gamma_{12}}{\zeta_{111}}\bigg] A_{mn}^2 e^{-iK[(m+\frac{1}{2})\delta_x+(n+\frac{1}{2})\delta_y]} 
        \\
        =& - \sum_{\substack{(m',n')\in\mathcal{B} \\\ (m',n')\ne(m,n)}} \frac{\omega_1 \alpha_{111}}{2 \zeta_{111}} \cos (\theta_{mn} - \theta_{m'n'}) 
        \bigg \lbrace \bigg[ \xi_{\substack{m-m'\\n-n'}}^1 + \xi_{\substack{m-m'\\n-n'}}^2 \frac{\alpha_{121}}{\alpha_{111}} e^{-iK[(m-m')\delta_x+(n-n')\delta_y]} \bigg] A_{m'n'}^1 
        \\
        & + \bigg[ \xi_{\substack{m-m'\\n-n'}}^1 \frac{\alpha_{112}}{\alpha_{111}} e^{-iK[(m'+\frac{1}{2})\delta_x+(n'+\frac{1}{2})\delta_y]} + \xi_{\substack{m-m'\\n-n'}}^2 \frac{\alpha_{122}}{\alpha_{111}} e^{-iK[(m+\frac{1}{2})\delta_x+(n+\frac{1}{2})\delta_y]} \bigg] A_{m'n'}^2
        \bigg \rbrace 
    \end{aligned}
    \label{eq:BilayerMasterEquation4p}
\end{equation}

Because the envelope wavefunctions $\Theta_1(z-z_1)$ and $\Theta_2(z-z_2)$ are strongly localized inside the regions $|z-z_1|<h_1/2$ and $|z-z_2|<h_2/2$ and exponentially decay outside these regions, we can estimate the $\alpha$ integrals as follows:
$\alpha_{121} \ll \alpha_{112} \approx \alpha_{122} \ll \alpha_{111}$.
Also, $\zeta_{112} \approx \zeta_{122} \ll \zeta_{111}$.
In addition, $(\omega_1^2-\omega_2^2)/(2\omega_1) \approx \omega_1 - \omega_2 \ll \omega_1/2$.
For this reason, the term $(\omega_1^2-\omega_2^2) \zeta_{122}/(2\omega_1 \zeta_{111})$ is negligible compared to the other terms in the left-hand side of \eqref{eq:BilayerMasterEquation4p}.
Therefore, we approximate the previous equation as follows:
\begin{equation}
    \begin{aligned}
        & \bigg[ \omega - \omega_1 + \frac{\omega_1 \zeta_{121}}{2 \zeta_{111}} - \frac{c^2(\mathbf{k}_{mn}^2-\mathbf{g}^2)}{2\omega_1\zeta_{111}} \bigg] A_{mn}^1 + 
        \bigg[ 
        \frac{\omega_1}{2} \frac{\zeta_{112}}{\zeta_{111}} - \frac{c^2(\mathbf{k}_{mn}^2-\mathbf{g}^2)}{2\omega_1 \zeta_{111}} \gamma_{12} \bigg] A_{mn}^2 e^{-iK[(m+\frac{1}{2})\delta_x+(n+\frac{1}{2})\delta_y]}
        \\
        =& - \sum_{\substack{(m',n')\in\mathcal{B} \\ (m',n')\ne(m,n)}} \frac{\omega_1 \alpha_{111}}{2 \zeta_{111}} \xi_{\substack{m-m'\\n-n'}}^1
        \cos (\theta_{mn}-\theta_{m'n'})  A_{m'n'}^1
    \end{aligned}
    \label{eq:BilayerMasterEquation5}
\end{equation}

Similarly, equation \eqref{eq:BilayerMasterEquation4} leads to: 
\begin{equation}
    \begin{aligned}
        & \bigg[ \frac{\omega_2}{2} \frac{\zeta_{221}}{\zeta_{222}} - \frac{c^2 (\mathbf{k}_{mn}^2-\mathbf{g}^2)}{2\omega_2\zeta_{222}} \gamma_{21} \bigg] A_{mn}^1  e^{i[K(m+\frac{1}{2})\delta_x+(n+\frac{1}{2})\delta_y]} 
        + \bigg[ \omega - \omega_2 + \frac{\omega_2}{2} \frac{\zeta_{212}}{\zeta_{222}} - \frac{c^2 (\mathbf{k}_{mn}^2-\mathbf{g}^2)}{2\omega_2\zeta_{222}} \bigg] A_{mn}^2  
        \\
        =& - \sum_{\substack{(m',n')\in\mathcal{B} \\ (m',n')\ne(m,n)}} \frac{\omega_2 \alpha_{222}}{2 \zeta_{222}} \xi_{\substack{m-m'\\n-n'}}^2 \cos (\theta_{mn}-\theta_{m'n'}) A_{m'n'}^2
    \end{aligned}
    \label{eq:BilayerMasterEquation6}
\end{equation}

Here we recover the direct coupling constants of the monolayer in section \ref{sec:SingleLayerHamiltonian}.
$\zeta_{111}$ and $\zeta_{222}$ play the role of $\zeta_{\varepsilon}$. 
Correspondingly, $\alpha_{111}$ and $\alpha_{222}$ play the role of $\alpha_{\varepsilon}$.
We define the layer group velocities: 
\begin{equation}
    \begin{aligned}
        v_1 =& \frac{\sqrt{2}K c^2}{2\omega_1\zeta_{111}} \\
        v_2 =& \frac{\sqrt{2}K c^2}{2\omega_2 \zeta_{222}}
    \end{aligned}
\end{equation}

Next, we define the energy:
\begin{equation}
    \begin{aligned}
        \omega_1^{(0)} =& \omega_1 \left( 1 - \frac{\zeta_{121}}{2\zeta_{111}} \right)
        \\
        \omega_2^{(0)} =& \omega_2 \left( 1 - \frac{\zeta_{212}}{2\zeta_{222}} \right)
    \end{aligned}
\end{equation}

From the definition of $\zeta_{abc}$, we see that $\zeta_{121}$ and $\zeta_{212}$ are real. 
This implies that $\omega_1^{(0)}$ and $\omega_2^{(0)}$ are real.
For a sake of simplicity, hereafter and in the main text, we refer $\omega_1^{(0)}$ and $\omega_2^{(0)}$ as $\omega_1$ and $\omega_2$, respectively.
We can regroup \eqref{eq:BilayerMasterEquation5} and \eqref{eq:BilayerMasterEquation6} into a $8\times 8$ Hamiltonian of four $4 \times 4$ blocks:
\begin{equation}
    \mathcal{H}_{bilayer} = 
    \begin{pmatrix}
        \Delta_1 & \Omega_{12} \\
        \Omega_{21} & \Delta_2
    \end{pmatrix}
\end{equation}

 By taking into account coupling between intralayer basic modes via higher-order modes, the diagonal blocks have the same form as in section \ref{sec:SingleLayerHamiltonian}.
 The two off-diagonal blocks are \textit{not} accurately Hermitian conjugate to each other.
 There matrix elements depend on both the genuine momenta $k_x$, $k_y$ and the relative displacement $\delta_x$, $\delta_y$ between the two slabs.
 Although the integrals $\zeta_{112}$, $\zeta_{221}$, $\gamma_{12}$ and $\gamma_{21}$ are all \textit{complex}, we see that they all exponentially decay as functions of the interlayer distance $d$, so all those matrix elements have a factor $e^{-d/d_0}$.
 In the limit where the two slabs are slightly different from each other, $\Omega_{12}$ and $\Omega_{21}$ become approximately Hermitian conjugate of each other.
 For a sake of simplicity, we consider them to have the following form:
 \begin{equation}
 \begin{aligned}
     \Omega_{12} =& \Omega \\ 
     \Omega_{21} =& \Omega^{\dagger}
\end{aligned}
 \end{equation}
 where 

 \begin{equation}
 \begin{aligned}
     \Omega =& e^{-d/d_0} \times 
     \text{diag} 
     \bigg \lbrace 
        \bigg[ V + \frac{\beta}{\sqrt{2}} (k_x+k_y) + \frac{\beta}{\sqrt{2}K} (k_x^2+k_y^2) \bigg] e^{-iK(\delta_x+\delta_y)/2}
        , 
        \bigg[ V + \frac{\beta}{\sqrt{2}} (k_x-k_y) + \frac{\beta}{\sqrt{2}K} (k_x^2+k_y^2) \bigg] e^{-iK(\delta_x-\delta_y)/2}
        ,
        \\
        & \bigg[ V + \frac{\beta}{\sqrt{2}} (-k_x+k_y) + \frac{\beta}{\sqrt{2}K} (k_x^2+k_y^2) \bigg] e^{-iK(-\delta_x+\delta_y)/2}
        ,
        \bigg[ V + \frac{\beta}{\sqrt{2}} (-k_x-k_y) + \frac{\beta}{\sqrt{2}K} (k_x^2+k_y^2) \bigg] e^{-iK(-\delta_x-\delta_y)/2}
     \bigg \rbrace
\end{aligned}
 \end{equation} 

\subsection{Synthetic momenta} 
We consider the case where the lower layer is shifted along the diagonal of the upper layer.
We define the shift $\delta$ along the diagonal as $\delta = \sqrt{2} \delta_x = \sqrt{2} \delta_y$.
Here $0 \le \delta_x, \delta_y \le a$, so $0 \le \delta \le \sqrt{2}a$.
We define the \textit{synthetic momentum} $q$ as follows:
\begin{equation}
\boxed{ 
    q =\frac{\delta}{\sqrt{2}a} - \frac{1}{2}   
}
\end{equation}

We notice that as $0 \le \delta_x, \delta_y \le a$, the synthetic momentum $q$ varies in the range $-\frac{1}{2} \le q \le \frac{1}{2}$.
Along the line $k_x = k_y$, we define the genuine momentum:
\begin{equation}
\boxed{ 
    k = \frac{k_x+k_y}{\sqrt{2}}
    }
\end{equation} 

For $-\pi / a \le k_x, k_y \le \pi / a$, the synthetic momentum lies on the range: $-\sqrt{2} \pi / a \le k \le \sqrt{2} \pi /a$.
The single-layer Hamiltonian becomes:
\begin{equation}
    \mathcal{H}^{1layer} = \omega_0 + \frac{v}{\sqrt{2}K} k^2 + 
    \begin{pmatrix}
        \eta + vk & W & W & U(1+\alpha) \\ 
        W & - \eta & U(1-\alpha) & W \\ 
        W & U(1-\alpha) & - \eta & W \\ 
        U(1+\alpha) & W & W & \eta -vk 
    \end{pmatrix}
\end{equation}

The interlayer coupling Hamiltonian becomes:
\begin{equation}
     \Omega = e^{-d/d_0} \times 
     \text{diag} 
     \bigg \lbrace 
        -\bigg( V + \beta k + \frac{\beta}{\sqrt{2} K} k^2 \bigg) e^{-i 2 \pi q}
        , 
        V + \frac{\beta}{\sqrt{2} K} k^2 
        ,
        V + \frac{\beta}{\sqrt{2} K} k^2
        ,
        -\bigg( V - \beta k + \frac{\beta}{\sqrt{2} K} k^2 \bigg) e^{i 2 \pi q}
     \bigg \rbrace
 \end{equation}

\subsection{Final expressions for monolayer and bilayer}

Finally, we change the energy, velocity, momentum to dimensionless unit $\omega \rightarrow \omega a /(2 \pi c)$, $v \rightarrow v/c$, $k \rightarrow ka/(2\pi)$. 
The coupling constants also change in the same way as $\omega$.
The single-layer Hamiltonian is rewritten as:
\begin{equation}
    \mathcal{H}^{1layer} (\omega_0,\eta,v,U,W,\alpha) = \omega_0 + \frac{v}{\sqrt{2}} k^2 + 
    \begin{pmatrix}
        \eta + vk & W & W & U(1+\alpha) \\ 
        W & - \eta & U(1-\alpha) & W \\ 
        W & U(1-\alpha) & - \eta & W \\ 
        U(1+\alpha) & W & W & \eta -vk 
    \end{pmatrix}
\end{equation}

The bilayer Hamiltonian is given by: 
\begin{equation}
    \mathcal{H}^{2bilayer} = 
    \begin{pmatrix}
        \Delta_1 & \Omega \\ 
        \Omega^{\dagger} & \Delta_2 
    \end{pmatrix}
\end{equation}
where the intralayer Hamiltonian of the layer $\ell$ ($\ell = 1,2$) is:
\begin{equation}
    \Delta_{\ell} = \mathcal{H}^{1layer} (\omega_{\ell},\eta_{\ell},v_{\ell},U_{\ell},W_{\ell},\alpha_{\ell})
\end{equation}
and the interlayer coupling Hamiltonian is:
\begin{equation}
     \Omega = e^{-d/d_0} \times 
     \text{diag} 
     \bigg \lbrace 
        - \bigg( V + \beta k + \frac{\beta}{\sqrt{2}} k^2 \bigg) e^{-i 2 \pi q}
        , 
        V + \frac{\beta}{\sqrt{2}} k^2 
        ,
        V + \frac{\beta}{\sqrt{2}} k^2
        ,
        - \bigg( V - \beta k + \frac{\beta}{\sqrt{2}} k^2 \bigg) e^{i 2 \pi q} 
     \bigg \rbrace
 \end{equation} 

\begin{figure}
    \centering
    \includegraphics[width=0.75\textwidth]{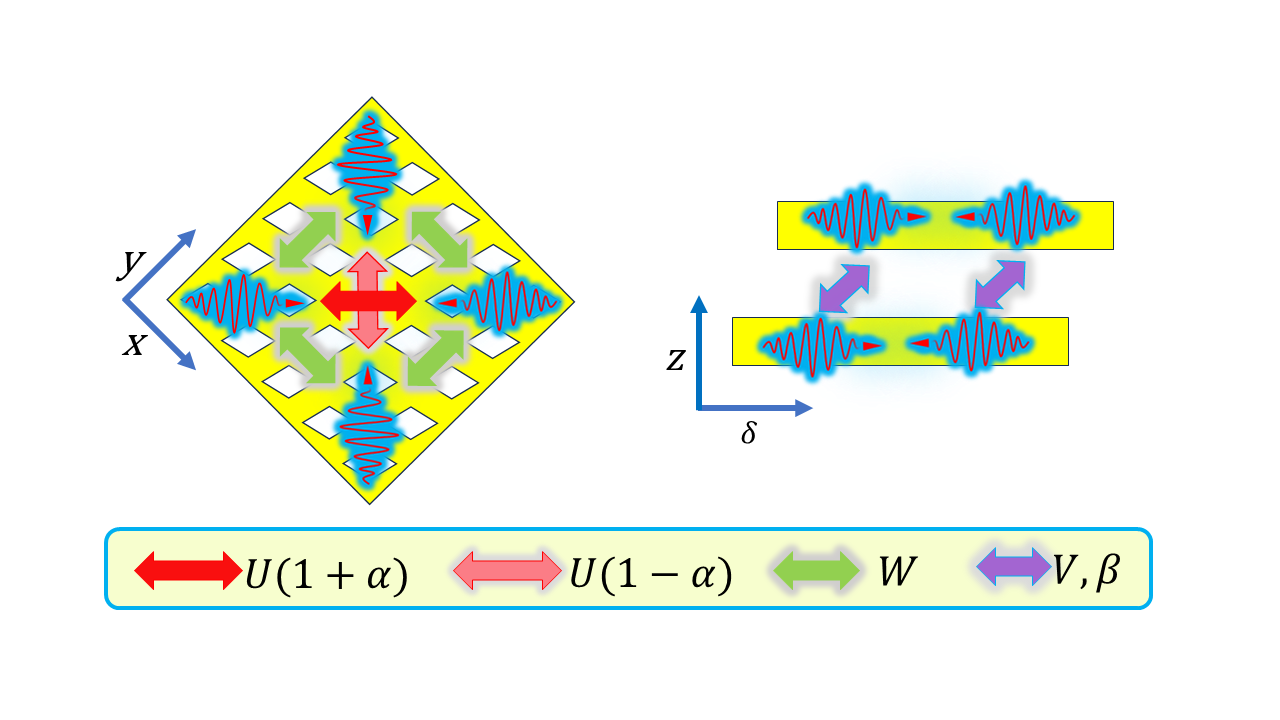}
    \caption{\textbf{Parameters of the effective model.} (a) The unit cell of the photonic crystal slab with broken $C_4$ symmetry. The lengths of the diagonals of the rhombus hole are $d_1$ and $d_2$. (b) The intralayer coupling strengths: $U$, $W$ and the symmetry-breaking parameter $\alpha$. (c) The interlayer coupling strength $V$.}
    \label{fig:EffectiveModel}
\end{figure}

This Hamiltonian is written in the basis $(A_{00}^1,A_{0-1}^1,A_{-10}^1,A_{-1-1}^1,A_{00}^2,A_{0-1}^2,A_{-10}^2,A_{-1-1}^2)$. 
The components of the eigenstates are \textbf{synthetic-momentum independent}.
In the paper, we work in this basis without doing any gauge transformation.
We summarize the intralayer and interlayer coupling strengths $U$, $W$ and $V$ in Fig.~\ref{fig:EffectiveModel}.

\subsection{Electric field profile}
The electric field profile of an eigenstate having components $(A_{00}^1, A_{0-1}^1, A_{-10}^1, A_{-1-1}^1, A_{00}^2, A_{0-1}^2, A_{-10}^2, A_{-1-1}^2)$ is approximatively given by: 
\begin{equation}
        \mathbf{E}_{\mathbf{k}\boldsymbol{\delta}} (\mathbf{r}) = \sum_{(m,n) \in \mathcal{B}} \mathbf{\hat{e}}_{mn} 
        \bigg \lbrace 
        A_{mn}^1 \Theta_1(z-z_1) + A_{mn}^2 \Theta_2(z-z_2) e^{-iK[(m+\frac{1}{2})\delta_x + (n+\frac{1}{2})\delta_y]} 
        \bigg \rbrace 
        e^{i(k_{mx}x+k_{ny}y)}
        \label{eq:ElectricFieldProfile1}
\end{equation}

The relationships $\delta = \sqrt{2} \delta_x = \sqrt{2} \delta_y$ and $q = \delta / \sqrt{2} - 0.5$ imply that:
\begin{equation}
    K \left[ \left(m+\frac{1}{2}) \delta_x + (n+\frac{1}{2} \right) \delta_y \right] = 2\pi (m+n+1) \left( q + \frac{1}{2} \right)
\end{equation}

For the case where the two slabs are moved with respect to each other along the main diagonal of their unit cells, we can express the electric field profile as follows:
\begin{equation}
        \mathbf{E}_{\mathbf{kq}} (\mathbf{r}) = \sum_{(m,n) \in \mathcal{B}} \mathbf{\hat{e}}_{mn} 
        \bigg \lbrace 
        A_{mn}^1 \Theta_1(z-z_1) + A_{mn}^2 \Theta_2(z-z_2) e^{-i2\pi (m+n+1) \left( q + \frac{1}{2} \right)} 
        \bigg \rbrace 
        e^{i(k_{mx}x+k_{ny}y)} 
        \label{eq:ElectricFieldProfile2} 
\end{equation}

We define a new coordinate system that is obtained by rotating the Cartesian coordinate system by an angle $\pi / 4$:
\begin{equation}
    \begin{aligned}
        \rho =& \frac{x+y}{\sqrt{2}}
        \\ 
        \bar{\rho} =& \frac{x-y}{\sqrt{2}}
    \end{aligned}
\end{equation}

Equivalently, we can write $x$ and $y$ in terms of $\rho$ and $\bar{\rho}$ as follows:
\begin{equation}
    \begin{aligned}
        x =& \frac{\rho + \bar{\rho}}{\sqrt{2}} \\ 
        y =& \frac{\rho - \bar{\rho}}{\sqrt{2}} 
    \end{aligned}
\end{equation}

In the case where $k_x = k_y = k / \sqrt{2}$, we have: 
\begin{equation}
    k_{mx} x + k_{ny} y = (m+n+1) \frac{K}{\sqrt{2}} \rho + (m-n) \frac{K}{\sqrt{2}} \bar{\rho} + k \rho 
\end{equation}

We can write the electric field profile in the new coordinate system as: 
\begin{equation}
    \mathbf{E}_{kq} (\rho, \bar{\rho}, z) = \sum_{(m,n) \in \mathcal{B}} \mathbf{\hat{e}}_{mn} 
        \bigg \lbrace 
        A_{mn}^1 \Theta_1(z-z_1) + A_{mn}^2 \Theta_2(z-z_2) e^{-i2\pi (m+n+1) \left( q + \frac{1}{2} \right)} 
        \bigg \rbrace 
        e^{i(m+n+1)K\rho/\sqrt{2}} e^{ik\rho} e^{i(m-n)K\bar{\rho}/\sqrt{2}} 
        \label{eq:ElectricFieldProfile3}
\end{equation}

By changing to the dimensionless units $\rho \rightarrow \rho / a$, $\bar{\rho} \rightarrow \bar{\rho}/a$, $k \rightarrow ka/(2\pi)$, we rewrite the electric field profile as follows:
\begin{equation}
    \mathbf{E}_{kq} (\rho, \bar{\rho}, z) = \sum_{(m,n) \in \mathcal{B}} \mathbf{\hat{e}}_{mn} 
        \bigg \lbrace 
        A_{mn}^1 \Theta_1(z-z_1) + A_{mn}^2 \Theta_2(z-z_2) e^{-i2\pi (m+n+1) \left( q + \frac{1}{2} \right)} 
        \bigg \rbrace 
        e^{i \pi \sqrt{2} (m+n+1) \rho} e^{i2\pi k\rho} e^{i \pi \sqrt{2} (m-n) \bar{\rho}}    
        \label{eq:ElectricFieldProfile4}
\end{equation}

We write the eigenstates under the form of a spinor: 
\begin{equation}
    \psi = 
    \begin{pmatrix}
        C_1 & C_2 & C_3 & C_4 & C_5 & C_6 & C_7 & C_8
    \end{pmatrix}^T
    = 
    \begin{pmatrix}
        A_{00}^1 & A_{0-1}^1 & A_{-10}^1 & A_{-1-1}^1 &
        A_{00}^2 & A_{0-1}^2 & A_{-10}^2 & A_{-1-1}^2 
    \end{pmatrix}^T
\end{equation}

In this notation, the electric field profiles of the \textbf{\textit{basic wave functions}} are given by:
\begin{equation}
    \begin{aligned}
        \mathbf{E}_1 =& \Theta_1 (z-z_1) e^{i\sqrt{2}\pi \rho} e^{i2\pi k\rho} \mathbf{\hat{e}}_{00} \\
        \mathbf{E}_2 =& \Theta_1 (z-z_1) e^{i\sqrt{2}\pi \bar{\rho}} e^{i2\pi k\rho} \mathbf{\hat{e}}_{0-1} \\
        \mathbf{E}_3 =& \Theta_1 (z-z_1) e^{-i\sqrt{2}\pi \bar{\rho}} e^{i2\pi k\rho} \mathbf{\hat{e}}_{-10} \\ 
        \mathbf{E}_4 =& \Theta_1 (z-z_1) e^{-i\sqrt{2}\pi \rho} e^{i2\pi k\rho} \mathbf{\hat{e}}_{-1-1} \\
        \mathbf{E}_5 =& -\Theta_2 (z-z_2) e^{i\sqrt{2}\pi \rho} e^{i2\pi k\rho} e^{-i2\pi q} \mathbf{\hat{e}}_{00} \\
        \mathbf{E}_6 =& \Theta_2 (z-z_2) e^{i\sqrt{2}\pi \bar{\rho}} e^{i2 \pi k\rho} \mathbf{\hat{e}}_{0-1} \\
        \mathbf{E}_7 =& \Theta_2 (z-z_2) e^{-i\sqrt{2}\pi \bar{\rho}} e^{i2\pi k\rho} \mathbf{\hat{e}}_{-10} \\ 
        \mathbf{E}_8 =& -\Theta_2 (z-z_2) e^{-i\sqrt{2}\pi \rho} e^{i2\pi k\rho} e^{i2\pi q} \mathbf{\hat{e}}_{-1-1} 
    \end{aligned}
\end{equation}

All the eight wave functions can be considered as plane waves with genuine momentum $k$, whereas only $\mathbf{E}_5$ and $\mathbf{E}_8$ are plane wave with respect to the synthetic momentum $q$.
By writing these eight vectors in matrix form:
\begin{equation}
    \boldsymbol{\mathcal{E}} (k,q;\rho,\bar{\rho}) = 
    \begin{pmatrix}
        \mathbf{E}_1 & \mathbf{E}_2 & \mathbf{E}_3 & \mathbf{E}_4 & \mathbf{E}_5 & \mathbf{E}_6 & \mathbf{E}_7 & \mathbf{E}_8 
    \end{pmatrix}^T
\end{equation}
we can rewrite \eqref{eq:ElectricFieldProfile4} as follows:
\begin{equation}
    \mathbf{E}_{kq}(\rho,\bar{\rho},z) = \psi^T \boldsymbol{\mathcal{E}} (k,q;\rho,\bar{\rho})
\end{equation}

Here $\boldsymbol{\mathcal{E}} (k,q;\rho,\bar{\rho})$ means $\boldsymbol{\mathcal{E}}$ is a function of two variables $\rho$ and $\bar{\rho}$ and parameters $k$ and $q$.
  
\section{Parameter retrievals for the effective model} 
\subsection{PWE band structure calculations}

\begin{figure}[h]
    \centering
    \includegraphics[width=\textwidth]{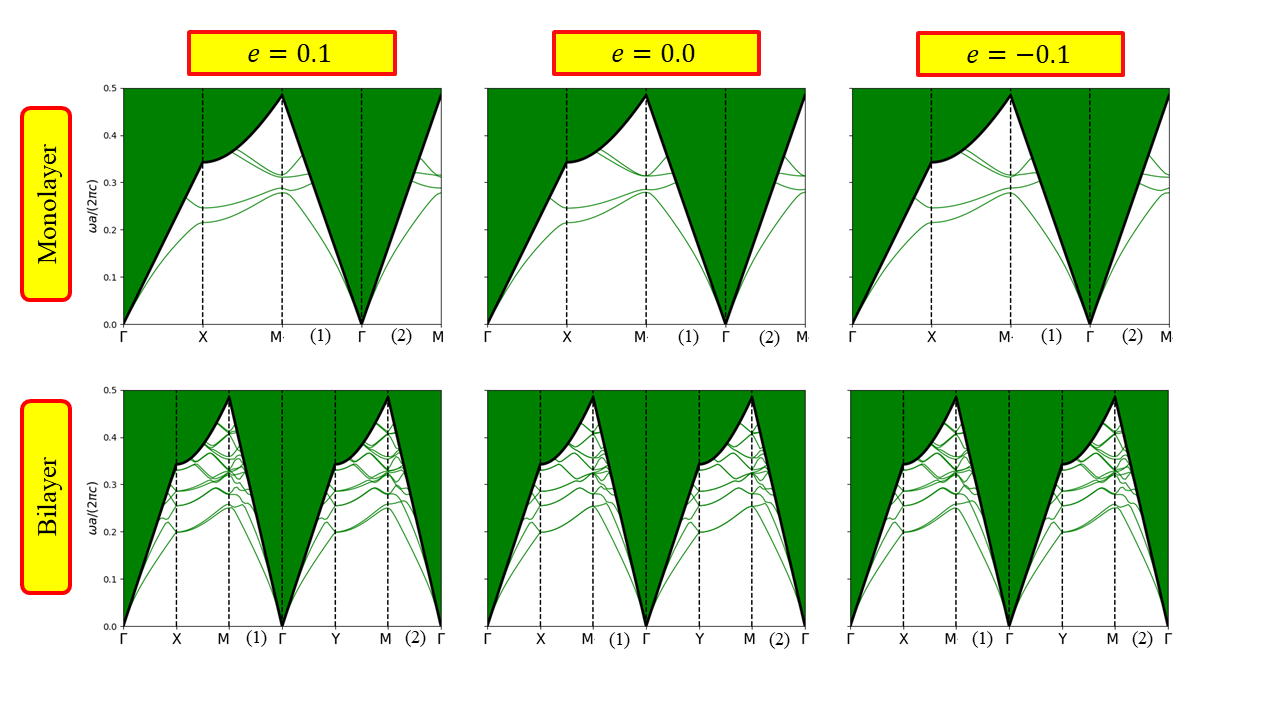}
    \caption{Photonic band structures of monolayer (first row) and bilayer with interlayer distance $d = 0.1a$ and shift $\delta = a/\sqrt{2}$(second row). The parameters of the structures are $h/a = 0.35$, $b/a = 0.38$, and $e = 0.1$ (first column), $e = 0$ (second column), $e = -0.1$ (third column). The band structures are calculated by employing the Plane Wave Expansion (PWE) method using the software MIT Photonic Bands (MPB). The path $\Gamma$-1-M corresponds to the line $k_x = k_y$ and the path $\Gamma$-2-M corresponds to the line $k_x = -k_y$ (see Fig.~\ref{fig:Monolayer}c).}
    \label{fig:MPB-Band-Structures}
\end{figure}

We perform the first-principal calculations of the photonic band structures using the Plane Wave Expansion (PWE) method by employing the MIT Photonic Bands (MPB) software.  
We calculate the band structure of even modes (with respect to the $z$-symmetric plane) of the monolayers of size $h = 0.35a$ and $b = 0.38a$ for three cases: $e = 0.1$, $e = 0$ and $e = -0.1$.
The band structures are shown in the first row of Fig.~\ref{fig:MPB-Band-Structures}.
When $e = 0$, the hole at the center of the unit cell is a square, the third and the fourth modes at the M point are degenerate, due to the C\textsubscript{4} symmetry of the structure.
When $e \ne 0$, this degeneracy point is no longer at the M point, but moves to either the line $k_x = k_y$ ($e = -0.1$) or $k_x = -k_y$ ($e = 0.1$), because the C\textsubscript{4} symmetry is broken.

Next, we calculate the photonic band structures for bilayers containing two identical photonic crystal slabs with interlayer distance $d = 0.1a$ and the shift along the main diagonal $\delta = \sqrt{2}a/2$.
Because the structure no longer has symmetry with respect to the $z$-mirror plane, one needs to perform the band structure calculation for light with arbitrary polarization instead of the calculating the dispersion relation of the $z$-even modes, which are ill-defined.
We remark that each of the four lowest pairs of bands is doubly degenerate at the M point. 

\subsection{Extract parameters for the effective model from PWE calculations}

\paragraph{C\textsubscript{4} limit:}
In this section, we extract the parameters of the effective models from the PWE calculations.
First, we extract the parameters $\omega_M$, $v$, $U$ and $W$ from PWE band structure of the 2D photonic crystal slab monolayer with square hole.
We vary each variable: the slab thickness $h$ and the hole edge $b$ by fixing the other variable at either $b = 0.38a$ and $h = 0.35a$, respectively.
In this case, $\alpha = 0$ and $\eta = 0$. 
After that, we fix $h = 0.35a$ and $b = 0.38a$ and vary $e$ to fit the symmetry-breaking parameters $\alpha$ and $\eta$ as functions of the diagonal deformation parameter $e$. 
Finally, we extract the parameters $V_0$ and $d_0$ by fitting the energy of the 8 lowest bands of the bilayer with respect to the interlayer distance $d$.

By setting $k_x = k_y = 0$, the energy eigenvalues of the monolayer at the M point are: 
\begin{equation}
    \begin{aligned}
        \Omega_1 =& \omega_M + U - 2W \\ 
        \Omega_2 =& \omega_M + U + 2W \\
        \Omega_3 =& \omega_M - U \\
        \Omega_4 =& \omega_M - U  
    \end{aligned}
    \label{eq:Monolayer_Eigenvalues_M}
\end{equation}

The corresponding eigenvectors are:
\begin{equation}
     \psi_1 = 
    \begin{pmatrix}
        1 \\ -1 \\ -1 \\ 1 
    \end{pmatrix}
    ,
    \psi_2 = 
    \begin{pmatrix}
        1 \\ 1 \\ 1 \\ 1 
    \end{pmatrix}
    ,
    \psi_3 = 
    \begin{pmatrix}
        1 \\ 0 \\ 0 \\ -1
    \end{pmatrix}
    , 
    \psi_4 = 
    \begin{pmatrix}
        0 \\ 1 \\ -1 \\ 0 
    \end{pmatrix}
\end{equation}

\begin{figure}
    \centering
    \includegraphics[width=0.5\linewidth]{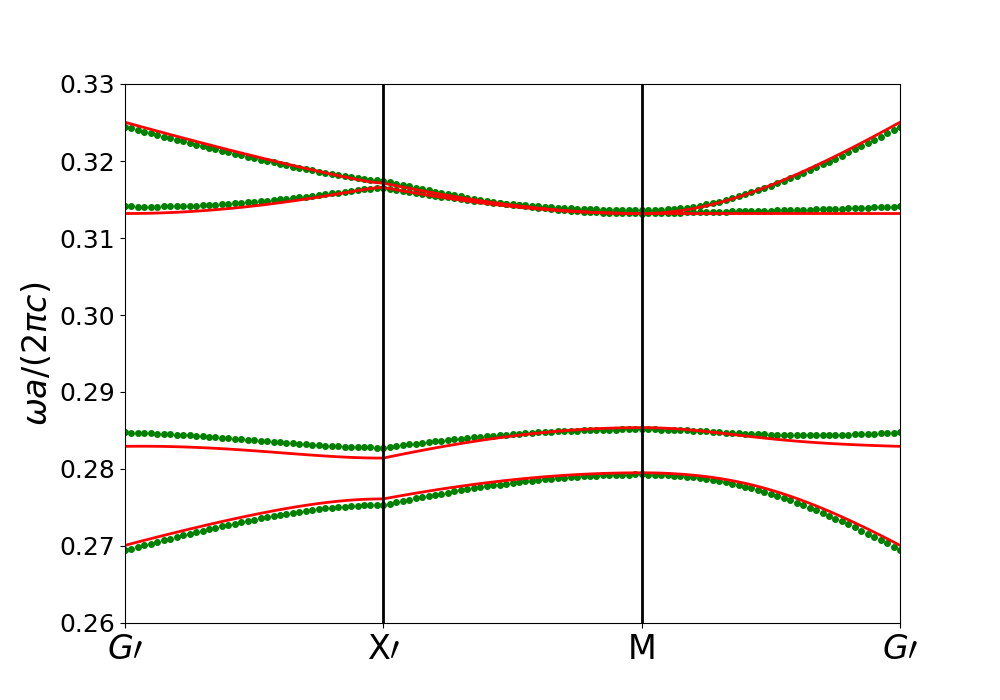}
    \caption{The photonic band structure in the vicinity of the M point. The parameters of the structure are: $h = 0.35a$, $b = 0.38a$, $e = 0$ (square hole). The PWE band structure is represented by the green dots. The effective model band structure is represented by the red lines. The coordinates of the points: G'(0.45,0.45), X'(0.5,0.45), M(0.5,0.5). The band structure by the effective model is represented by the solid lines. The parameters extracted from the PWE calculations are: $\omega_M = 0.2978$, $U = -0.01537$, $W = 0.001466$ and $v = 0.3170$. Here the energy is expressed in unit $2\pi c/a$, the velocity in unit $c$ and the momentum in unit $2\pi/a$.}
    \label{fig:FitBand}
\end{figure}

\begin{figure}
    \centering
    \includegraphics[width=0.8\textwidth]{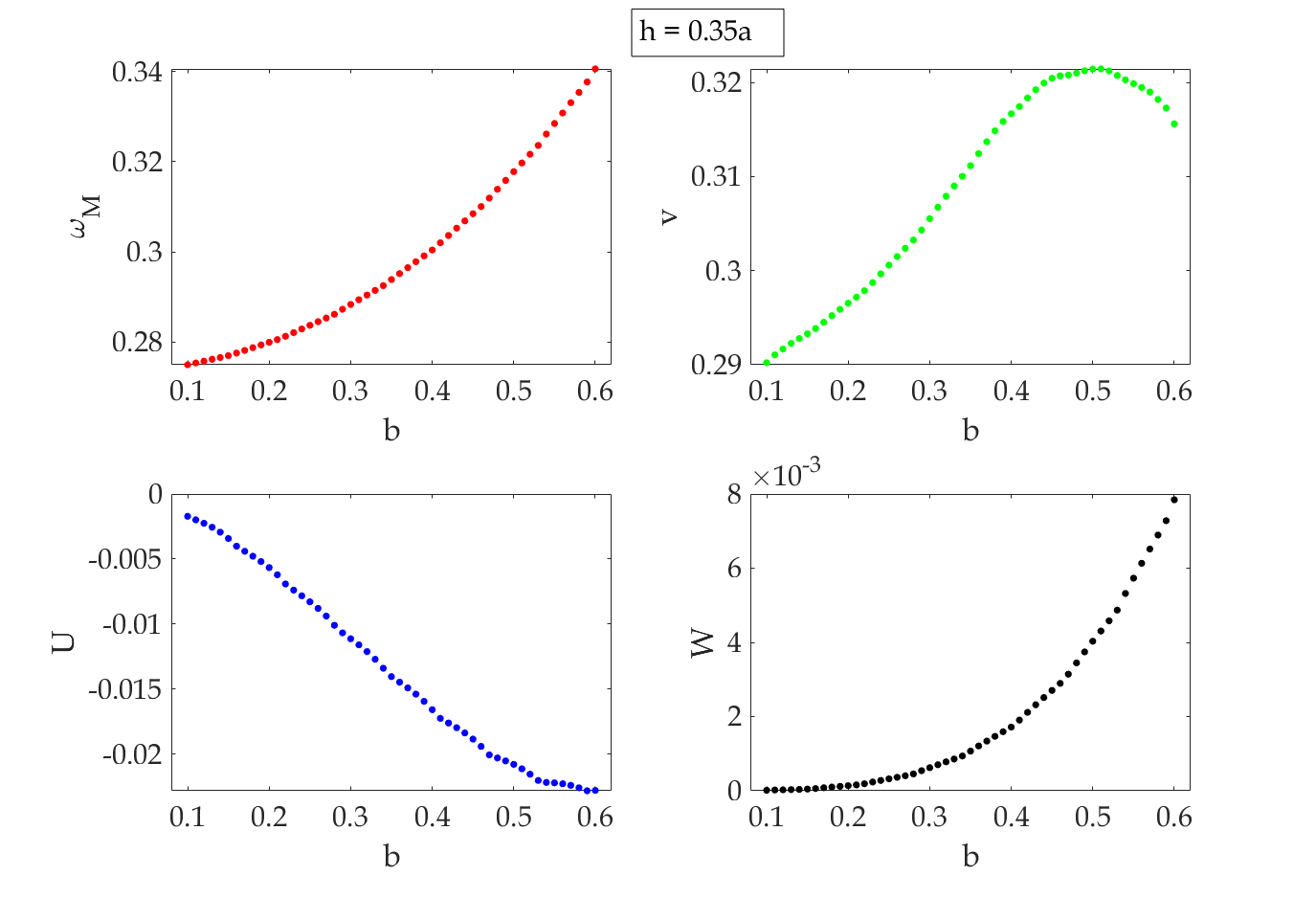}
    \caption{The evolution of the parameters $\omega_M$, $U$, $W$ and $v$ with respect to the edge length $b$ of the square hole when we fix $h = 0.35a$.}
    \label{fig:Fix_h-Scan-b}
\end{figure}

\begin{figure}
    \centering
    \includegraphics[width=0.8\textwidth]{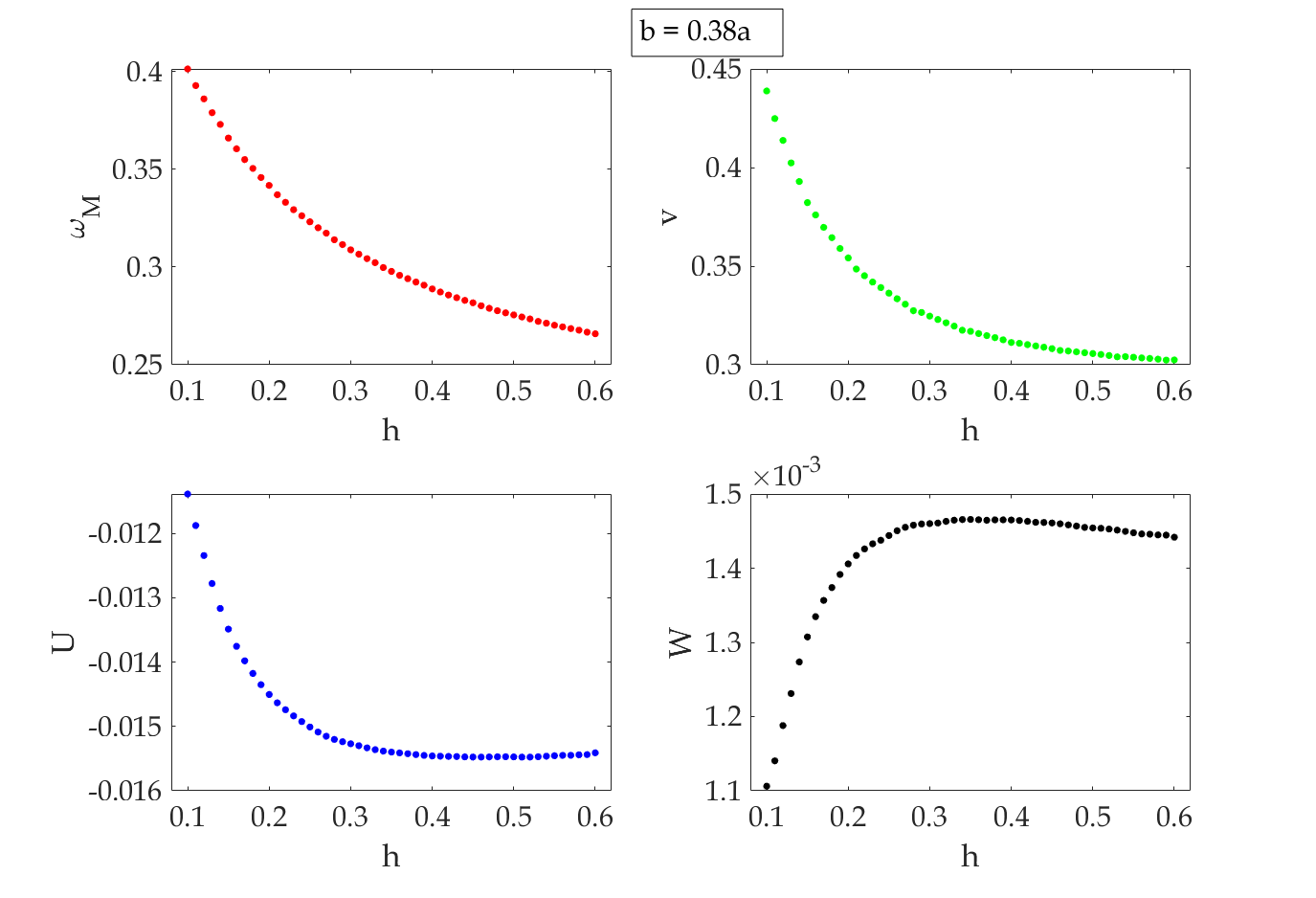}
    \caption{The evolution of the parameters $\omega_M$, $U$, $W$ and $v$ with respect to the thickness $h$ of the slab when we fix $b = 0.38a$.}
    \label{fig:Fix_b-Scan-h}
\end{figure}

The bands 3 and 4 are degenerate at the M point, in agreement with the C\textsubscript{4} symmetry of the structure.
The formulae \eqref{eq:Monolayer_Eigenvalues_M} allow us to directly extract the parameters $\omega_M$, $U$ and $W$ from the PWE values of the energy eigenvalues $\Omega_1$, $\Omega_2$, $\Omega_3$ and $\Omega_4$ at the M point.
The group velocity $v$ is extracted by fitting the energy of the four bands along the path G'-X'-M-G' where the coordinates of G' and X' are G'(0.45,0.45) and X'(0.5,0.45) (unit: $2\pi /a$).
For the structure of interest with $h = 0.35a$ and $b = 0.38a$, the extracted parameters are ($\omega_M$, $U$ and $W$ are in unit of $2\pi c/a$): 
$\omega_M = 0.2978$,
$U = -0.01537$,
$W = 0.001466$, 
and $v = 0.3170c$ (Fig.~\ref{fig:FitBand}).
We also fix the thickness $h = 0.35a$ and vary the hole edge length $b$ from $0.1a$ to $0.6a$; 
and vary the thickness $h$ from $0.1a$ to $0.6a$ while fixing the hole edge $b = 0.38a$.
The corresponding extracted parameters $\omega_M$, $U$, $W$ and $v$ are shown in Figs.~\ref{fig:Fix_h-Scan-b} and ~\ref{fig:Fix_b-Scan-h}.
Fig.~\ref{fig:Fix_h-Scan-b} shows that when we vary $b$ while fixing $h$, the parameters $U$ and $W$ change with opposite trends, implying that $\Delta U/U$ and $\Delta W/W$ have opposite signs (Fig.~\ref{fig:Fix_h-Scan-b}). 


\paragraph{Broken C\textsubscript{4} symmetry:}
Next, we fit the symmetry-breaking parameters $\alpha$ and $\eta$ with respect to the deformation parameter $e$. 
At $k_x = k_y = 0$, the energy eigenvalues of the monolayer with broken C\textsubscript{4} symmetry are given by:
\begin{equation}
    \begin{aligned}
        \Omega_1 =& \omega_M + U - \sqrt{\eta^2 + 2\alpha \eta U + \alpha^2 U^2 + 4W^2}
        \\
        \Omega_2 =& \omega_M + U + \sqrt{\eta^2 + 2\alpha \eta U + \alpha^2 U^2 + 4W^2}
        \\
        \Omega_3 =& \omega_M + \eta - U - \alpha U \\
        \Omega_4 =& \omega_M - \eta - U + \alpha U 
    \end{aligned}
\end{equation}

The bands 3 and 4 are split under $\eta \ne 0$ and $\alpha \ne 0$.
However, there are 4 equations for 5 variables. 
We cannot definitely extract all the 5 variables from the values of $\Omega_1$, $\Omega_2$, $\Omega_3$, $\Omega_4$ at the M point. 
Therefore, we substitute $\omega_M$, $U$ and $W$ retrieved from the case of the square hole and extract the parameters $\eta$ and $\alpha$ from the resulting equations:
\begin{equation}
    \begin{aligned}
        \eta - \alpha U =& \frac{1}{2} (\Omega_3 - \Omega_4) \\
        (\eta + \alpha U)^2 =& \left( \frac{\Omega_1-\Omega_2}{2} \right)^2 - 4W^2
    \end{aligned}
    \label{eq:Equation-alpha-eta}
\end{equation}

The second equation in \eqref{eq:Equation-alpha-eta} gives two cases corresponding to either 
$\eta + \alpha U = \sqrt{\left( \dfrac{\Omega_1-\Omega_2}{2} \right)^2 - 4W^2}$ or 
$\eta + \alpha U = -\sqrt{\left( \dfrac{\Omega_1-\Omega_2}{2} \right)^2 - 4W^2}$.
The following values of $\eta$ and $\alpha$ are consistent with the MPB results:
\begin{equation}
    \begin{aligned}
        \eta =& \frac{1}{2} \text{sign} (e) \bigg| \frac{1}{2} (\Omega_3-\Omega_4) - \sqrt{\left(\frac{\Omega_1-\Omega_2}{2}\right)^2 - 4W^2} \bigg| 
        \\
        \alpha =& - \frac{1}{2|U|} \text{sign} (e) \bigg| \frac{1}{2} (\Omega_3-\Omega_4) + \sqrt{\left(\frac{\Omega_1-\Omega_2}{2} \right)^2 - 4W^2} \bigg| 
    \end{aligned}
\end{equation}

\begin{figure}
    \centering
    \includegraphics[width=0.4\linewidth]{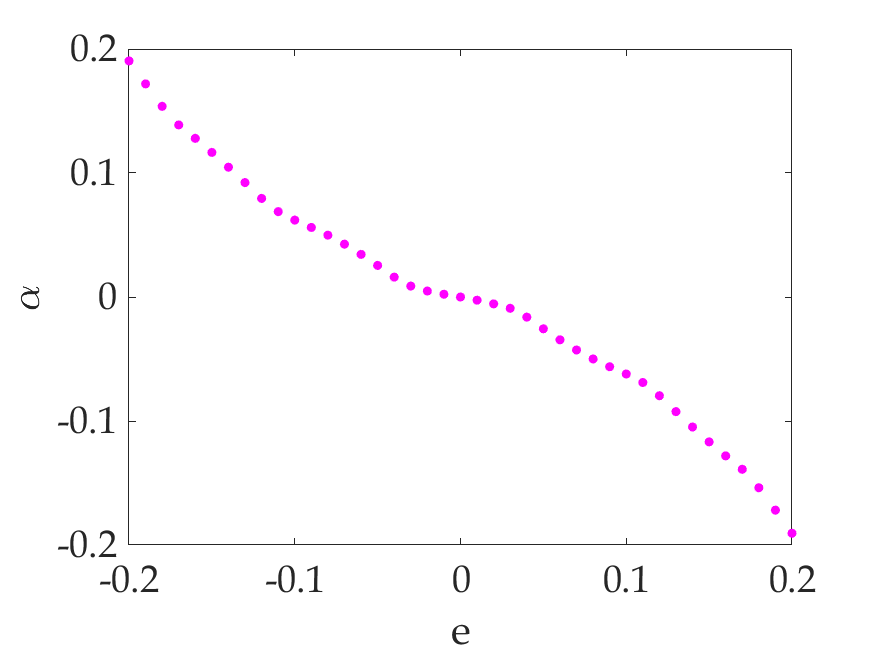}
    \includegraphics[width=0.4\linewidth]{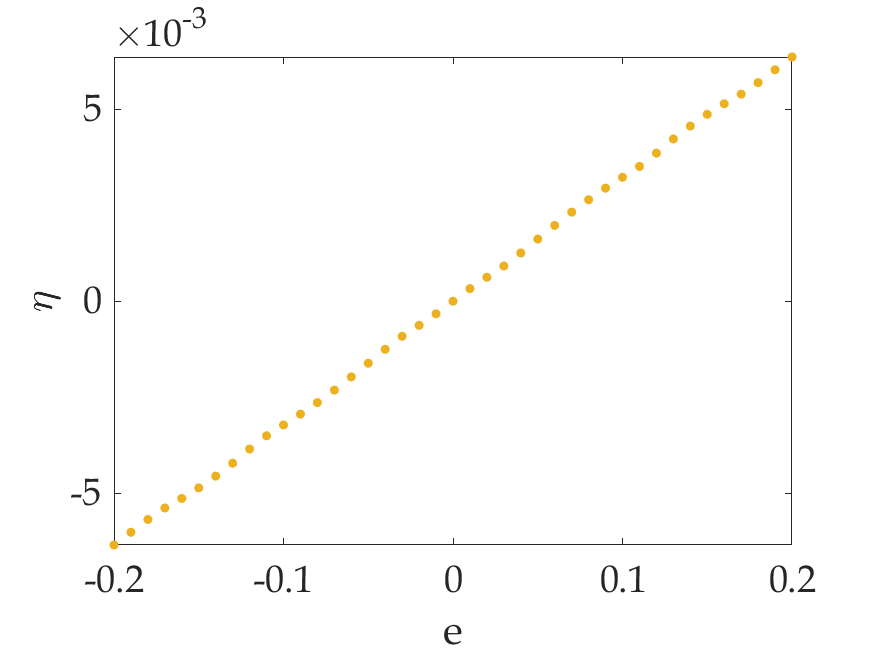}
    \caption{The symmetry-breaking parameters $\alpha$ (left) and $\eta$ (right) as functions of the deformation $e$.}
    \label{fig:alpha-eta-e}
\end{figure}

\begin{figure}
    \centering
    \includegraphics[width=0.4\linewidth]{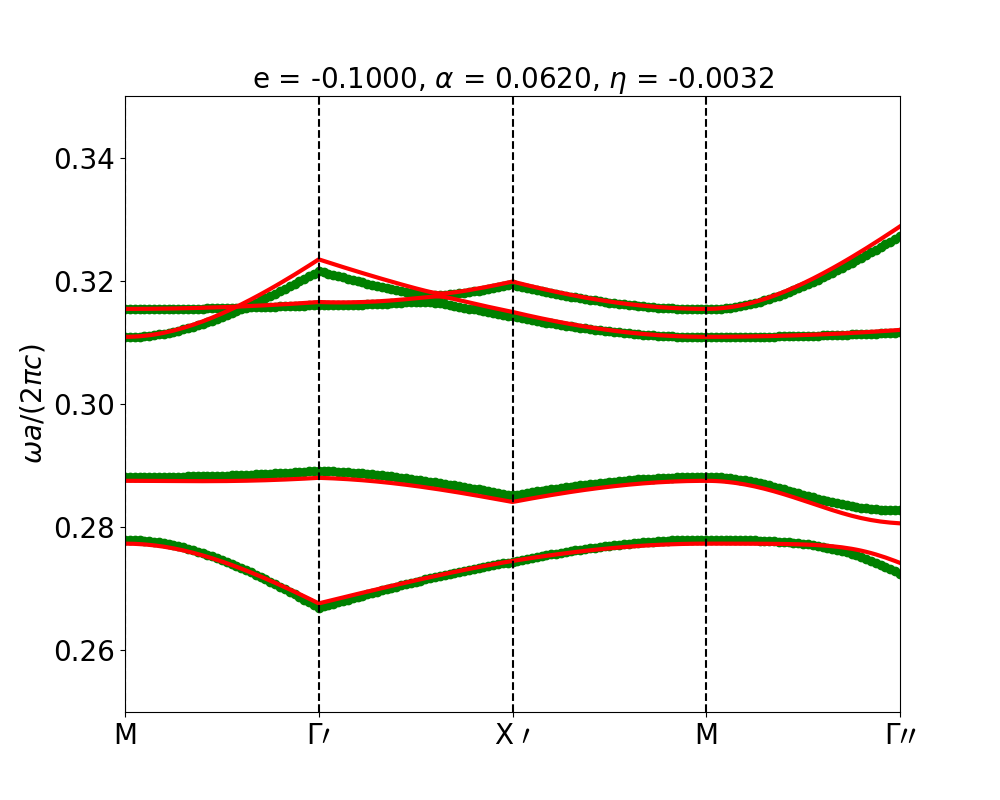}
    \includegraphics[width=0.4\linewidth]{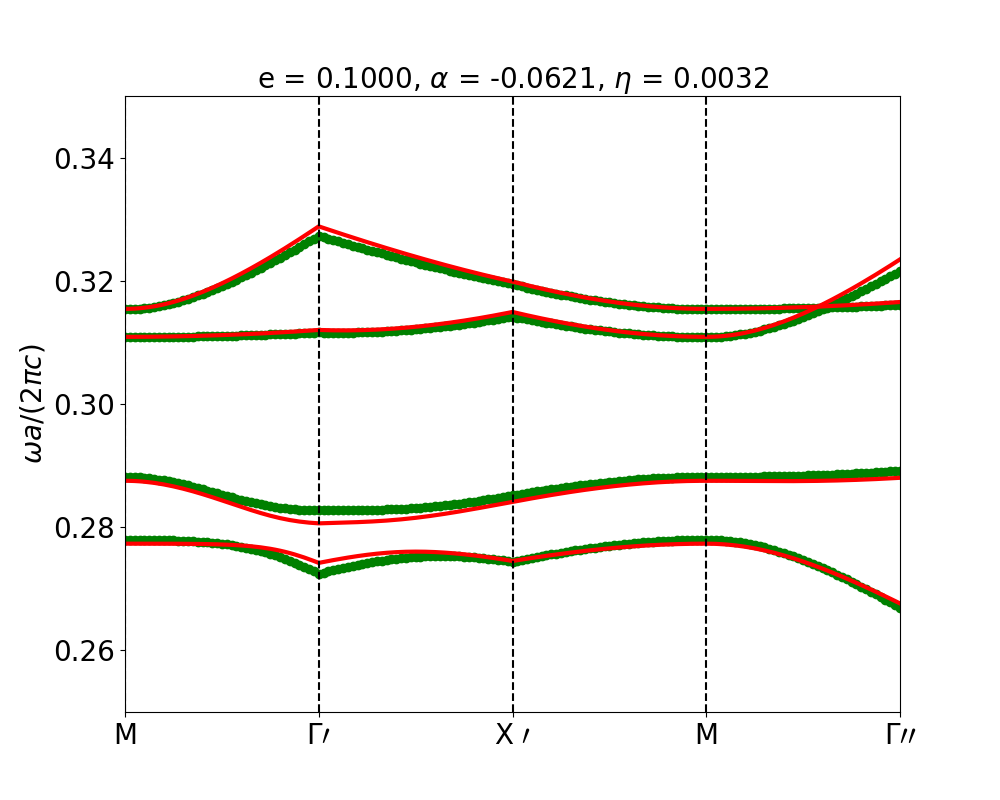}
    \caption{The photonic band structure of single layer slab with $h = 0.35a$, $b = 0.38a$ for: (left) $e = -0.1$, $\alpha = 0.062$, $\eta = -0.0032$ and (right) $e = 0.1$, $\alpha = -0.0621$, $\eta = 0.0032$. The green dots show the MPB results and red solid lines represent the results by the effective model. The point $\Gamma \prime$ is located at (0.45,0.45). The point X' is located at (0.5,0.45). The point $\Gamma \prime \prime$ is located at (0.55,0.45). It allows us to see the band structure along the lines $k_x=k_y$ (M-$\Gamma$') and $k_x=-k_y$ (M-$\Gamma$'').}
    \label{fig:2DSlab1L-RHoleP-abse_0.1}
\end{figure}

We show $\alpha$ and $\eta$ as functions of $e$ in Fig.~\ref{fig:alpha-eta-e}.
Both $\alpha$ and $\eta$ vary monotonically as function of $e$.
$\alpha$ is an decreasing function of $e$ whereas $\eta$ is a decreasing function of $e$.
When $e = 0$, $\alpha = 0$ and $\eta = 0$.
The band structures for the two cases $e = -0.1$ and $e = 0.1$ are shown in Fig.~\ref{fig:2DSlab1L-RHoleP-abse_0.1}.
The effective model reproduces the PWE calculations along the high-symmetry lines, and in particular, both of them show the difference in the band structure along the lines $k_x=k_y$ and $k_x=-k_y$.
For $e<0$, the degenerate point moves from the M point to the line $k_x=k_y$.
By contrast, it moves to the line $k_x=-k_y$ for $e>0$.
This is consistent to the fact that two structures with opposite $e$ are image of each other by the rotation of angle 90 degrees, which interchanges the two high-symmetry lines $k_x=k_y$ and $k_x=-k_y$.
 
\begin{figure}
    \centering
    \includegraphics[width=0.5\textwidth]{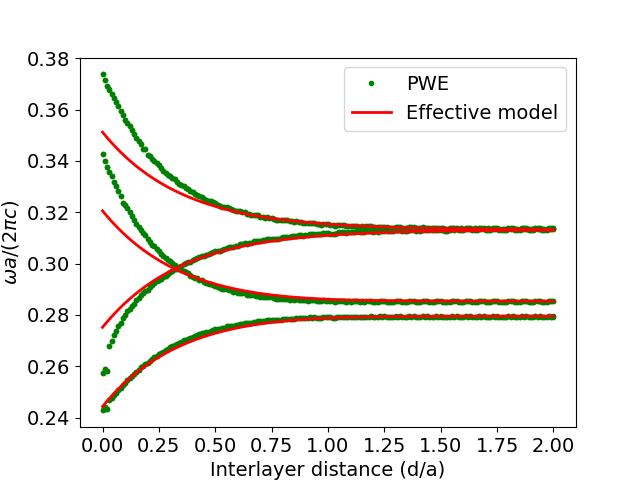}
    \caption{Fitting the interlayer coupling strength $V_0$ and the characteristic length $d_0$ for the bilayer with identical slabs of size $h = 0.35a$ and $b = 0.38a$. The shift along the diagonal direction is $\delta = \sqrt{2}a/2$. The fitted parameters are $V_0 = 0.04573 \times 2\pi c/a$ and $d_0 = 0.2790a$. Notice that each band is doubly degenerate in both PWE calculations and the effective model.}
    \label{fig:2DSlab2L-Fit-dvar}
\end{figure}

\paragraph{Bilayer:}
We fit the interlayer coupling strength $V_0$, the characteristic length $d_0$ and the parameter $\beta$.
We calculate the energy eigenvalues at the M point of the bilayer with identical slabs of size $h = 0.35a$ and having square holes of edge length $b = 0.38a$.
The shift along the diagonal direction is $\delta = \sqrt{2}a/2$ or equivalently $\delta_x = \delta_y = a/2$. 
That means $k = q = 0$ or $k_x = k_y = q_x = q_y = 0$.
The fitted parameters are: 
$V_0 = 0.038 \times 2\pi c / a$ and $d_0 = 0.35 a$. 
We show the fitted results in Fig.~\ref{fig:2DSlab2L-Fit-dvar}.
Finally, by setting $k \ne 0$, we can fit with MPB results for $e = 0, \pm 1$ and see that $\beta = -0.3$.

\paragraph{Broken-inversion symmetry:} We make the two layers to be no longer identical, but having $h_1=h_2=h$ and $b_1 \ne b_2$. 
That breaks the inversion symmetry and make $\omega_1 \ne \omega_2$, $U_1 \ne U_2$, $W_1 \ne W_2$ and $v_1 \ne v_2$. 
By fitting the effective model with the PWE results, we realize that the values of $\alpha$ and $\eta$ remain unchanged in the broken-inversion symmetry case compared to the case where the two layers are identical.

The parameters of the bilayer Hamiltonian of can now be written as:
\begin{equation}
    \begin{aligned}
        \omega_1 =& \omega (1 + m_{\omega}), & \omega_2 =& \omega (1 - m_{\omega}) \\ 
        U_1 =& U (1 + m_U), & U_2 =& U(1 - m_U) \\ 
        W_1 =& W (1 + m_W), & W_2 =& W(1 - m_W) \\
        v_1 =& v (1 + m_v), & v_2 =& v(1 - m_v) 
    \end{aligned}
\end{equation}

\begin{figure}
    \centering
    \includegraphics[width=\linewidth]{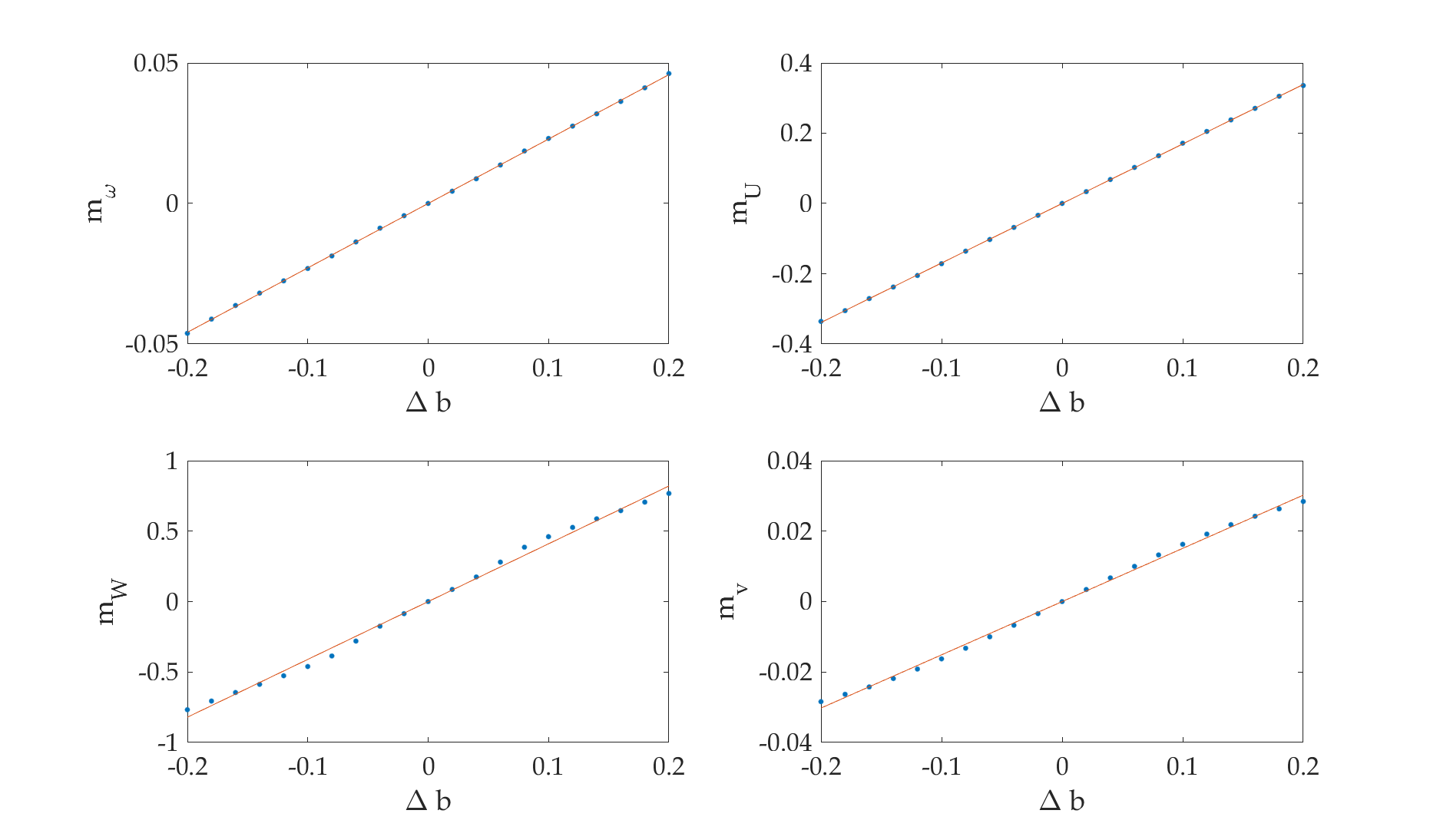}
    \caption{Linear fitting of $m_U$, $m_W$, $m_{\omega}$ and $m_v$ as a function of $\Delta b$.}
    \label{fig:Fit_UWomegav}
\end{figure}

The case where two layers are identical happens when $m_{\omega} = m_U = m_W = m_v = 0$.
We investigate the variation of $m_{\omega}$, $m_U$ and $m_W$ as $b_1$ and $b_2$ vary in the vicinity of $b = 0.38a$ such that $(b_1+b_2)/2 = 0.38a$. 
We define $\Delta b = b_1 - b_2$ and examine the cases where $-0.2a \le \Delta b \le 0.2a$.
Fitting gives $m_{\omega}$, $m_U$ and $m_W$ as linear functions of $\Delta b$ (Fig.~\ref{fig:Fit_UWomegav}):
\begin{equation}
    \begin{aligned}
        m_{\omega} =& 0.2292 \Delta b \\
        m_U =& 1.6941 \Delta b \\ 
        m_W =& 4.1021 \Delta b \\   
        m_v =& 0.1512 \Delta b
    \end{aligned}
\end{equation}

Consequently, we define $m = m_U$ and obtain:
$m_{\omega} = r_{\omega} m$, $m_W = r_W m$ and $m_v = r_v m$ with $r_{\omega}=0.1353$, $r_W=2.4215$ and $r_v = 0.0892$.
The bilayer Hamiltonian is dependent on the parameter $m$, which represents the difference between the two layers.
The parameter $m$ tells us if the system is gapped or gapless.
If $m = 0$, the two layers are identical, the system is gapless and the touching points are either two Dirac cones or a quadratic point.
If $m \ne 0$, the two layers are different from each other, the system becomes gapped.
We notice that the cases $m$ and $-m$ correspond to a same structure, but with the two layers swapped with respect to each other.
We call $m$ the \textbf{gap parameter}.

With the gap parameter, we denote the two layers $\ell = \pm$ instead of $\ell = 1,2$.
Here $\ell = +$ is the upper layer and $\ell = -$ is the lower layer.
By adding the gap parameter $m$ to the two-dimensional hybrid momentum space $(k,q)$, we form a three-dimensional parameter space $(k,q,m)$.
The Berry monopoles live in this three-dimensional parameter space.
Notice that $m$ does \textit{not} play the role of a momentum, unlike $k$ and $q$.
We denote the bilayer Hamiltonian as:
\begin{equation}
    \mathcal{H} (k,q,m) = 
    \begin{pmatrix}
        \Delta_+ & \Omega \\ 
        \Omega^{\dagger} & \Delta_-
    \end{pmatrix}
\end{equation}

We rewrite the single-layer parameters in a more concise form:
\begin{equation}
    \begin{aligned}
        \omega_{\pm} =& \omega ( 1 \pm r_{\omega} m ) \\
        v_{\pm} =& v (1 \pm r_v m) \\
        U_{\pm} =& U (1 \pm m) \\
        W_{\pm} =& W (1 \pm r_W m) \\
    \end{aligned}
\end{equation}
where $\omega = \dfrac{\omega_1 + \omega_2}{2}$, $v = \dfrac{v_1 + v_2}{2}$, $U = \dfrac{U_1 + U_2}{2}$ and $W = \dfrac{W_1 + W_2}{2}$.

\subsection{Dispersion relation of the bilayer}
\begin{figure}
    \centering
    \includegraphics[width=\textwidth]{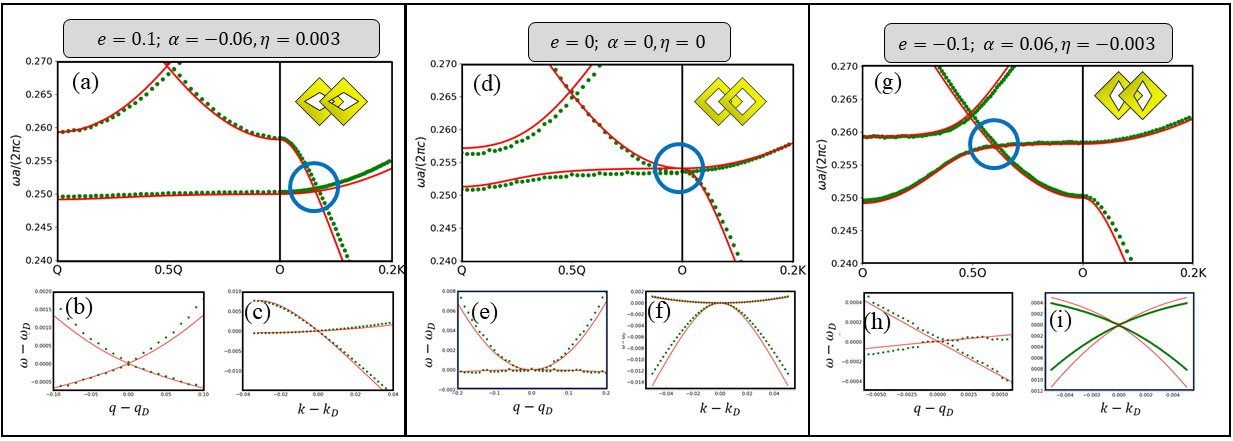}
    \caption{(a,d,g) Dispersion relation in the hybrid momentum space for the structures with $e = 0.1$, $e = 0$ and $e = -0.1$. (b,e,h) The dispersion curves in the vicinity of the Dirac poins along the $q$ direction. (c,f,i) The dispersion curves in the vicinity of the Dirac points along the $k$ direction. The band degeneracy points are located at $(k_D,q_D)$.}
    \label{fig:Dirac_Dispersion}
\end{figure}

The dispersion relation of the bilayer photonic crystal slab with $m = 0$ for 3 cases: $e = 0.1$, $e = 0$ and $e = -0.1$ is shown in Fig.~\ref{fig:Dirac_Dispersion}. For $e = 0.1$ and $e = -0.1$, the bands have linear dispersion along both the $k$ and $q$ directions, forming tilted type-I Dirac cones. For $e = 0$, both bands have parabolic dispersion along both $k$ and $q$ directions at the point O, showing that this point is a quadratic touching point.  

\subsection{Time-reversal symmetry operator}
We now define the time-reversal operation for the effective Hamiltonian used in the manuscript. In the folded-wave basis $\left(
|00\rangle,\ |0,-1\rangle,\ |-1,0\rangle,\ |-1,-1\rangle
\right)$,
time reversal complex conjugates the wave amplitudes and reverses the genuine Bloch momentum. Around the high-symmetry point considered here, this operation exchanges the counter-propagating folded waves as
\begin{equation}
|00\rangle \leftrightarrow |-1,-1\rangle,\qquad
|0,-1\rangle \leftrightarrow |-1,0\rangle .
\end{equation}

The unitary part of the time-reversal operator for one layer can therefore be written as
\begin{equation}
\mathcal P =
\begin{pmatrix}
0&0&0&1\\
0&0&1&0\\
0&1&0&0\\
1&0&0&0
\end{pmatrix}.
\end{equation}
For the full bilayer Hamiltonian $\mathcal H(k,q,m)=
\begin{pmatrix}
\Delta_+ & \Omega\\
\Omega^\dagger & \Delta_-
\end{pmatrix}$, we define
\begin{equation}
\mathcal U_T=
\begin{pmatrix}
\mathcal P&0\\
0&\mathcal P
\end{pmatrix},
\qquad
\mathcal T=\mathcal U_T\mathcal K,
\end{equation}
where $\mathcal K$ denotes complex conjugation. Physical time reversal then gives
\begin{equation}
\mathcal U_T \mathcal H^*(k,q,m)\mathcal U_T^\dagger
=
\mathcal H(-k,q,m).
\end{equation}
This relation expresses the reciprocity of each fixed-$q$ bilayer structure. Importantly, however, it is not the time-reversal condition for a two-dimensional Bloch Hamiltonian in the synthetic momentum space $(k,q)$. Such a condition would require $\mathcal U_T \mathcal H^*(k,q,m)\mathcal U_T^\dagger = \mathcal H(-k,-q,m)$. This latter relation is not satisfied, because $q$ is a geometrical displacement parameter and is even under physical time reversal. Therefore, the family of reciprocal bilayer structures realizes an effectively time-reversal-broken Hamiltonian in the synthetic $(k,q)$ space. This distinction is the reason why the system can exhibit a finite Berry curvature $\mathcal F_{kq}$ and nonzero Chern numbers, although no magnetic or gyrotropic material is involved.

\section{Berry curvature calculations}

The Berry curvature in the three-dimensional parameter space $(k,q,m)$ is given by the formulae~\cite{Xiao2010,Girvin_Yang_2019}: 
\begin{equation}
\begin{aligned}
    \mathcal{F}_{kq}^{(n)} (k,q,m) 
    =& i \left( \bigg \langle \frac{\partial n(k,q,m)}{\partial k} \bigg| \frac{\partial n(k,q,m)}{\partial q} \bigg \rangle - \bigg \langle \frac{\partial n(k,q,m)}{\partial q} \bigg| \frac{\partial n(k,q,m)}{\partial k} \bigg \rangle   \right)  
    \\ 
    =& i \sum_{l \ne n} 
    \frac{\langle n | \partial \mathcal{H} / \partial k |l\rangle \langle l | \partial \mathcal{H} / \partial q | n \rangle - c.c.}{(E_n-E_l)^2}
    \\
    \mathcal{F}_{qm}^{(n)}(k,q,m) =& i \left( \bigg \langle \frac{\partial n(k,q,m)}{\partial q} \bigg| \frac{\partial n(k,q,m)}{\partial m} \bigg \rangle - \bigg \langle \frac{\partial n(k,q,m)}{\partial m} \bigg | \frac{\partial n (k,q,m)}{\partial q} \bigg \rangle \right) 
    \\
    =& i \sum_{l \ne n} \frac{\langle n | \partial \mathcal{H} / \partial q | l \rangle \langle l | \partial \mathcal{H} / \partial m | n \rangle - c.c. }{(E_n - E_l)^2}
    \\
    \mathcal{F}_{mk}^{(n)}(k,q,m) =& i \left( \bigg \langle \frac{\partial n(k,q,m)}{\partial m} \bigg | \frac{\partial n(k,q,m)}{\partial k} \bigg \rangle - \bigg \langle \frac{\partial n(k,q,m)}{\partial k} \bigg | \frac{\partial n(k,q,m)}{\partial m} \bigg \rangle \right) 
    \\ 
    =& i \sum_{l \ne n} \frac{\langle n | \partial \mathcal{H} / \partial q | l \rangle \langle l | \partial \mathcal{H} / \partial m | n \rangle - c.c.}{(E_n - E_l)^2} 
\end{aligned}
\end{equation}
where $|n\rangle$ is the eigenstate of the bilayer Hamiltonian $\mathcal{H}_{\textbf{bilayer}}$ corresponding to the eigenvalue $E_n$.

The derivative of the Hamiltonian with respect to the genuine momentum $k$ is given by: 

\begin{equation}
    \frac{\partial \mathcal{H}}{\partial k} = 
    \begin{pmatrix}
        \dfrac{\partial \Delta_1}{\partial k} & \dfrac{\partial \Omega}{\partial k} \\
        \dfrac{\partial \Omega^{\dagger}}{\partial k} & \dfrac{\partial \Delta_2}{\partial k}
    \end{pmatrix}
\end{equation}

Here for $\ell = \pm$:
\begin{equation}
    \frac{\partial \Delta_{\ell} }{\partial k} = 
    \begin{pmatrix}
        v_{\ell} (1+\sqrt{2}k) & 0 & 0 & 0 \\ 
        0 & \sqrt{2} v_{\ell} k & 0 & 0 \\ 
        0 & 0 & \sqrt{2} v_{\ell} k & 0 \\ 
        0 & 0 & 0 & v_{\ell}(-1+\sqrt{2}k)
    \end{pmatrix}
\end{equation}
and 
\begin{equation}
    \frac{\partial \Omega}{\partial k} = e^{-d/d_0} 
    \text{diag} \lbrace 
    - \beta (1+\sqrt{2}k) e^{-i2\pi q} ,
    \sqrt{2} \beta k , 
    \sqrt{2} \beta k , 
    \beta (1 - \sqrt{2}k) e^{i2\pi q}
    \rbrace 
\end{equation}

The derivative of the Hamiltonian with respect to the synthetic momentum is given by: 
\begin{equation}
    \frac{\partial \mathcal{H}}{\partial q} = 
    \begin{pmatrix}
        0 & \dfrac{\partial \Omega}{\partial q} \\
        \dfrac{\partial \Omega^{\dagger}}{\partial q} & 0
    \end{pmatrix}
\end{equation}
where 
\begin{equation}
    \frac{\partial \Omega}{\partial q} = e^{-d/d_0} 
    \text{diag} \bigg\lbrace i2\pi \left( V + \beta k + \frac{\beta}{\sqrt{2}} k^2 \right) e^{-i2\pi q} ,
    0, 0,
    -i2\pi \left( V - \beta k + \frac{\beta}{\sqrt{2}} k^2 \right) e^{i2\pi q}
    \bigg\rbrace 
\end{equation}

The derivative of the Hamiltonian with respect to the mass parameter $m$ is given by: 
\begin{equation}
    \frac{\partial \mathcal{H}}{\partial m} = 
    \begin{pmatrix}
        \dfrac{\partial \Delta}{\partial m} & 0 \\ 
        0 & - \dfrac{\partial \Delta}{\partial m}
    \end{pmatrix}
\end{equation}
where 
\begin{equation}
        \frac{\partial \Delta}{\partial m} = \omega r_{\omega} + \frac{v r_v}{\sqrt{2}} k^2 + 
        \begin{pmatrix}
            v r_v k & W r_W & W r_W & U(1+\alpha) \\ 
            W r_W & 0 & U(1-\alpha) & W r_W \\ 
            W r_W & U(1-\alpha) & 0 & W r_W \\ 
            U(1+\alpha) & W r_W & W r_W & - v r_v k 
        \end{pmatrix} 
\end{equation}

\section{Berry monopole strength}
Knowing the Berry curvature profile in the three-dimensional parameter space $(k,q,m)$, we calculate the Berry monopole strength, by integrating over a closed surface $\Sigma$ surrounding the monopole:
\begin{equation} 
    g = \frac{1}{4\pi} \oiint_{\Sigma} \boldsymbol{\mathcal{F}}^{(1)} \cdot d \mathbf{S}  
\end{equation}

Here the three components of the field emitted by the monopole are the Berry curvatures: $\boldsymbol{\mathcal{F}} = (\mathcal{F}_{qm},\mathcal{F}_{mk},\mathcal{F}_{kq})$.
To do the integration, assume that the monopole is located at the point $(k_0,q_0,0)$, choose the closed surface $\Sigma$ containing the monopole to be the surface of the box of center $(k_0,q_0,0)$ and edges $L_k$,$L_q$,$L_m$ along the $k$,$q$,$m$ directions, respectively. 
The surface $\Sigma$ contains 6 faces:
\begin{equation}
 \begin{aligned}
  \Sigma_1:& k_0 - \frac{L_k}{2} \le k \le k_0 + \frac{L_k}{2} ;& q_0 - \frac{L_q}{2} \le q \le q_0 + \frac{L_q}{2} ;& m = \frac{L_m}{2} \\
  \Sigma_2:& k_0 - \frac{L_k}{2} \le k \le k_0 + \frac{L_k}{2} ;& q_0 - \frac{L_q}{2} \le q \le q_0 + \frac{L_q}{2} ;& m = -\frac{L_m}{2} \\
  \Sigma_3:& k_0 - \frac{L_k}{2} \le k \le k_0 + \frac{L_k}{2} ;& q = q_0 + \frac{L_q}{2} ;& - \frac{L_m}{2} \le m \le \frac{L_m}{2} \\
  \Sigma_4:& k_0 - \frac{L_k}{2} \le k \le k_0 + \frac{L_k}{2} ;& q = q_0 - \frac{L_q}{2} ;& - \frac{L_m}{2} \le m \le \frac{L_m}{2} \\
  \Sigma_5:& k = k_0 + \frac{L_k}{2} ;& q_0 - \frac{L_q}{2} \le q \le q_0 + \frac{L_q}{2} ;& - \frac{L_m}{2} \le m \le \frac{L_m}{2} \\
  \Sigma_6:& k = k_0 - \frac{L_k}{2} ;& q_0 - \frac{L_q}{2} \le q \le q_0 + \frac{L_q}{2} ;& - \frac{L_m}{2} \le m \le \frac{L_m}{2} \\
 \end{aligned}
\end{equation}

The monopole strength is given by:
\begin{equation}
\begin{aligned} 
 g =& \frac{1}{4\pi} \int_{k_0-L_k/2}^{k_0+L_k/2} dk \int_{q_0-L_q/2}^{q_0+L_q/2} dq 
 \left[ \mathcal{F}_{kq}\left( k,q,\frac{m}{2} \right) - \mathcal{F}_{kq} \left( k,q,-\frac{m}{2} \right) \right] 
 \\
 & + \frac{1}{4\pi} \int_{q_0-L_q/2}^{q_0+L_q/2} dq \int_{-L_m/2}^{L_m/2} dm \left[ \mathcal{F}_{qm}\left(k_0+\frac{L_k}{2},q,m\right) - \mathcal{F}_{qm}\left(k_0-\frac{L_k}{2},q,m \right) \right]
 \\
 & + \frac{1}{4\pi} \int_{-L_m/2}^{L_m/2} dm \int_{k_0-L_k/2}^{k_0+L_k/2} dk \left[ \mathcal{F}_{mk} \left( m, k, q_0 + \frac{L_q}{2} \right) - \mathcal{F}_{mk} \left(m, k, q_0 - \frac{L_q}{2} \right) \right] 
\end{aligned}
\end{equation}

To do the integration, we discretize the box into $N_k \times N_q \times N_m$ boxes.
Surface integration using code written in the MATLAB programming language gives the monopole strength as follows:
\begin{itemize}
 \item For $e = 0.1$: two Berry monopoles of monopole strength $g = \dfrac{1}{2}$ are located on the $k$-axis.
 We evaluate their monopole strengths using boxes of center $(k_0,q_0,m_0) = (\pm 0.0325,0,0)$.
 \item For $e = 0.0$: the Berry monopole of strength $g = 1$ is located at the point $(k0,q0,m0) = (0,0,0)$, confirming that two monopoles meet each other at this point.
 \item For $e = -0.1$: two Berry monopoles of monopole strength $g = \dfrac{1}{2}$ are located on the $q$-axis.
 We evaluate their monopole strength using boxes of center $(k_0,q_0,0) = (0,0,\pm 0.1965)$. 
\end{itemize}

The monopole strength and the total flux passing through the box are independent of the size of the box, in agreement with the Gauss law.
We also get the charge at the collision to be exactly equal to the sum of the two charges of the two Berry monopoles before and after collision, showing a conservation of Berry monopole not only before and after collision, but also at collision.
Overall, \textbf{\textit{the total charge of the monopoles is conserved during the collision.}}

We finish this section by a comment about the value of the monopole strength of the Dirac points.
Standard textbooks (for example the book by Bernevig and Hughes~\cite{BernevigHughes2013}, section 2.4) show that the Berry monopole in spin-$s$ systems equals to the spin $s$ of the elementary particle in the system.
One can argue that photon is spin-1 particle so the Berry monopole should be 1 instead of $\dfrac{1}{2}$.
This point of view is not true in the context of the present work.
The answer lies on the way we define the \textit{spin}.
Indeed, the proof in section 2.4 of Ref.~\cite{BernevigHughes2013} lies on considering the Hamiltonian $H = \mathbf{k} \cdot \mathbf{S} = k_x S_x + k_y S_y + k_z S_z$ where $S_x$, $S_y$ and $S_z$ are $(2s + 1) \times (2s + 1)$ matrices representing the corresponding spin operators for spin-$s$ systems.
When $\mathbf{k} = \boldsymbol{0}$, the Hamiltonian has $2s + 1$ degenerate eigenstates of energy $E = 0$, that means it is the degeneracy point of $2s+1$ bands.
Each of the bands has a Berry monopole placed at this degeneracy point, but these monopole strengths (or charges) are not identical.
Rather, the monopole strength equals $s$, $s-1$, $\dots$, $-(s-1)$, $-s$ for the bands arranged in increasing order of energy. 
At the beginning, the proof is for the case of a particle of spin-$s$ in three spatial dimension under a magnetic field.
However, the result applies not only for the specific case of spin-$s$ under magnetic field.
In general, the proof illustrates the case of $2s+1$ degenerate bands and the magnetic field in this proof plays the role of any set of three parameters, for example the three-dimensional momentum $\mathbf{k} = (k_x,k_y,k_z)$ or our parameter space $(k,q,m)$. 
For any case, we can do a unitary transformation to transform the Hamiltonian in the vicinity of the degneracy point to the form $H = \mathbf{k}\cdot\mathbf{S}$ and \textbf{\textit{the band degree of freedom plays the role of the pseudospin}}, while the intrinsic spin degree of freedom of the particle of interest (electron, photon, phonon) does not play any role here.
In the present work, the Dirac points are \textbf{two-band} degeneracy points, corresponding to a pseudospin-$\dfrac{1}{2}$, so the Berry monopoles located there have strength $g = \dfrac{1}{2}$.

\section{Edge states}
\subsection{Effective model calculation of edge states}
We consider a junction made of two systems described by two Hamiltonian $\mathcal{H}_L(k,q)$ and $\mathcal{H}_R(k,q)$. Both of them are quadratic in genuine momentum $k$ and periodic in synthetic momentum $q$. They can be represented by $8\times 8$ matrices. The eigenstates can be written in forms of $8$-component spinor:
\begin{equation}
    \psi_{\alpha m} = 
    \begin{pmatrix}
        C_{1\alpha m} & C_{2\alpha m} & C_{3\alpha m} & C_{4\alpha m} & C_{5\alpha m} & C_{6\alpha m} & C_{7\alpha m} & C_{8\alpha m}
    \end{pmatrix}^T 
    e^{i k \rho}
\end{equation}

Here $\alpha = L,R$ indicates the side, $m \in \lbrace 1, 2, \dots , 8 \rbrace$ means the band index. The spatial coordinate $\rho = (x+y)/\sqrt{2}$ means the coordinate along the $x = y$ direction. The bulk Hamiltonian of the side $\alpha$ satisfies the eigenvalue equation:
\begin{equation}
    \mathcal{H}_{\alpha} (k,q) \psi_{\alpha m}(k,q) = E_{\alpha m}(k,q) \psi_{\alpha m} (k,q)
\end{equation}

We consider the exponentially localized states located at the junction, which is located at $\rho = 0$. The wave number of these localized states has complex value  $k = k_r + i k_i$ relative to the direction perpendicular to the junction, due to the breaking of the translational symmetry. For the states to be exponentially localized at the heterojunction, $k_i > 0$ for $\rho > 0$ and $k_i < 0$ for $\rho < 0$. For each value of the energy $E$, these wavenumbers $k$ are solutions of the secular equation $\text{det} (\mathcal{H}_{\alpha}(k,q) - E) = 0$. Since the Hamiltonian is quadratic in $k$, we can instead write $\mathcal{H}_{\alpha}(k,q) = h_{\alpha}(q) + H_1(q) k + H_2(q) k^2$ and solve the quadratic eigenvalue problem~\cite{Tisseur2000,Tisseur2001,Dedieu2003,Guttel2017}:
\begin{equation}
    [(h_{\alpha}(q) - E_{\alpha}) + H_1(q) k + H_2(q) k^2] \psi_{\alpha m}(E,q) = 0
    \label{eq:PolyEigEquation}
\end{equation}

Here:
\begin{equation}
\begin{aligned}
    h_{\alpha} =& 
    \begin{pmatrix}
        \Delta_{\alpha 1} & \Omega_{\alpha} \\ 
        \Omega_{\alpha}^{\dagger} & \Delta_{\alpha 2}
    \end{pmatrix}
    \\ 
    \Delta_{\alpha \ell} =& 
    \begin{pmatrix}
        \omega_{\ell} + \eta_{\ell} & W_{\ell} & W_{\ell} & U_{\ell}(1+\alpha_{\ell}) \\
        W_{\ell} & \omega_{\ell} - \eta_{\ell} & U_{\ell}(1-\alpha_{\ell}) & W_{\ell} \\ 
        W_{\ell} & U_{\ell}(1-\alpha_{\ell}) & \omega_{\ell} - \eta_{\ell} & W_{\ell} \\
        U_{\ell}(1+\alpha_{\ell}) & W_{\ell} & W_{\ell} & \omega_{\ell} + \eta_{\ell}
    \end{pmatrix} 
    (\ell = 1,2)
    \\
    \Omega_{\alpha} =& e^{-d/d_0} \text{diag}
    \lbrace -Ve^{-i2\pi q}, V, V, -Ve^{i2\pi q} \rbrace 
\end{aligned}
\end{equation}

\begin{equation}
    \begin{aligned}
        H_1 =& 
        \begin{pmatrix}
            \Delta_{11} & \Omega_{1} \\
            \Omega_1^{\dagger} & \Delta_{12}
        \end{pmatrix}
    \\
    \Delta_{1\ell} =& \text{diag} \lbrace v_{\ell},0,0,-v_{\ell} \rbrace (\ell=1,2)
    \\
    \Omega_1 =& e^{-d/d_0} \text{diag} \lbrace -\beta e^{-i2\pi q}, 0, 0, \beta e^{i2\pi q} \rbrace 
    \end{aligned}
\end{equation}

\begin{equation}
    \begin{aligned}
        H_2 =& 
        \begin{pmatrix}
            \Delta_{21} & \Omega_2 \\ 
            \Omega_2^{\dagger} & \Delta_{22}
        \end{pmatrix}
        \\ 
        \Delta_{2\ell} =& \frac{1}{\sqrt{2}} \text{diag} \lbrace v_{\ell}, 0, 0, -v_{\ell} \rbrace (\ell=1,2)
        \\
        \Omega_2 =& \frac{1}{\sqrt{2}} e^{-d/d_0} \text{diag} \lbrace -\beta e^{-i2\pi q}, \beta, \beta, -\beta e^{i2\pi q} \rbrace 
    \end{aligned}
\end{equation}

The quadratic eigenvalue equation \eqref{eq:PolyEigEquation} is solved using the function {\fontfamily{lmss}\selectfont polyeig} of the Matlab/Octave programming language~\cite{Tisseur2000,Tisseur2001,Dedieu2003}. It has \textit{sixteen} eigenvalues, \textit{eight} of them have non-negative imaginary parts, and the remaining \textit{eight} have non-positive imaginary parts. 
Among them, one is strictly positive pure real, and one is strictly negative pure real. They belong to the part of the dispersion curve corresponding to uncoupled photonic modes and having linear dispersion, so they are non-physical and do not contribute to the photonic modes of the heterojunction. 
If we rearrange and label the complex eigenvalues in order of increasing argument (take the arguments to be in the range $0 \le \varphi < 2\pi$), the wavefunction of the localized eigenstates are given by:
\begin{equation}
    \begin{aligned}
        \psi_{L,E,q} (\rho) =& \sum_{m=10}^{16} A_m \psi_{Lm} (E,q) 
        \\
        \psi_{R,E,q} (\rho) =& \sum_{m=2}^8 B_m \psi_{Rm} (E,q)  
    \end{aligned}
\end{equation}

Here we do not employ the $m = 1$ mode and the $m = 10$ mode having positive and negative pure real $k$, respectively. The wavefunction should be continuous at the heterojunction interface ($\rho = 0$):
\begin{equation}
    \psi_{L,E,q}(0) = \psi_{R,E,q}(0)
\end{equation}

The current density should be continuous at the heterojunction interface: 
\begin{equation}
    J_{L,E,q} (0) = J_{R,E,q} (0)
\end{equation}
This is equivalent to:
\begin{equation}
    \psi_{L,E,q}^T \frac{\partial H_L(E,q)}{\partial k} \psi_{L,E,q} = \psi_{R,E,q}^T \frac{\partial H_R(E,q)}{\partial k} \psi_{R,E,q}
\end{equation}
leading to 
\begin{equation}
    \frac{\partial H_L(E,q)}{\partial k} \psi_{L,E,q} = \frac{\partial H_R(E,q)}{\partial k} \psi_{R,E,q}
\end{equation}

Overall, the continuity conditions at the heterojunction interface are given by:
\begin{equation}
\begin{aligned}
    \psi_{L,E,q}(0) =& \psi_{R,E,q}(0) 
    \\
    \frac{\partial H_L(E,q)}{\partial k} \psi_{L,E,q} =& \frac{\partial H_R(E,q)}{\partial k} \psi_{R,E,q}
\end{aligned}
\label{eq:ContinuityAtJunction}
\end{equation}

The first condition in \eqref{eq:ContinuityAtJunction} implies that:
\begin{equation}
    \sum_{m=10}^{16} A_m \psi_{Lm}(E,q) = \sum_{m=2}^8 B_m \psi_{Rm} (E,q))
    \label{eq:JunctionCondition1}
\end{equation}

The second condition in \eqref{eq:ContinuityAtJunction} is written as:
\begin{equation}
    \sum_{m=10}^{16} A_m \frac{\partial H_L(E,q)}{\partial k} \psi_{Lm}(E,q) = \sum_{m=2}^8 B_m \frac{\partial H_R(E,q)}{\partial k} \psi_{Rm} (E,q))
    \label{eq:JunctionCondition2}
\end{equation}

The two conditions \eqref{eq:JunctionCondition1} and \eqref{eq:JunctionCondition2} give us a homogeneous system of 16 linear equations of 14 variables $A_{10},\dots,A_{16},B_2,\dots,B_8$:
\begin{equation}
    \begin{aligned}
        \sum_{m=10}^{16} A_m \psi_{Lm}(E,q) - \sum_{m=2}^8 B_m \psi_{Rm} (E,q)) =& 0 
        \\
        \sum_{m=10}^{16} A_m \frac{\partial H_L(E,q)}{\partial k} \psi_{Lm}(E,q) - \sum_{m=2}^8 B_m \frac{\partial H_R(E,q)}{\partial k} \psi_{Rm} (E,q)) =& 0 
    \end{aligned} 
    \label{eq:JunctionConditionSystem}
\end{equation}



The $16 \times 14$ coefficient matrix $\mathbf{C}$ of this system of linear equations is expressed in block form as:
\begin{equation}
    \mathbf{C} = 
    \begin{pmatrix}
        \mathbf{C}_{11} & \mathbf{C}_{12} \\ 
        \mathbf{C}_{21} & \mathbf{C}_{22} 
    \end{pmatrix}
\end{equation}

The $8 \times 7$ blocks are given by adding the eigenstate matrix columwise: 
\begin{equation}
\begin{aligned}
    \mathbf{C}_{11} =& \lbrace \psi_{L10} , \dots , \psi_{L16} \rbrace \\
    \mathbf{C}_{12} =& \lbrace -\psi_{R2}, \dots , -\psi_{R8} \rbrace \\
    \mathbf{C}_{21} =& \bigg \lbrace \frac{\partial H_L (k_{10})}{\partial k} \psi_{L10} , \dots , \frac{\partial H_L (k_{16})}{\partial k} \psi_{L16} \bigg \rbrace 
    \\ 
    \mathbf{C}_{22} =& \bigg \lbrace - \frac{\partial H_R(k_2)}{\partial k} \psi_{R2} , \dots , - \frac{\partial H_R (k_8)}{\partial k} \psi_{R8} \bigg \rbrace 
\end{aligned}
\end{equation}

The system of linear equations \eqref{eq:JunctionConditionSystem} has the trivial solution: $A_{10} = \dots = A_{16} = B_2 = \dots = B_8 = 0$. 
The localized edge states exist if and only if the system \eqref{eq:JunctionConditionSystem} has infinitely many solutions, that means if and only if the rank of the coefficient matrix $\mathbf{C}$ is less than the number of variables:
\begin{equation}
    \text{rank} (\mathbf{C}) < 14
\end{equation}

We finish the discussion on the calculation of the edge state using the effective model by remarking that the number of variables taken to establish the coefficient matrix is not always 14. The coefficient matrices for the chiral edge states have size $16 \times 14$. We examine that these chiral edge states are superposition of 14 eigenstates, the wavenumber $k$ of 4 among them have finite real part. Although they exponentially decays from the heterojunction boundary, the nonvanishing real part of their wave number $k$ allows them to reach the other end of the heterojunction and are detectable by transmission measurements. By contrast, the coefficient matrices of the trivial M-shaped edge states have size $16 \times 10$. These edge states are superposition of 10 eigenstates, whose wavenumber $k$ are purely imaginary. Therefore they are more strongly localized at the heterojunction interface and cannot be detectable in the FDTD transmission spectra. The strong localization of these edge states is shown in the integrated field profile Fig.~\ref{fig:e_-01_delta_071_fit} in subsection \ref{subsec:field-profile}.

\subsection{FDTD simulation of edge states}

\begin{figure}
    \centering
    \includegraphics[width=0.5\textwidth]{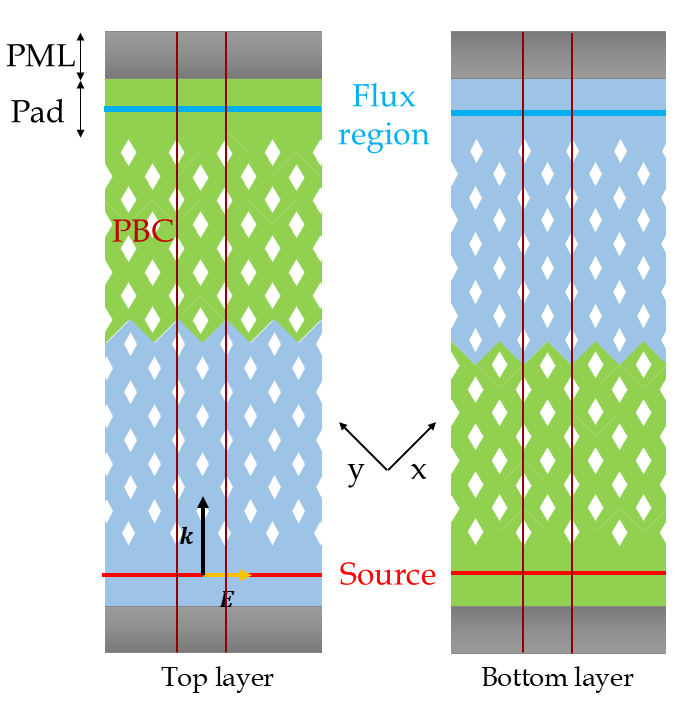}
    \caption{Sketch of the heterojunction between the one bilayer and the same bilayer swapped by interchanging the two layers.}
    \label{fig:Heterojunction-0}
\end{figure}

\begin{figure}
    \centering
    \includegraphics[width=\linewidth]{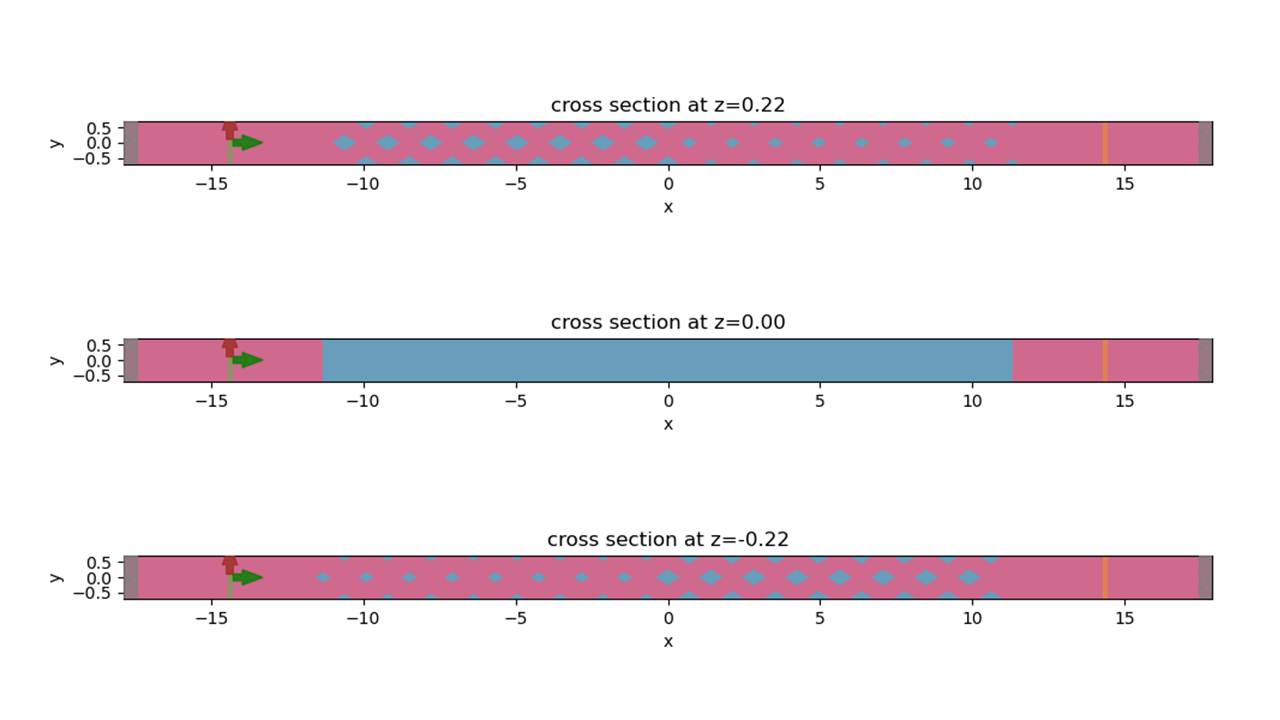}
    \caption{Cross-sections of the simulation cell in the planes parallel to the $xy$ plane with $z = (h+d)/2 = 0.22a$, $z = 0$ and $z = -(h+d)/2=-0.22a$. The source is the brown line. Green arrow: propagation direction of ligth. Red arrow: light polarization. The monitor is the orange line. The PML layers are the grey layers on the left and on the right boundaries. The figures are plotted using the Tidy3D software of FlexCompute.}
    \label{fig:Heterojunction-xy}
\end{figure}

\begin{figure} 
    \centering
    \includegraphics[width=\linewidth]{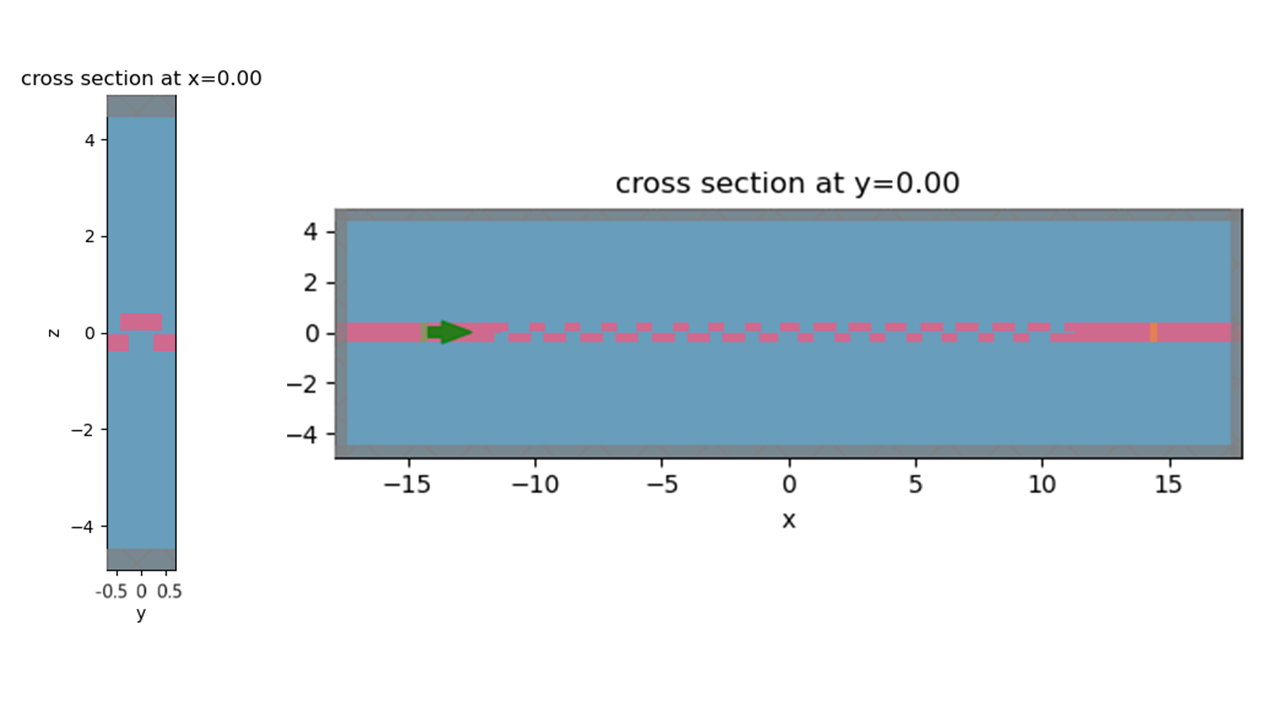}
    \caption{Cross-sections of the simulation cell in the planes $x=0$ and $y=0$. The PML layers are the grey layers. The figures are plotted using the Tidy3D software of FlexCompute.}
    \label{fig:Heterojunction-xz-yz}
\end{figure}

We simulate the edge states of the real heterojunctions between photonic crystal slab bilayers using the open-source package MEEP~\cite{Oskooi2010} (Fig.~\ref{fig:Heterojunction-0}). The two sides of the heterojunction are identical photonic crystal slabs, but are swapped with respect to each other. To facilitate future fabrication, the thickness of the upper (lower) part of both layers are identical for both sides. Both slabs have thickness $h_1 = h_2 = 0.35a$ ($a$ is the edge of the square primitive cell) and are separated apart by an interlayer distance $d = 0.1a$. Technically, the gap between band 1 and band 2 needs to be large enough so that we can clearly see the edge states inside it. Therefore, we increase the difference between $b_1$ and $b_2$. The upper slab of the left-hand side and the lower slab of the right-hand side have $b_1 = 0.46a$, whereas the lower slab of the left-hand side and the upper slab of the right-hand side have $b_2 = 0.30a$. Indeed, the two sides have the same deformation parameter $e$, which we varies in order to control the parameters $\alpha$ and $\eta$ and see the transformation of the edge states. We simulate the heterojunctions for the following values of $e = 0.1, 0.075, 0.05, 0.025, 0, -0.025, -0.05, -0.075, -0.1$.

Because we simulate the wave propagation at the M point, the electromagnetic wave propagates along the diagonal of the primitive cell. For this region, the source and the flux regions need to be inclined by an angle 45 degrees so that they are parallel to the heterojunction. At the best of our knowledge, the MEEP software does not allow to specify a source and a flux region that are inclined with respect to the $x$ and $y$ axes of the simulation cell. Therefore, we rotate the structure by an angle 90 degrees so that the diagonals of the primitive cell lie along the $x$ and $y$ directions and place the source and the flux region parallel to the $xz$ plane. We apply the PML boundary condition~\cite{Oskooi2011} for the $x$ and $z$ directions, and the periodic boundary condition for the $y$ direction. That means the heterojunction is made finite along the $x$ direction, but infinitely expands along the $y$ direction. Two padding blocks of length $3a$ are placed between the heterojunction and the PML boundaries along the $x$-direction. The source and the flux regions are placed in the middle of the two padding regions, as shown in Figs.~\ref{fig:Heterojunction-xy} and ~\ref{fig:Heterojunction-xz-yz}.

An interested reader may wonder if this setup leads to band folding because the primitive cells are rotated by an angle 90 degrees and the periodicity along the $x$ and $y$ directions is now $\sqrt{2}a$ instead of $a$. In other words, the lattice vectors for this simulation setup is $\sqrt{2}a \mathbf{\hat{x}}$ and $\sqrt{2}a\mathbf{\hat{y}}$. In his or her opinion, the unit cell is a square of edge $\sqrt{2}a$ and contains two primitive cells, leading to band-folding and brings the physics at the M-point to the $\Gamma$-point, that means in a non-Hermitian regime. However, although this issue appears when calculating the photonic band structure with MPB, this is not an issue for the MEEP simulations. MPB applies the periodic boundary condition so that the simulation domain is infinitely replicated along the $x$, $y$ and $z$ directions. The user clearly defines the lattice vectors and the Bloch wavevectors are given in the basis of reciprocal lattice vectors. If the user defines larger lattice vectors than the primitive ones, band folding happens. By contrast, in the framework of the FDTD method, the boundary is in general not periodic, but can be reflective or absorbing (like the case of the PML and the adiabatic absorbing boundary condition~\cite{Oskooi2008}). In MEEP, the user does not have to define the lattice vectors and reciprocal lattice vectors. For these region, we can be confident that our simulation results do not have band-folding issues.

\begin{figure}
    \centering
    \includegraphics[width=\linewidth]{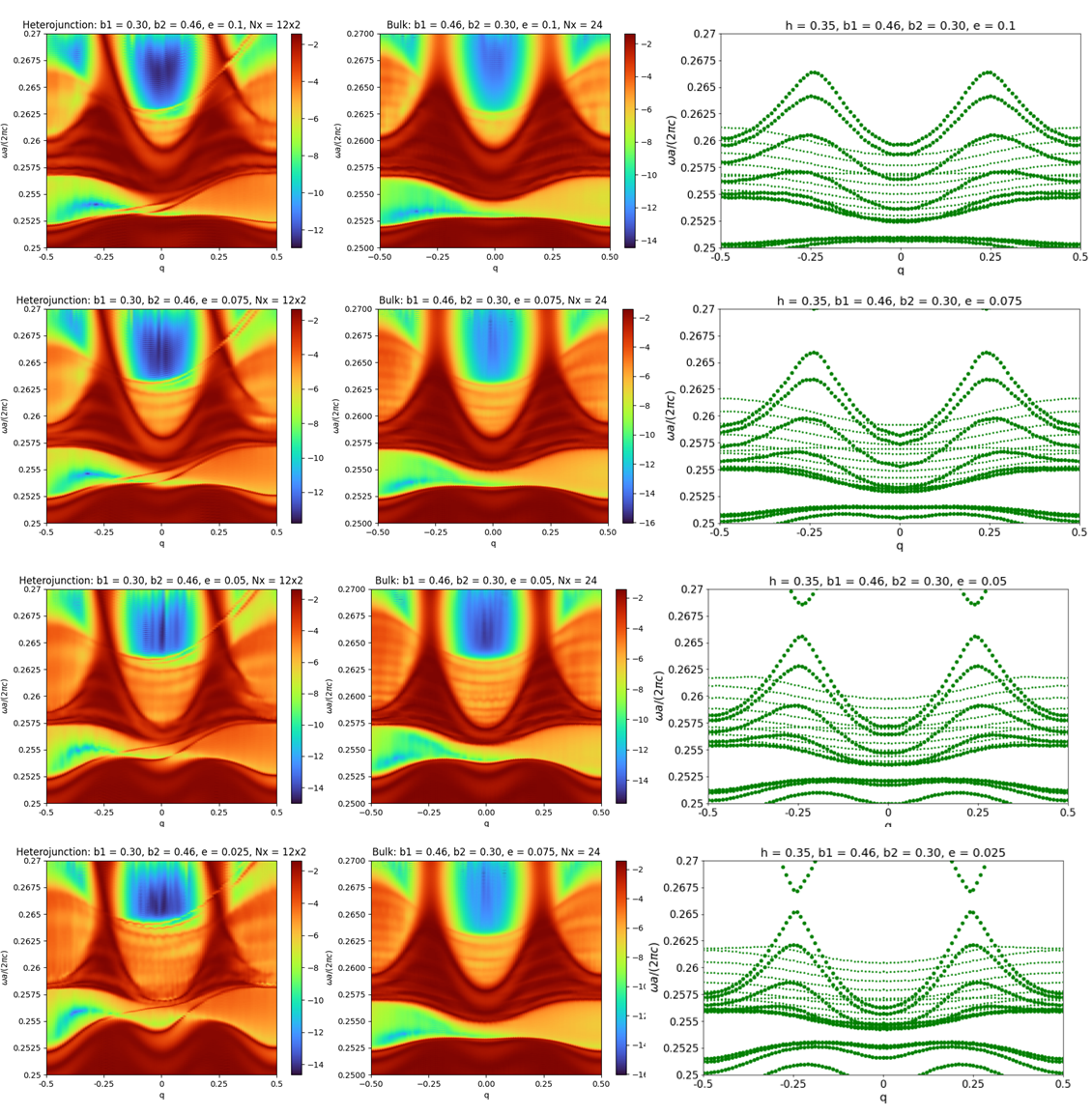}
    \caption{1\textsuperscript{st} column: the transmission spectra of the heterojunctions, each side of the heterojunction contains 9 periods. 2\textsuperscript{nd} column: the transmission spectra of a 2D photonic crystal slab bilayer containing 24 periods. 3\textsuperscript{rd} column: the MPB band structure of the bulk. The parameters are: $h_1=h_2=0.35a$,$b_1=0.46a$,$b_2=0.30a$. The deformation parameter $e = 0.1,0.075,0.05,0.025$.}
    \label{fig:Spectra1}
\end{figure}

\begin{figure}
    \centering
    \includegraphics[width=\linewidth]{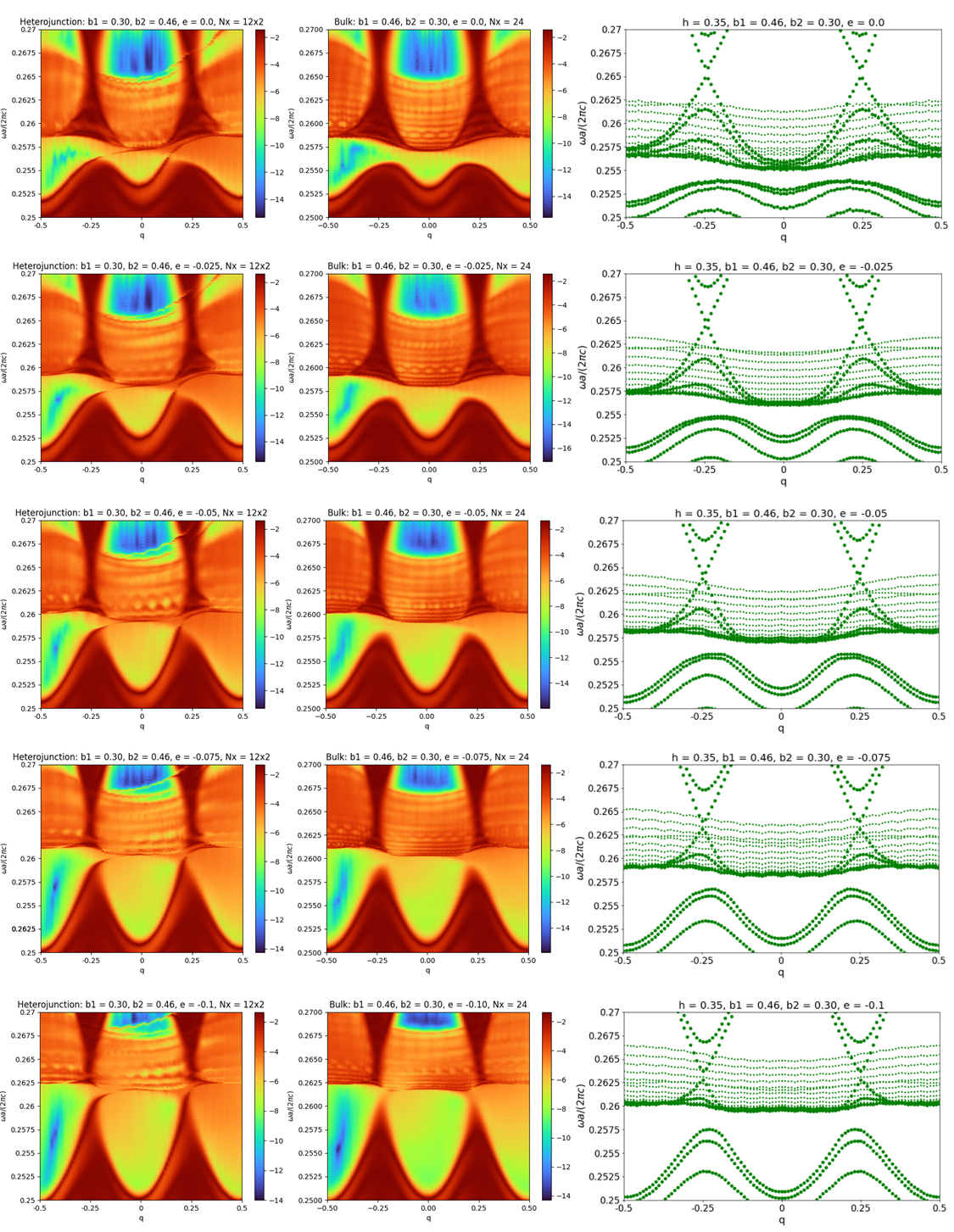}
    \caption{1\textsuperscript{st} column: the transmission spectra of the heterojunctions, each side of the heterojunction contains 9 periods. 2\textsuperscript{nd} column: the transmission spectra of a photonic crystal slab bilayer containing 24 periods. 3\textsuperscript{rd} column: the MPB band structure of the bulk. The parameters are: $h_1=h_2=0.35a$,$b_1=0.46a$,$b_2=0.30a$. The deformation parameter $e = 0,-0.025,-0.05,-0.075,-0.1$.}
    \label{fig:Spectra2}
\end{figure}

Figs.~\ref{fig:Spectra1} and ~\ref{fig:Spectra2} show the transmission spectra of the heterojunction and the 2D photonic crystal slab and the band structure of the bulk for different values of the deformation parameter $e$.\
Each side of the heterojunction has 9 periods while the 2D photonic crystal slab contains 24 periods. 
The parameters are: $h_1 = h_2 = 0.35a$, $b_1 = 0.46a$, $b_2 = 0.30a$.
We recognize that the bulk bands of heterojunctions are perfectly similar to the bands of the 2D photonic crystal slab bilayers in terms of band edge frequency, for all values of $e$.  
The transmission spectra of the heterojunction  clearly show two edge states inside the band gap, which connect the first and the second bulk bands.
The transmission spectra of the 2D photonic crystal slab bilayer are also in agreement with the MPB results.

\subsection{Field profile}
\label{subsec:field-profile}
\begin{figure}
    \centering
    \includegraphics[width=\textwidth]{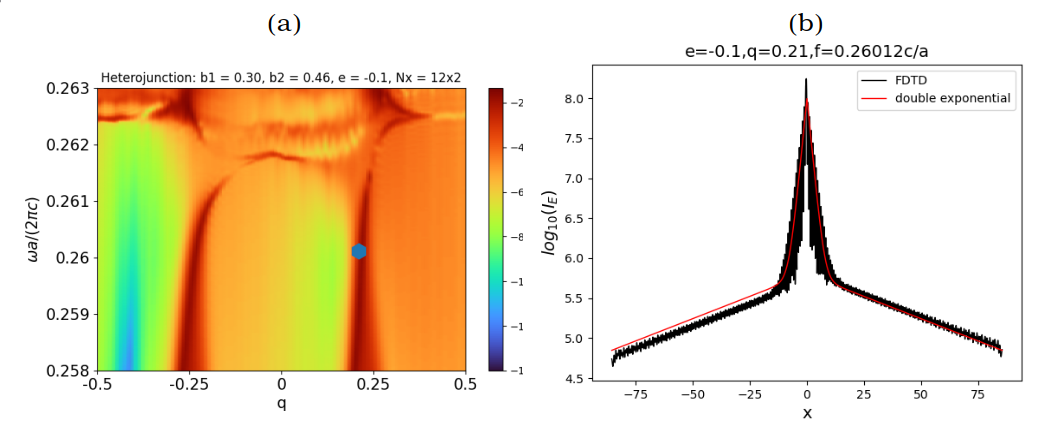}
    \caption{ (a) The transmission spectrum of the structure with $e = -0.1$. (b) Integrated field profile of the structure with $e = -0.1$ with $N_x = 60$ unit cells/side at $q = 0.21$ and $f = 0.26012c/a$, corresponding to the state at the blue hexagonal in figure (a). The black line is the results done by FDTD simulations, and the red line is the fitted with the double exponential function.}
    \label{fig:e_-01_delta_071_fit}
\end{figure}

In this subsection, we discuss the field profile of the edge states. To calculate the edge states field profile, we perform FDTD simulations of the structures of size up to 60 unit cells per side. Here we show the integrated field profile of the structure with $e = -0.1$ and 60 unit cells per side, at $q = 0.21$, excited by a gaussian source of center $fa/c = 0.26012$ and width $\Delta f a / c = 0.0004$. We present the result of the case $e = -0.1$ because its two chiral edge states are totally separated in synthetic momenta so for a fixed value of relative displacement (equivalently a fixed value of synthetic momentum $q$), a gaussian light pulse excites only a topological edge state instead of both of them. For the cases of $e = 0.1$ and $e = 0$, such gaussian pulse may excite both chiral edge states, causes interference so it is impossible to identify the field profile of each individual edge state. To express the decay of the edge state from the interface, we consider the integrated electric field as function of the distance $x$ from the interface, as defined by:

\begin{equation}
    \mathcal{I}_E (x) = \int_{\mathcal{S}} | E_y (x,y,z) |^2 dy dz 
\end{equation}
where $\mathcal{S}$ is the cross-section of the simulation domain spanning in the $yz$-plane. The integrated field profile is shown in Fig.~\ref{fig:e_-01_delta_071_fit}, showing that the integrated field profile exponentially decays from the interface $x = 0$. That means the edge state is localized at the interface.


Next, we determine the decay length of the edge state. We recognize that the shape of the integrated field profile is not purely an exponential function of the distance $x$ from the interface, but rather a superposition of two exponential functions with different decay lengths $d_1$ and $d_2$:
\begin{equation}
    \mathcal{I}_E (x) = A \text{exp} \left( - \frac{|x|}{d_1} \right) + B \text{exp} \left( - \frac{|x|}{d_2} \right)   
\end{equation}

By fitting with the FDTD simulation results, we obtain $A = 10^8 V^2$, $d_1 = 1.5a$, $B = 10^{5.8} V^2$, $d_2 = 39a$. The fitted function is shown as the red line in Fig.~\ref{fig:e_-01_delta_071_fit}. It can be interpreted as the total field is the superposition of two modes: one mode has decay length $d_1 = 1.5a$ and the other has decay length $d_2 = 39a$. Both of them are localized at the interface and exponentially decays as we go further from the interface. One mode has a very short decay length and cannot couple to the flux monitor placed at the end of the heterojunction, so it does not appear in the transmission spectrum. We identify it as the trivial edge mode obtained from the effective model. Notice that according to the analytical model, the trivial edge state has a purely imaginary wave number and is more strongly localized at the interface than the two topological edge states (having complex wave number). At the current stage, we claim having found possible signature of the trivial edge state in the FDTD simulations. A more careful treatment should be conducted to confirm this proposition, as well as giving the full picture. This analysis is beyond the scope of the present work.

\section{Equivalent tight-binding model}
\begin{figure}
    \centering
    \includegraphics[width=0.8\linewidth]{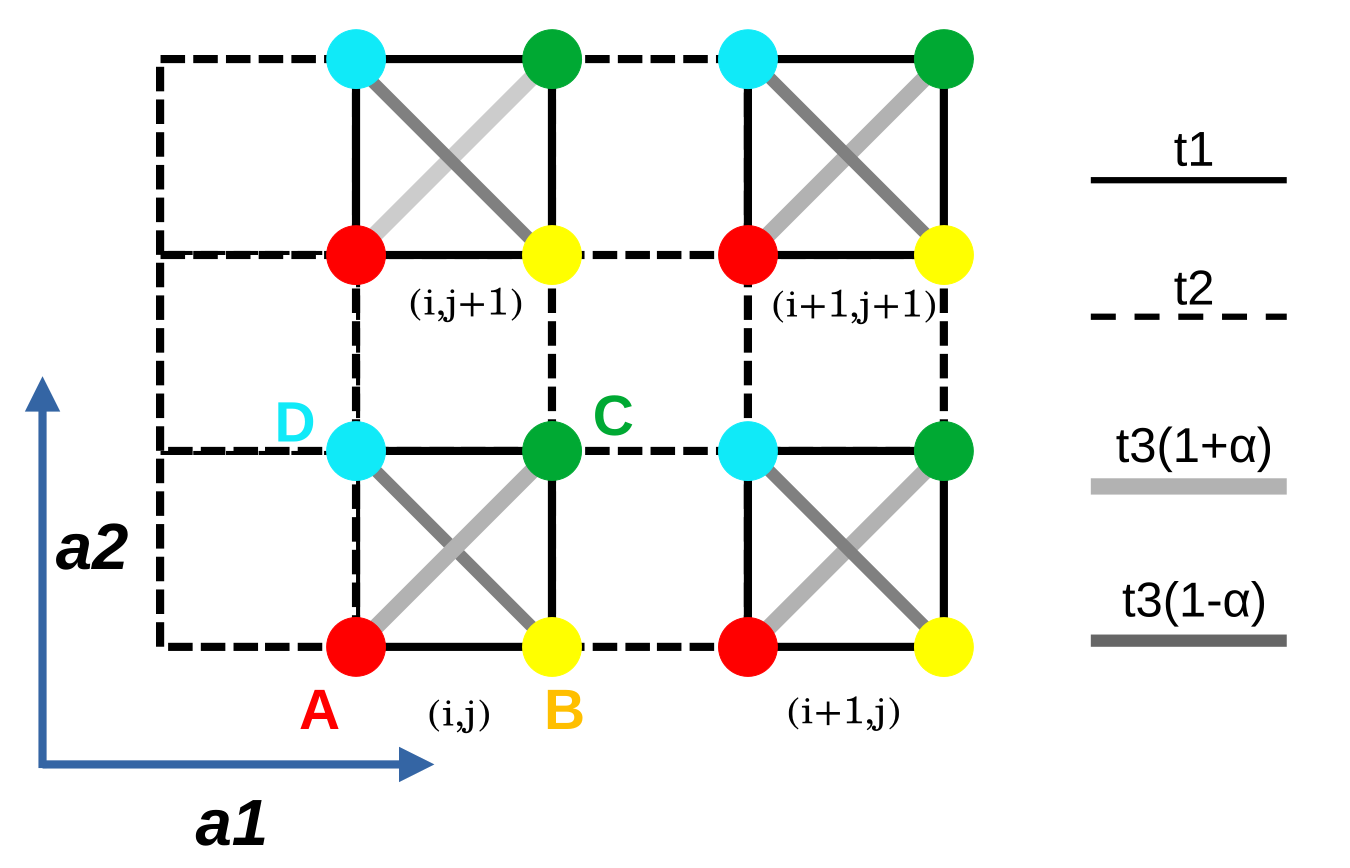}
    \caption{Monolayer with square lattice and 4 sites A,B,C,D. The coupling constants are $t_1$,$t_2$,$t_3$. The parameter $\alpha$ represents the degree of breaking of the C\textsubscript{4} symmetry.}
    \label{fig:TB_Monolayer_lattice}
\end{figure}

In this section, we establish a tight-binding model of a bilayer that exhibits the phenomenon of Berry monopole merging and splitting. First of all, consider a monolayer having a square lattice of lattice parameter $a$ (Fig.~\ref{fig:TB_Monolayer_lattice}) with primitive lattice vectors $\mathbf{a}_1 = (a,0)$ and $\mathbf{a}_2 = (0,a)$. The primitive unit cell contains 4 sites A$(0,0)$, B$(d_1,0)$, C$(d_1,d_2)$, D$(0,d_2)$ ($0 \le d_1,d_2 \le a$). Inside each primitive cell, the particle hops along the edge with hopping parameter $t_1$ and along the diagonal with hopping parameter $t_3$. Particle at each site hops to the nearest-neighbor in the closest primitive cell with hopping parameter $t_2$. We break the C\textsubscript{4} symmetry of this square lattice by making the hopping integrals $t_{AC}$ and $t_{BD}$ to be different from each other, with $\alpha$ be the C\textsubscript{4} symmetry breaking parameter. The case $\alpha = 0$ is reminiscent to a photonic crystal slab monolayer with square hole ($e = 0$), while the case $\alpha \ne 0$ is analogous to a photonic crystal slab with rhombus hole ($ e \ne 0$). Explicitly, we write:
\begin{equation}
    \begin{aligned}
    & t_{AB} = t_{BC} = t_{CD} = t_{DA} = t_1 \\  
    & t_{BA}' = t_{CD}' = t_{CB}' = t_{DA}' = t_2 \\
    & t_{AC} = (1-\alpha) t_3 \\ 
    & t_{BD} = (1+\alpha) t_3
    \end{aligned}
\end{equation}

We denote $\mathbf{R} = i a \mathbf{\hat{x}} + j a \mathbf{\hat{y}}$, $\mathbf{d}_A = (0,0)$, $\mathbf{d}_B = (d_1,0)$, $\mathbf{d}_C = (d_1,d_2)$, $\mathbf{d}_D = (0,d_2)$. The full Hamiltonian of this monolayer is given by:
\begin{equation}
    \begin{aligned}
        H_{\text{1layer}} &= \sum_{\mathbf{R}} & t_1 ( c_{B \mathbf{R}}^{\dagger} c_{A \mathbf{R}} + c_{C \mathbf{R}}^{\dagger} c_{B \mathbf{R}} + c_{D \mathbf{R}}^{\dagger} c_{C \mathbf{R}} + c_{A \mathbf{R}}^{\dagger} c_{D \mathbf{R}}  \\ 
        & & + c_{A \mathbf{R}}^{\dagger} c_{B \mathbf{R}} + c_{B \mathbf{R}}^{\dagger} c_{C \mathbf{R}} + c_{C \mathbf{R}}^{\dagger} c_{D \mathbf{R}} + c_{D \mathbf{R}}^{\dagger} c_{A \mathbf{R}} ) \\
        & + \sum_{\mathbf{R}} & t_2 ( c_{A \mathbf{R}}^{\dagger} c_{B \mathbf{R}-a\mathbf{\hat{x}}} + c_{A \mathbf{R}}^{\dagger} c_{D,\mathbf{R}-a\mathbf{\hat{y}}} + c_{B \mathbf{R}}^{\dagger} c_{C \mathbf{R}-a\mathbf{\hat{y}}} + c_{B \mathbf{R}}^{\dagger} c_{A \mathbf{R}+a\mathbf{\hat{x}}} 
        \\
        & & + c_{C \mathbf{R}}^{\dagger} c_{D \mathbf{R}+a\mathbf{\hat{x}}} + c_{C \mathbf{R}}^{\dagger} c_{B \mathbf{R}+a\mathbf{\hat{y}}} + c_{D \mathbf{R}}^{\dagger} c_{C \mathbf{R}-a\mathbf{\hat{x}}} + c_{D \mathbf{R}}^{\dagger} c_{A \mathbf{R}+a\mathbf{\hat{y}}} )
        \\ 
        &+ \sum_{\mathbf{R}} & [ t_3 (1-\alpha) ( c_{A \mathbf{R}}^{\dagger} c_{C \mathbf{R}} + c_{C \mathbf{R}}^{\dagger} c_{A \mathbf{R}} ) \\
        & & + t_3 (1+\alpha) (c_{B \mathbf{R}}^{\dagger} c_{D \mathbf{R}} + c_{D \mathbf{R}}^{\dagger} c_{B \mathbf{R}}) ] 
    \end{aligned}    
\end{equation}

We remark that in the $t_2$-sum, both $c_{A \mathbf{R}}^{\dagger}c_{B \mathbf{R}-a\mathbf{\hat{x}}}$ and $c_{A \mathbf{R}+a\mathbf{\hat{x}}}^{\dagger}c_{B \mathbf{R}}$ correspond to the hopping from the B-site of one unit cell to the A-site of the next adjacent unit cell to the right-hand side of the initial unit cell. The same argument applies for the similar terms. Consequently, the $t_2$-sum contains only 8 terms (to avoid double-counting). 
We do the Fourier transform for the creation and annihilation operators ($\alpha = $A,B,C,D):

\begin{equation}
    \begin{aligned}
        c_{\alpha\mathbf{k}}^{\dagger} =& \frac{1}{\sqrt{N}} \sum_{\mathbf{R}} e^{i \mathbf{k} \cdot ( \mathbf{R} + \mathbf{d}_{\alpha} ) } c_{\alpha \mathbf{R}}^{\dagger} 
        \\
        c_{\alpha\mathbf{k}} =& \frac{1}{\sqrt{N}} \sum_{\mathbf{R}} e^{-i \mathbf{k} \cdot (\mathbf{R} + \mathbf{d}_{\alpha}) } c_{\alpha \mathbf{R}}
    \end{aligned}
\end{equation}

The inverse Fourier transforms are given by ($\alpha = $A,B,C,D):
\begin{equation}
    \begin{aligned}
        c_{\alpha \mathbf{R}}^{\dagger} =& \frac{1}{\sqrt{N}} \sum_{\mathbf{k}} e^{-i\mathbf{k} \cdot (\mathbf{R} + \mathbf{d}_{\alpha}) } c_{\alpha \mathbf{k}}^{\dagger} 
        \\
        c_{\alpha \mathbf{R}} =& \frac{1}{\sqrt{N}} \sum_{\mathbf{k}} e^{i\mathbf{k} \cdot (\mathbf{R} + \mathbf{d}_{\alpha} ) } c_{\alpha \mathbf{k}}
    \end{aligned}
\end{equation}

The Hamiltonian is rewritten in momentum space as follows:

\begin{equation}
    \begin{aligned}
        H_{\text{1layer}} =& \sum_{\mathbf{k}} t_1 ( e^{-ik_xd_1} c_{B\mathbf{k}}^{\dagger} c_{A\mathbf{k}} + e^{-ik_yd_2} c_{C\mathbf{k}}^{\dagger} c_{B\mathbf{k}} + e^{ik_xd_1} c_{D\mathbf{k}}^{\dagger} c_{C\mathbf{k}} + e^{ik_yd_2} c_{A\mathbf{k}}^{\dagger} c_{D\mathbf{k}} 
        \\
        &+ e^{ik_xd_1} c_{A\mathbf{k}}^{\dagger} c_{B\mathbf{k}} + e^{ik_yd_2} c_{B\mathbf{k}}^{\dagger} c_{C\mathbf{k}} + e^{-ik_xd_1} c_{C\mathbf{k}}^{\dagger} c_{D\mathbf{k}} + e^{-ik_yd_2} c_{D\mathbf{k}}^{\dagger} c_{A\mathbf{k}} ) 
        \\
        & + t_2 (  e^{ik_x(d_1-a)} c_{A\mathbf{k}}^{\dagger} c_{B\mathbf{k}} + e^{ik_y(d_2-a)} c_{A\mathbf{k}}^{\dagger} c_{D\mathbf{k}} + e^{ik_y(d_2-a)} c_{B\mathbf{k}}^{\dagger} c_{C\mathbf{k}} + e^{-ik_x(d_1-a)} c_{B\mathbf{k}}^{\dagger} c_{A\mathbf{k}} \\
        & + e^{-ik_x(d_1-a)} c_{C\mathbf{k}}^{\dagger} c_{D\mathbf{k}} + e^{-ik_y(d_2-a)} c_{C\mathbf{k}}^{\dagger} c_{B\mathbf{k}} + e^{ik_x(d_1-a)} c_{D\mathbf{k}}^{\dagger} c_{C\mathbf{k}} + e^{-ik_y(d_2-a)} c_{D\mathbf{k}}^{\dagger} c_{A\mathbf{k}} ) 
        \\
        &+ t_3(1-\alpha) ( e^{i(k_xd_1+k_yd_2)} c_{A\mathbf{k}}^{\dagger} c_{C\mathbf{k}} + e^{-i(k_xd_1+k_yd_2)} c_{C\mathbf{k}}^{\dagger} c_{A\mathbf{k}} ) 
        \\
        &+ t_3(1+\alpha) ( e^{i(-k_xd_1+k_yd_2)} c_{B\mathbf{k}}^{\dagger} c_{D\mathbf{k}} + e^{i(k_xd_1-k_yd_2)} c_{D\mathbf{k}}^{\dagger} c_{B\mathbf{k}} )
    \end{aligned}
\end{equation}

The Hamiltonian of a monolayer is given in the basis (A,B,C,D) by:

\begin{equation}
\begin{aligned}
    & H_{\text{1layer}}(t_1,t_2,t_3,\alpha;\mathbf{k}) = \\ 
    &= \begin{pmatrix}
        0 & t_1 e^{ik_xd_1} + t_2 e^{ik_x(d_1-a)} & t_3 (1-\alpha) e^{i(k_xd_1+k_yd_2)} & t_1 e^{ik_yd_2} + t_2 e^{ik_y(d_2-a)} \\ 
        t_1e^{-ik_xd_1} + t_2 e^{-ik_x(d_1-a)} & 0 & t_1e^{ik_yd_2} + t_2 e^{ik_y(d_2-a)} & t_3(1+\alpha) e^{i(-k_xd_1+k_yd_2)} \\ 
        t_3(1-\alpha) e^{-i(k_xd_1+k_yd_2)} & t_1 e^{-ik_yd_2} + t_2 e^{-ik_y(d_2-a)} & 0 & t_1 e^{-ik_xd_1} + t_2 e^{-ik_x(d_1-a)} \\ 
        t_1 e^{-ik_yd_2} + t_2 e^{-ik_y(d_2-a)} & t_3(1+\alpha) e^{i(k_xd_1-k_yd_2)} & t_1 e^{ik_xd_1}+t_2 e^{ik_x(d_1-a)} & 0 
    \end{pmatrix}
\end{aligned}
\end{equation}

\begin{figure}
    \centering
    \includegraphics[width=0.3\linewidth]{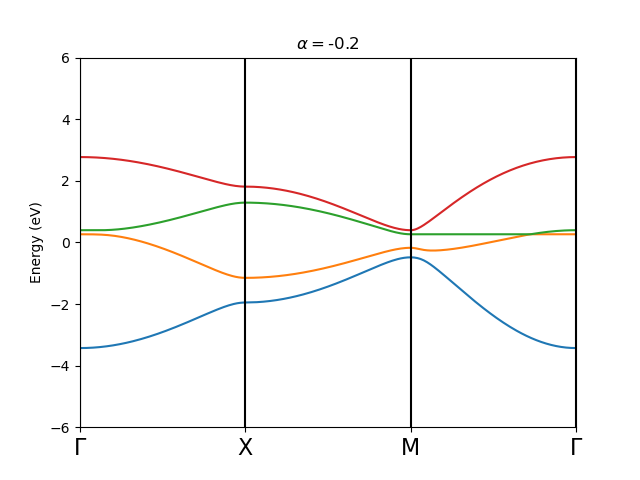}
    \includegraphics[width=0.3\linewidth]{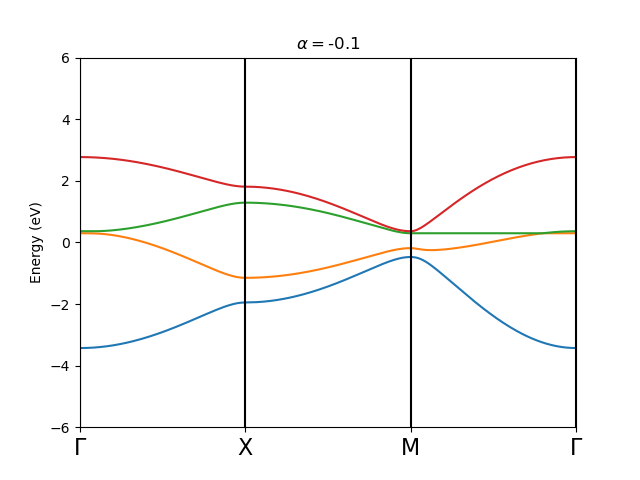}
    \includegraphics[width=0.3\linewidth]{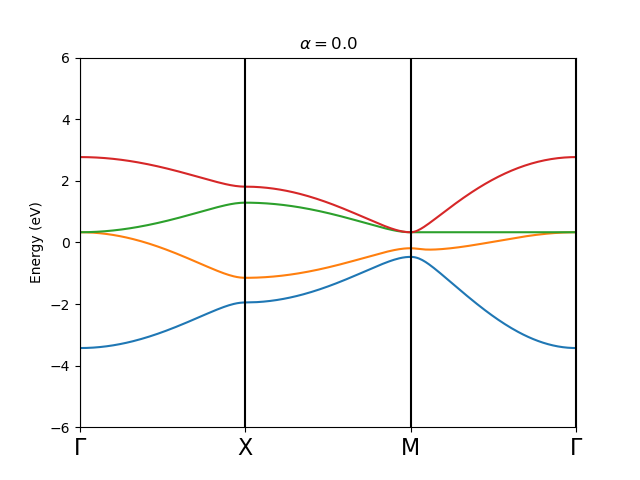}
    \includegraphics[width=0.3\linewidth]{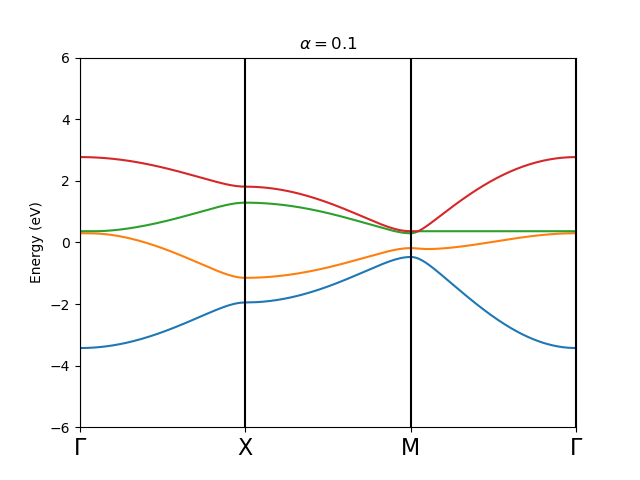}
    \includegraphics[width=0.3\linewidth]{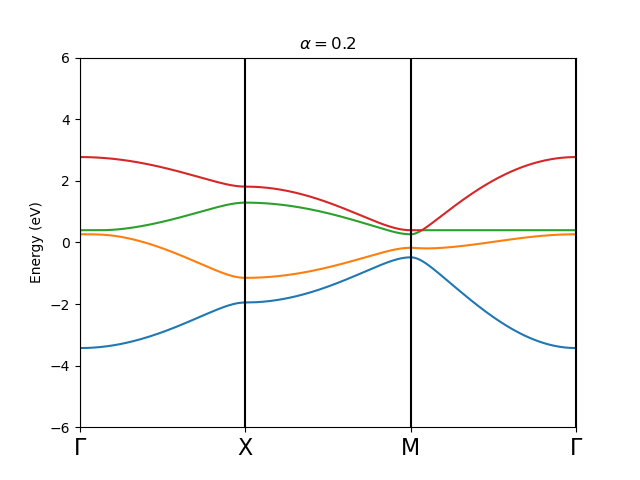}
    \caption{Band structure of monolayer with $\alpha$ varies from -0.2 to 0.2.}
    \label{fig:TB-monolayer-band-structure}
\end{figure}

We plot the band structure of the monolayer with parameters $t_1 = -0.81$ eV, $t_2 = -0.74$ eV, $t_3 = -0.33$ eV in Fig.~\ref{fig:TB-monolayer-band-structure} for $\alpha = -0.2,-0.1,0,0.1,0.2$. For $\alpha = 0$, the bands 2 and 3 are degenerate at the $\Gamma$ and M-points. These degeneracies are lifted when $\alpha \ne 0$.

Now, we consider a bilayer consisting of two square lattices sharing the same lattice parameter $a$, as well as the lattice primitive vectors $\mathbf{a}_1 = (a,0)$ and $\mathbf{a}_2 = (0,a)$. To construct the bilayer Hamiltonian, we apply the method that Ref.~\cite{Do2021} uses for the case of sliding bilayer graphene. Each lattice has 4 sites A,B,C,D. We denote $t_1^l$,$t_2^l$,$t_3^l$ the hopping parameters of the layer $l$ ($l = +$ for upper layer and $l = -$ for lower layer) by noting that $t_k^+$ and $t_k^-$ may not be the same ($k=1,2,3$). However, the symmetry breaking paremeter is the same for both layers $\alpha_+ = \alpha_- = \alpha$. The two layers are placed parallel to each other and the upper layer $l = +$ is displaced by a displacement vector $\boldsymbol{\Delta} + \boldsymbol{\delta} = \left( \dfrac{a}{2}+\delta_x,\dfrac{a}{2}+\delta_y \right)$ ($-\dfrac{a}{2} \le \delta_x,\delta_y \le \dfrac{a}{2}$). Let $D$ be the interlayer distance. The positions of the 8 atomic sites are as follows: 
A\textsubscript{+}$\left(\dfrac{a}{2}+\delta_x,\dfrac{a}{2}+\delta_y,\frac{D}{2}\right)$,
B\textsubscript{+}$\left(\dfrac{a}{2}+d_1+\delta_x,\dfrac{a}{2}+\delta_y,\frac{D}{2}\right)$,
C\textsubscript{+}$\left(\dfrac{a}{2}+d_1+\delta_x,\dfrac{a}{2}+d_2+\delta_y,\frac{D}{2}\right)$,
D\textsubscript{+}$\left(\dfrac{a}{2}+\delta_x,\dfrac{a}{2}+d_2+\delta_y,\frac{D}{2}\right)$,
A\textsubscript{-}$\left(0,0,-\frac{D}{2}\right)$,
B\textsubscript{-}$\left(d_1,0,-\frac{D}{2}\right)$,
C\textsubscript{-}$\left(d_1,d_2,-\frac{D}{2}\right)$,
D\textsubscript{-}$\left(0,d_2,-\frac{D}{2}\right)$.

Consider the atom $\alpha$ at site $\mathbf{R}_i$ in the upper layer and the atom $\beta$ at site $\mathbf{R}_j$ in the lower layer ($\alpha$,$\beta$ = A,B,C,D). We define the projection on the $xy$-plane of the vector pointing from $\beta$ to $\alpha$, \textit{whithout taking into accoung the vector} $\boldsymbol{\Delta}$:
\begin{equation}
    \mathbf{d}_{\alpha\beta}^{ij} = \left( \mathbf{R}_i + \mathbf{d}_{\alpha} + \boldsymbol{\delta} \right) - \left( \mathbf{R}_j + \mathbf{d}_{\beta} \right)  
\end{equation}

Let 
$d_{\alpha\beta}^{ij} = \| \mathbf{d}_{\alpha \beta}^{ij} + \boldsymbol{\Delta} \| $ (please be aware that $d_{\alpha\beta}^{ij}$ is not the length of $\boldsymbol{d}_{\alpha\beta}^{ij}$), the interlayer hopping parameters depend only on the distance between the sites involved and are given (in eV) by:

\begin{equation}
\begin{aligned}
    t_{\alpha\beta}^{ij} =& V_{pp\pi}(d) \left[ 1 - \left( \frac{\mathbf{d}}{d} \cdot \hat{\mathbf{z}} \right)^2 \right] + V_{pp\sigma}(d) \left( \frac{\mathbf{d}}{d} \cdot \hat{\mathbf{z}} \right)^2  
    \\ 
    =& e^{ - \dfrac{ (d_{\alpha\beta}^{ij} )^2+D^2}{a^2} } \times \left( 2.5 \dfrac{D^2}{ (d_{\alpha \beta}^{ij})^2 + D^2} - 2 \right)   
\end{aligned} 
\end{equation}

\textit{Here $\mathbf{d} = \mathbf{d}_{\alpha \beta}^{ij} + \boldsymbol{\Delta} \pm D \hat{\mathbf{z}}$ is the vector joining the two atoms $(i,\alpha)$ and $(j,\beta)$ with $d = \sqrt{ (d_{ij}^{\alpha\beta})^2 + D^2}$. The coefficients are $V_{pp\pi}(d) = V_{pp\pi}^0 \text{exp} \left( - \dfrac{(d_{\alpha\beta}^{ij})^2+D^2}{a^2} \right)$ and $V_{pp\sigma}(d) = V_{pp\sigma}^0 \text{exp} \left( - \dfrac{(d_{\alpha\beta}^{ij})^2+D^2}{a^2} \right)$} with $V_{pp\pi}^0 = -2$eV and $V_{pp\sigma}^0 = 0.5$eV. 

The contribution from the interlayer hopping to the Hamiltonian is given by:
\begin{equation}
    H_{\text{interlayer}} = 
    \begin{aligned}
        \sum_{\mathbf{R}_i \mathbf{R}_j} \sum_{\alpha \beta} t_{\alpha\beta}^{ij} ( c_{\alpha \mathbf{R}_i +}^{\dagger} c_{\beta \mathbf{R}_j -} + c_{\beta \mathbf{R}_j -}^{\dagger} c_{\alpha \mathbf{R}_i +} )
    \end{aligned}
\end{equation}

The Fourier transforms of the creation and annihilation operators are:
\begin{equation}
    \begin{aligned}
        c_{\alpha \mathbf{R} +}^{\dagger} =& \frac{1}{\sqrt{N}} \sum_{\mathbf{k}} e^{ - i \mathbf{k} \cdot (\mathbf{R} + \mathbf{d}_{\alpha} + \boldsymbol{\delta} + \boldsymbol{\Delta} + \frac{D}{2} \mathbf{\hat{z}} ) } c_{\alpha \mathbf{k} +}^{\dagger} 
        \\
        c_{\alpha \mathbf{R} -}^{\dagger} =& \frac{1}{\sqrt{N}} \sum_{\mathbf{k}} e^{ - i \mathbf{k} \cdot (\mathbf{R} + \mathbf{d}_{\alpha} - \frac{D}{2} \mathbf{\hat{z}} ) } c_{\alpha \mathbf{k} -}^{\dagger}  
        \\
        c_{\alpha \mathbf{R} +} =& \frac{1}{\sqrt{N}} \sum_{\mathbf{k}} e^{ i \mathbf{k} \cdot (\mathbf{R} + \mathbf{d}_{\alpha} + \boldsymbol{\delta} + \boldsymbol{\Delta} + \frac{D}{2} \mathbf{\hat{z}} ) } c_{\alpha \mathbf{k} +} 
        \\
        c_{\alpha \mathbf{R} -} =& \frac{1}{\sqrt{N}} \sum_{\mathbf{k}} e^{i \mathbf{k} \cdot (\mathbf{R} + \mathbf{d}_{\alpha} - \frac{D}{2} \mathbf{\hat{z}} ) } c_{\alpha \mathbf{k} -} 
        \\
    \end{aligned}
\end{equation}

Because the vector $\mathbf{k}$ is an in-plane wave vector, the dot product $\mathbf{k} \cdot \frac{D}{2} \mathbf{\hat{z}} = 0$. Therefore, we can write the Fourier transforms as follows:
\begin{equation}
    \begin{aligned}
        c_{\alpha \mathbf{R} +}^{\dagger} =& \frac{1}{\sqrt{N}} \sum_{\mathbf{k}} e^{ - i \mathbf{k} \cdot (\mathbf{R} + \mathbf{d}_{\alpha} + \boldsymbol{\delta} + \boldsymbol{\Delta} ) } c_{\alpha \mathbf{k} +}^{\dagger} 
        \\
        c_{\alpha \mathbf{R} -}^{\dagger} =& \frac{1}{\sqrt{N}} \sum_{\mathbf{k}} e^{ - i \mathbf{k} \cdot (\mathbf{R} + \mathbf{d}_{\alpha} ) } c_{\alpha \mathbf{k} -}^{\dagger}  
        \\
        c_{\alpha \mathbf{R} +} =& \frac{1}{\sqrt{N}} \sum_{\mathbf{k}} e^{ i \mathbf{k} \cdot (\mathbf{R} + \mathbf{d}_{\alpha} + \boldsymbol{\delta} + \boldsymbol{\Delta} ) } c_{\alpha \mathbf{k} +}
        \\
        c_{\alpha \mathbf{R} -} =& \frac{1}{\sqrt{N}} \sum_{\mathbf{k}} e^{ i \mathbf{k} \cdot (\mathbf{R} + \mathbf{d}_{\alpha} ) } c_{\alpha \mathbf{k} -} 
        \\
    \end{aligned}
\end{equation}

Consequently, we have: 
\begin{equation}
    c_{\alpha \mathbf{R}_i +}^{\dagger} c_{\beta \mathbf{R}_j -} = \frac{1}{N} \sum_{\mathbf{k} \mathbf{k}'} e^{-i\mathbf{k}\cdot ( \mathbf{d}_{\alpha\beta}^{ij} + \boldsymbol{\Delta} ) } e^{ i (\mathbf{k}'-\mathbf{k}) \cdot (\mathbf{R}_j + \mathbf{d}_{\beta}) } c_{\alpha \mathbf{k} +}^{\dagger} c_{\beta \mathbf{k}' -} 
\end{equation}

Now, we sum over $i,j$. The sums over $i$ and $j$ independent to each other are equivalent to the sum over $j$ and then the sum over all values of $i$ given that $j$ is fixed to zero. Therefore, we have:
\begin{equation}
    \begin{aligned}
        & \sum_{\mathbf{R}_i \mathbf{R}_j} \sum_{\alpha \beta} t_{\alpha \beta}^{ij} c_{\alpha \mathbf{R}_i +}^{\dagger} c_{\beta \mathbf{R}_j -} 
        \\
        =& \sum_{\mathbf{k} \mathbf{k}'} \sum_{\alpha \beta} \sum_{\mathbf{R}_i \mathbf{R}_j} t_{\alpha\beta}^{ij} \frac{1}{N} e^{-i\mathbf{k}\cdot (\mathbf{d}_{\alpha\beta}^{ij} + \boldsymbol{\Delta} )} e^{ i (\mathbf{k}'-\mathbf{k}) \cdot (\mathbf{R}_j + \mathbf{d}_{\beta}) } c_{\alpha \mathbf{k} +}^{\dagger} c_{\beta \mathbf{k}' -}  
        \\
        =& \sum_{\mathbf{k} \mathbf{k}'} \sum_{\alpha \beta} \sum_{\mathbf{R}_j} \sum_{\mathbf{R}_i} t_{\alpha\beta}^{i0} \frac{1}{N} e^{-i\mathbf{k}\cdot (\mathbf{d}_{\alpha\beta}^{i0}+\boldsymbol{\Delta})} e^{ i (\mathbf{k}'-\mathbf{k}) \cdot \mathbf{R}_j } e^{ i (\mathbf{k}' - \mathbf{k}) \cdot \mathbf{d}_{\beta} } c_{\alpha \mathbf{k} +}^{\dagger} c_{\beta \mathbf{k}' -}  
        \\
        =& \sum_{\mathbf{k} \mathbf{k}'} \sum_{\alpha \beta} \left( \frac{1}{N} \sum_{\mathbf{R}_j} e^{i(\mathbf{k}'-\mathbf{k})\cdot\mathbf{R}_j} \right) \left( \sum_{\mathbf{R}_i} t_{\alpha\beta}^{i0} e^{-i\mathbf{k}\cdot(\mathbf{d}_{\alpha\beta}^{i0} + \boldsymbol{\Delta} )} \right) e^{i(\mathbf{k}'-\mathbf{k})\cdot \mathbf{d}_{\beta}} c_{\alpha \mathbf{k} +}^{\dagger} c_{\beta \mathbf{k}' -} 
        \\
        =& \sum_{\mathbf{k} \mathbf{k}'} \sum_{\alpha \beta} \delta_{\mathbf{k} \mathbf{k'}} \left( \sum_{\mathbf{R}_i} t_{\alpha\beta}^{i0} e^{-i\mathbf{k}\cdot (\mathbf{d}_{\alpha\beta}^{i0} + \boldsymbol{\Delta} )} \right) e^{i(\mathbf{k}'-\mathbf{k})\cdot\mathbf{d}_{\beta}} c_{\alpha\mathbf{k}+}^{\dagger} c_{\beta\mathbf{k}'-} 
        \\
        =& \sum_{\mathbf{k}} \sum_{\alpha \beta} h_{\alpha\beta} (\mathbf{k}) c_{\alpha \mathbf{k}+}^{\dagger} c_{\beta \mathbf{k}-} 
    \end{aligned}
\end{equation}

The matrix element $h_{\alpha\beta}(\mathbf{k})$ is: 
\begin{equation}
    h_{\alpha \beta} (\mathbf{k}) = \sum_{\mathbf{R}_i} t_{\alpha\beta}^{i0} e^{-i \mathbf{k} \cdot (\mathbf{d}_{\alpha \beta}^{i0} + \boldsymbol{\Delta})} 
\end{equation}

The interlayer hopping contribution to the Hamiltonian is therefore:
\begin{equation}
    H_{\text{interlayer}} = \sum_{\mathbf{k}} \sum_{\alpha \beta} ( h_{\alpha\beta} (\mathbf{k})  c_{\alpha \mathbf{k}+}^{\dagger} c_{\beta \mathbf{k}-} + h_{\alpha\beta}(\mathbf{k})^* c_{\beta \mathbf{k} -}^{\dagger} c_{\alpha \mathbf{k} +} ) 
\end{equation}

This interlayer hopping is written as the block $\Omega$ as follows:
\begin{equation}
    \Omega_{\alpha \beta} = h_{\alpha \beta} (\mathbf{k}) = \sum_{\mathbf{R}_i} t_{\alpha \beta}^{i0} e^{-i \mathbf{k} \cdot (\mathbf{d}_{\alpha \beta}^{i0} + \boldsymbol{\Delta} )} 
    = e^{-i \mathbf{k} \cdot \boldsymbol{\delta}} \sum_{\mathbf{R}_i} t_{\alpha \beta}^{i0} e^{-i \mathbf{k} \cdot (\mathbf{R}_i + \mathbf{d}_{\alpha} + \boldsymbol{\Delta} - \mathbf{d}_{\beta}) } 
\end{equation}

The full $8 \times 8$ Hamiltonian for the bilayer tight-binding lattice is given in the basis (A1,B1,C1,D1,A2,B2,C2,D2) by:
\begin{equation}
    H_{\text{bilayer}} (\mathbf{k}) =
    \begin{pmatrix}
        \Delta_+ & \Omega \\ 
        \Omega^{\dagger} & \Delta_- 
    \end{pmatrix}
\end{equation}
where $\Delta_l = H_{\text{1layer}} (t_1^l,t_2^l,t_3^l,\alpha;\mathbf{k})$ for $l = +/-$. In the following, we choose the \textit{interlayer distance} to be $D = 2.5$ nm. 

We remark that the shift $\boldsymbol{\delta}$ appears in the interlayer block $\Omega$. We define the synthetic momenta by:
\begin{equation}
    \begin{aligned}
        q_x =& \frac{\delta_x}{a} \\ 
        q_y =& \frac{\delta_y}{a} 
    \end{aligned}
\end{equation}

This tight-binding model differs from the continuous model in the main text by the following points:
\begin{itemize}
    \item The interlayer block $\Omega$ is not diagonal
    \item The appearance of a product between the genuine and synthetic momenta $\mathbf{k} \cdot \mathbf{q} a$  
\end{itemize}

\begin{figure}
    \centering
    \includegraphics[width=0.3\linewidth]{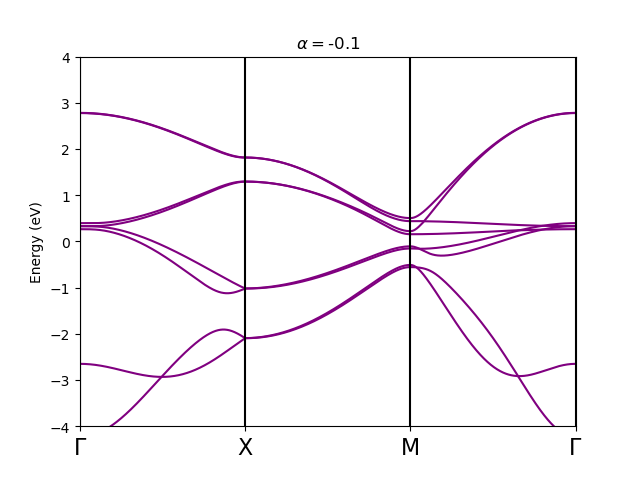}
    \includegraphics[width=0.3\linewidth]{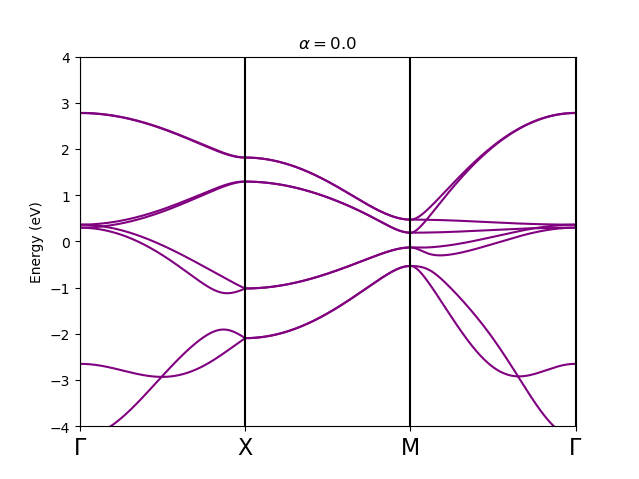}
    \includegraphics[width=0.3\linewidth]{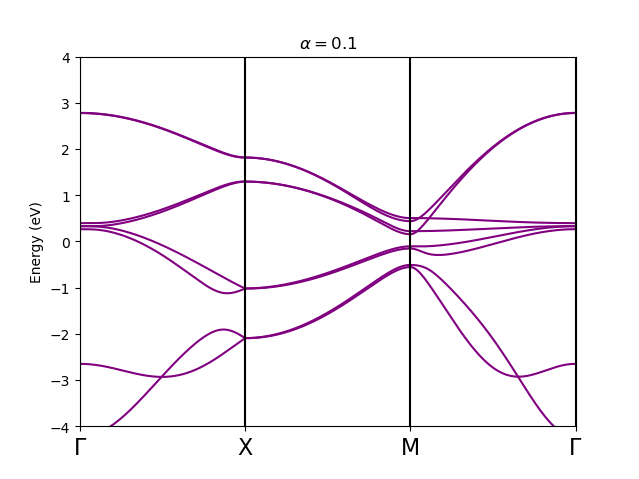}
    \caption{The band structures of the bilayer square lattice for $\alpha = -0.1, 0, 0.1$ in the $k_x$-$k_y$ genuine momentum space.}
    \label{fig:Bilayer-band-structure-kxky}
\end{figure}

\begin{figure}
    \centering
    \includegraphics[width=0.3\linewidth]{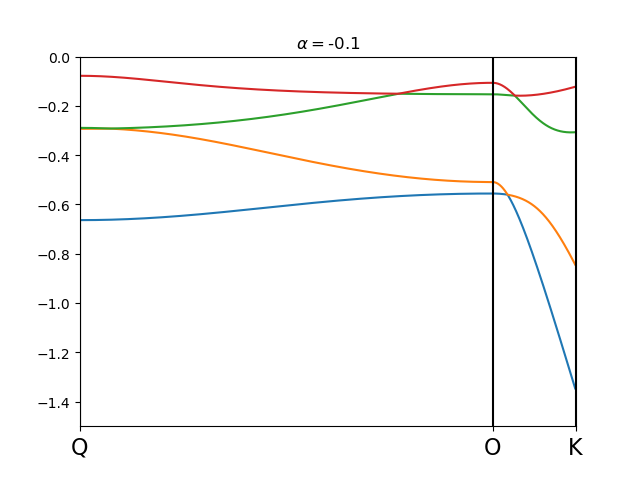}
    \includegraphics[width=0.3\linewidth]{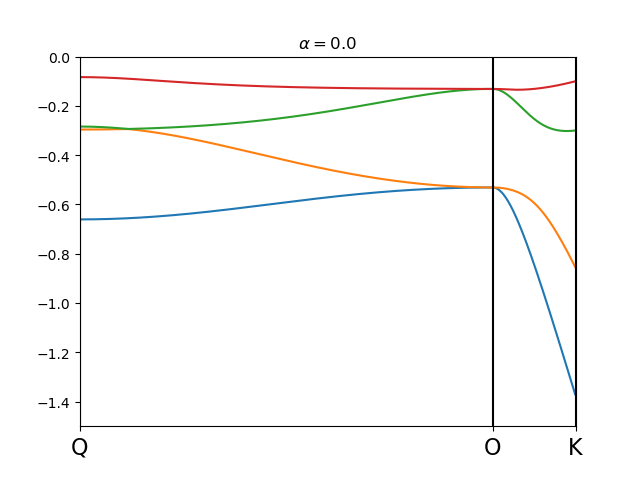}
    \includegraphics[width=0.3\linewidth]{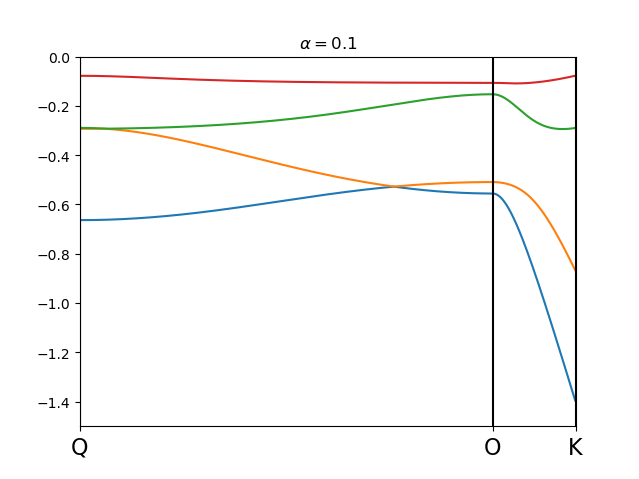}
    \caption{The band structure of the bilayer square lattice for $\alpha = -0.1, 0, 0.1$ in the $k$-$q$ hybrid momentum space.}
    \label{fig:Bilayer-band-structure-kq}
\end{figure}

We show the band structures of the bilayer square lattice for 3 cases $e = 0.1, 0, -0.1$ corresponding to 3 bilayer photonic crystals in Figs.~\ref{fig:Bilayer-band-structure-kxky} (genuine momentum space) and ~\ref{fig:Bilayer-band-structure-kq} (hybrid momentum space). Fig.~\ref{fig:Bilayer-band-structure-kq} shows that when $\alpha$ changes from -0.1 to 0.1, two Dirac cones merge and split from the $k$-axis to the $q$-axis, reproducing the phenomenon in the bilayer photonic crystals. By breaking the inversion symmetry between the two layers, the Dirac cones will open the gap and the bands will have nonzero Berry curvature accumulated around the position of the Dirac cones in the hybrid momentum space.

\section{2D bilayer photonic crystal slab with defects}

In this section, we investigate the transmission spectra of a heterojunction formed by two 2D bilayer photonic crystal slabs in the presence of defects or imperfections, which are unavoidable in experiments. We show that the edge states are robust despite the presence of a defect.

\begin{figure}
    \centering
    \includegraphics[width=\linewidth]{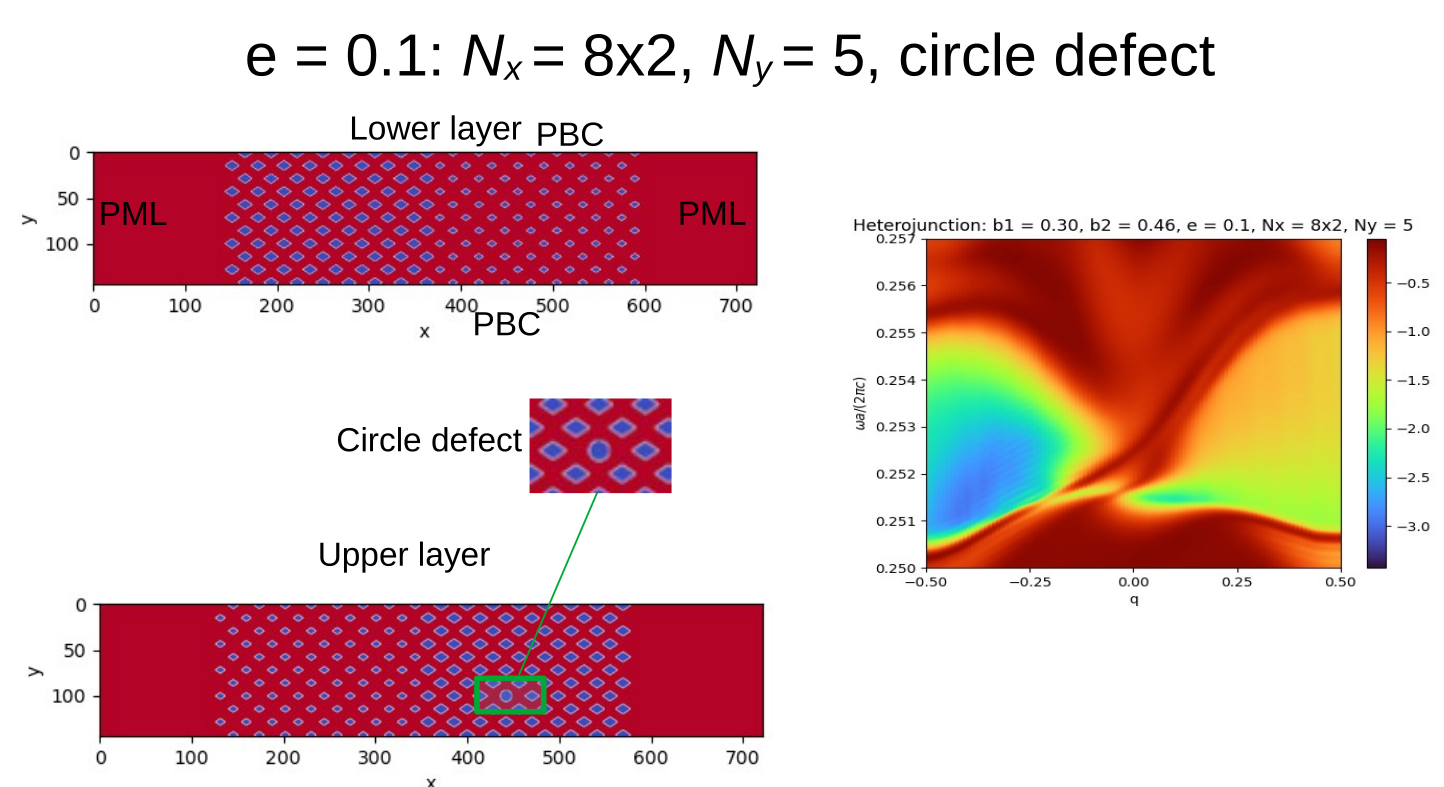}
    \caption{Simulated structure with one circular hole defect and transmission spectrum for the case where $e = 0.1$.}
    \label{fig:Defect_e_0.1}
\end{figure}

\begin{figure}
    \centering
    \includegraphics[width=\linewidth]{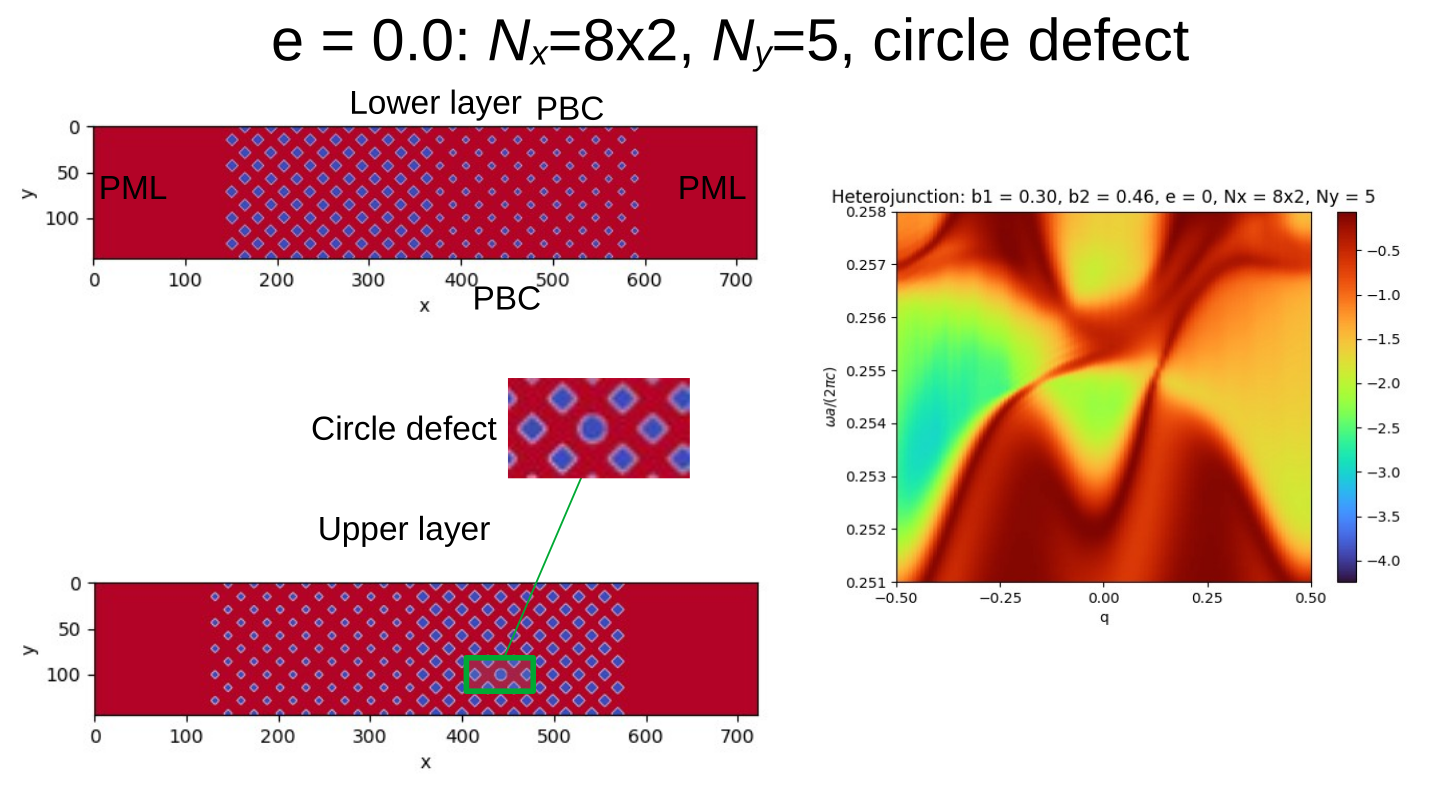}
    \caption{Simulated structure with one circular hole defect and transmission spectrum for the case where $e = 0$.}
    \label{fig:Defect_e_0.0}
\end{figure}

\begin{figure}
    \centering
    \includegraphics[width=\linewidth]{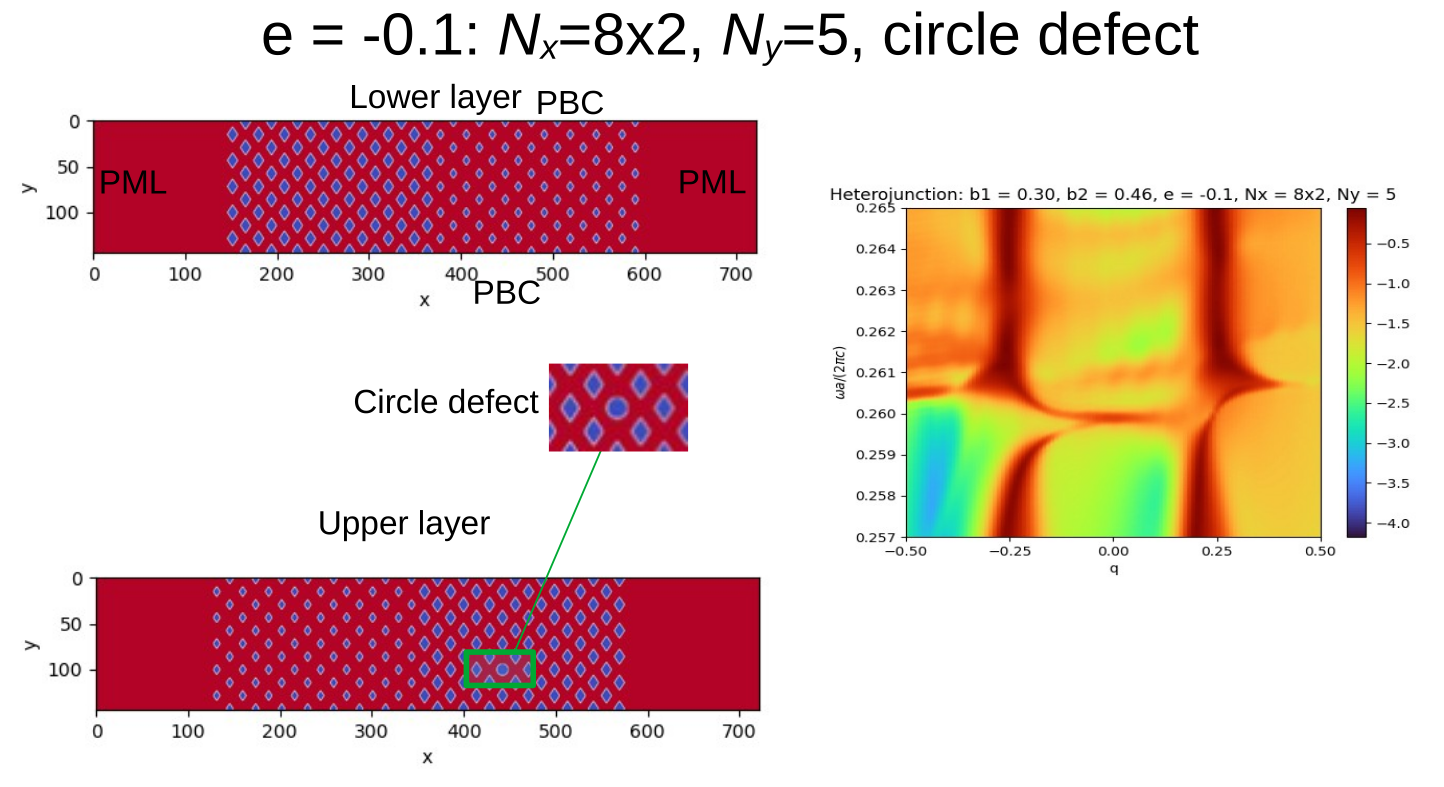}
    \caption{Simulated structure with one circular hole defect and transmission spectrum for the case where $e = -0.1$.}
    \label{fig:Defect_e_-0.1}
\end{figure}

Indeed, we consider a heterojunction between two 2D photonic crystal slabs with opposite Chern numbers. The geometrical parameters are identical to the ones in the main text. The system consists of $N_x = 8$ unit cells at each side along the longitudinal $x$-direction, and $N_y = 5$ unit cells along the transverse $y$-direction. Perfectly Matched Layer (PML) boundary conditions are applied along the longitudinal $x$-direction and Periodic Boundary Conditions (PBC) are imposed along the transverse $y$-direction. In the upper layer, we introduce one circular defect with radius $r = b_1 a = 0.30 a$ equal to the edge length of the undeformed small square. This circle defect is located 2 periods away from the interface as shown in Figs.~\ref{fig:Defect_e_0.1},~\ref{fig:Defect_e_0.0} and ~\ref{fig:Defect_e_-0.1}.

For all the 3 cases $e = 0.1, 0, -0.1$, we simulate the transmission spectra and find the the two edge states remain robust against the presence of the defect, confirming their topological nature. The setups and the corresponding spectra are shown in Figs.~\ref{fig:Defect_e_0.1},~\ref{fig:Defect_e_0.0} and ~\ref{fig:Defect_e_-0.1}. The transmissionss (a.u.) are plotted on a logarithm scale.

%% file: bib_reply.bib
@article{Tan2025,
  title = {Unilateral Asymmetric Radiation in Bilayer Metasurfaces},
  volume = {35},
  ISSN = {1616-3028},
  url = {http://dx.doi.org/10.1002/adfm.202501156},
  DOI = {10.1002/adfm.202501156},
  number = {28},
  journal = {Advanced Functional Materials},
  publisher = {Wiley},
  author = {Tan,  Junhao and Pan,  Ruhao and Xiang,  Yuan and Tang,  Zhiyang and Wang,  Bo and Li,  Junjie},
  year = {2025},
  month = apr 
}

@article{Wang2026,
  title = {Chiral orbital lasing in a twisted bilayer metasurface},
  volume = {17},
  ISSN = {2041-1723},
  url = {http://dx.doi.org/10.1038/s41467-026-69665-w},
  DOI = {10.1038/s41467-026-69665-w},
  number = {1},
  journal = {Nature Communications},
  publisher = {Springer Science and Business Media LLC},
  author = {Wang,  Mingjin and Lv,  Nianyuan and Zhang,  Zixuan and Chen,  Ye and Si,  Jiahao and Chen,  Jingxuan and Tang,  Chenyan and Yin,  Xuefan and Liu,  Zhen and Xin,  Dongxu and Yi,  Zhaozheng and Zheng,  Wanhua and Kivshar,  Yuri and Peng,  Chao},
  year = {2026},
  month = mar 
}

@article{saadi2025,
  title = {{Tailoring Flatband Dispersion in Bilayer Moiré Photonic Crystals}},
  ISSN = {1863-8899},
  DOI = {10.1002/lpor.202501038},
  journal = {Laser Photonics Rev.},
  publisher = {Wiley},
  author = {Saadi,  Chirine and Cueff,  Sébastien and Ferrier,  Lydie and Benamrouche,  Aziz and Gayrard,  Maxime and Drouard,  Emmanuel and Letartre,  Xavier and Nguyen,  Hai Son and Callard,  Ségolène},
  year = {2025},
  month = {jun},
  pages = {e01038},
  volume = {n/a}
}

@article{Choi2025,
author={Choi, Daegwang
and Lee, Ki Young
and Shin, Dong-Jin
and Yoon, Jae Woong
and Gong, Su-Hyun},
title={{Unidirectional guided resonance continuum of Dirac bands in WS$_2$ bilayer metasurfaces}},
journal={Nature Nanotechnology},
year={2025},
month={Jun},
day={04},
issn={1748-3395},
doi={10.1038/s41565-025-01945-w},
}

@article{Tang2025,
  title = {{An adaptive moiré sensor for spectro-polarimetric hyperimaging}},
  volume = {19},
  ISSN = {1749-4893},
  url = {http://dx.doi.org/10.1038/s41566-025-01650-z},
  DOI = {10.1038/s41566-025-01650-z},
  number = {5},
  journal = {Nat. Photon.},
  publisher = {Springer Science and Business Media LLC},
  author = {Tang,  Haoning and Lou,  Beicheng and Du,  Fan and Gao,  Guangqi and Zhang,  Mingjie and Ni,  Xueqi and Hu,  Evelyn and Yacoby,  Amir and Cao,  Yuan and Fan,  Shanhui and Mazur,  Eric},
  year = {2025},
  month = {Apr},
  pages = {463}
}

@article{Tang2023,
author = {Haoning Tang  and Beicheng Lou  and Fan Du  and Mingjie Zhang  and Xueqi Ni  and Weijie Xu  and Rebekah Jin  and Shanhui Fan  and Eric Mazur },
title = {{Experimental probe of twist angle–dependent band structure of on-chip optical bilayer photonic crystal}},
journal = {Sci. Adv.},
volume = {9},
number = {28},
pages = {eadh8498},
year = {2023},
doi = {10.1126/sciadv.adh8498}}

@book{BernevigHughes2013,
 ISBN = {9780691151755},
 author = {B. Andrei Bernevig and Taylor L. Hughes},
 publisher = {Princeton University Press},
 title = {{Topological Insulators and Topological Superconductors}},
 urldate = {2025-04-07},
 year = {2013},
doi = {10.1515/9781400846733}
}

@article{Do2021,
  title = {Proof for the electronic band crossing in sliding bilayer graphene},
  author = {Do, V. Nam},
  journal = {Phys. Rev. B},
  volume = {104},
  issue = {20},
  pages = {205121},
  numpages = {16},
  year = {2021},
  month = {Nov},
  publisher = {American Physical Society},
  doi = {10.1103/PhysRevB.104.205121},
  url = {https://link.aps.org/doi/10.1103/PhysRevB.104.205121}
}

@article{Fonseca2024,
  title = {{Weyl Points on Nonorientable Manifolds}},
  author = {Fonseca, Andr\'e Grossi and Vaidya, Sachin and Christensen, Thomas and Rechtsman, Mikael C. and Hughes, Taylor L. and Solja\ifmmode \check{c}\else \v{c}\fi{}i\ifmmode \acute{c}\else \'{c}\fi{}, Marin},
  journal = {Phys. Rev. Lett.},
  volume = {132},
  issue = {26},
  pages = {266601},
  numpages = {7},
  year = {2024},
  month = {Jun},
  publisher = {American Physical Society},
  doi = {10.1103/PhysRevLett.132.266601},
}

@article{Naselli2024,
  title = {{Stability of Weyl Node Merging Processes under Symmetry Constraints}},
  author = {Naselli, Gabriele and Frank, Gy\"orgy and Varjas, D\'aniel and Fulga, Ion Cosma and Pint\'{e}r, Gerg\'{\'{o}} and P\'alyi, Andr\'as and K\"onye, Viktor},
  journal = {Phys. Rev. Lett.},
  volume = {133},
  issue = {19},
  pages = {196602},
  numpages = {9},
  year = {2024},
  month = {Nov},
  publisher = {American Physical Society},
  doi = {10.1103/PhysRevLett.133.196602},
}

@article{Jorg2022,
author = {Jörg, Christina and Vaidya, Sachin and Noh, Jiho and Cerjan, Alexander and Augustine, Shyam and von Freymann, Georg and Rechtsman, Mikael C.},
title = {Observation of Quadratic (Charge-2) Weyl Point Splitting in Near-Infrared Photonic Crystals},
journal = {Laser \& Photonics Reviews},
volume = {16},
number = {1},
pages = {2100452},
doi = {https://doi.org/10.1002/lpor.202100452},
year = {2022}
}

@article{Gail2012, 
year = {2012}, 
title = {{Manipulation of Dirac points in graphene-like crystals}}, 
author = {Gail, R. de and Fuchs, J.-N. and Goerbig, M.O. and Pi\'{e}chon, F. and Montambaux, G.}, 
journal = {Physica B}, 
issn = {0921-4526}, 
doi = {10.1016/j.physb.2012.01.072}, 
pages = {1948--1952}, 
number = {11}, 
volume = {407}, 
}

@misc{SM,
    note = {{S}ee {S}upplemental {M}aterial at [URL], which includes details on 
(i) Derivation of the effective model: single layer, bilayer, and synthetic momenta; 
(ii). Parameter retrievals for the effective model; (iii) Berry curvature calculations;  (iv) Berry monopole strength; (v) Edge states; (vi) Equivalent tight-binding model; and (vii) 2D bilayer photonic crystal slab with defects.}}


%% file: bibliography.bib
@article{Li2017,
  title = {Weyl points and Fermi arcs in a chiral phononic crystal},
  volume = {14},
  ISSN = {1745-2481},
  url = {http://dx.doi.org/10.1038/nphys4275},
  DOI = {10.1038/nphys4275},
  number = {1},
  journal = {Nature Physics},
  publisher = {Springer Science and Business Media LLC},
  author = {Li,  Feng and Huang,  Xueqin and Lu,  Jiuyang and Ma,  Jiahong and Liu,  Zhengyou},
  year = {2017},
  month = oct,
  pages = {30–34}
}

@article{Acharya2022,
author={Acharya, B.
and others},
title={{Search for magnetic monopoles produced via the Schwinger mechanism}},
journal={Nature},
year={2022},
month={Feb},
day={01},
volume={602},
number={7895},
pages={63-67},
issn={1476-4687},
doi={10.1038/s41586-021-04298-1},
}

@article{Armitage2018,
  title = {{Weyl and Dirac semimetals in three-dimensional solids}},
  author = {Armitage, N. P. and Mele, E. J. and Vishwanath, Ashvin},
  journal = {Rev. Mod. Phys.},
  volume = {90},
  issue = {1},
  pages = {015001},
  numpages = {57},
  year = {2018},
  month = {Jan},
  publisher = {American Physical Society},
  doi = {10.1103/RevModPhys.90.015001} 
}

@article{Berry1984,
  title = {{Quantal phase factors accompanying adiabatic changes}},
  author = {Micheal Victor Berry},
  journal = {Proc. R. Soc. Lond. A},
  volume = {392},
  pages = {45-57},
  numpages = {13},
  year = {1984},
  month = {March},
  publisher = {The Royal Society Publishing},
  doi = {10.1098/rspa.1984.0023}
}

@article{Ozawa_2019,
    title={{Topological quantum matter in synthetic dimensions}}, 
    volume={1}, 
    ISSN={2522-5820}, 
    url={http://dx.doi.org/10.1038/s42254-019-0045-3}, 
    DOI={10.1038/s42254-019-0045-3}, 
    number={5}, 
    journal={Nat. Rev. Phys.}, 
    publisher={Springer Science and Business Media LLC}, 
    author={Ozawa, Tomoki and Price, Hannah M.}, 
    year={2019}, 
    month=apr, 
    pages={349} 
}

@article{Boada2012,
  title = {{Quantum Simulation of an Extra Dimension}},
  author = {Boada, O. and Celi, A. and Latorre, J. I. and Lewenstein, M.},
  journal = {Phys. Rev. Lett.},
  volume = {108},
  issue = {13},
  pages = {133001},
  numpages = {6},
  year = {2012},
  month = {Mar},
  publisher = {American Physical Society},
  doi = {10.1103/PhysRevLett.108.133001},
}

@article{Bouhiron2024,
author = {Jean-Baptiste Bouhiron  and Aurélien Fabre  and Qi Liu  and Quentin Redon  and Nehal Mittal  and Tanish Satoor  and Raphael Lopes  and Sylvain Nascimbene },
title = {{Realization of an atomic quantum Hall system in four dimensions}},
journal = {Science},
volume = {384},
number = {6692},
pages = {223-227},
year = {2024},
doi = {10.1126/science.adf8459}}

@article{Castelnovo2008,
author={Castelnovo, C.
and Moessner, R.
and Sondhi, S. L.},
title={{Magnetic monopoles in spin ice}},
journal={Nature},
year={2008},
month={Jan},
day={01},
volume={451},
number={7174},
pages={42-45},
issn={1476-4687},
doi={10.1038/nature06433},
}

@article{Chen2016,
author={Chen, Wen-Jie
and Xiao, Meng
and Chan, C. T.},
title={{Photonic crystals possessing multiple Weyl points and the experimental observation of robust surface states}},
journal={Nat. Commun.},
year={2016},
month={Oct},
day={05},
volume={7},
number={1},
pages={13038},
issn={2041-1723},
doi={10.1038/ncomms13038},
}

@article{Chong2008,
  title = {{Effective theory of quadratic degeneracies}},
  author = {Chong, Y. D. and Wen, Xiao-Gang and Solja\'{c}i\'{c}, Marin},
  journal = {Phys. Rev. B},
  volume = {77},
  issue = {23},
  pages = {235125},
  numpages = {6},
  year = {2008},
  month = {Jun},
  publisher = {American Physical Society},
  doi = {10.1103/PhysRevB.77.235125},
}

@dataset{Data_Zenodo,
  author       = {Le, Ngoc Duc},
  title        = {Berry monopole scattering in the synthetic
                   momentum space of a bilayer photonic crystal slab
                  },
  month        = jul,
  year         = 2025,
  publisher    = {Zenodo},
  doi          = {10.5281/zenodo.16277941},
  doi          = {10.5281/zenodo.16277941},
}

@article{Dedieu2003,
title = {{Perturbation theory for homogeneous polynomial eigenvalue problems}},
journal = {Linear Algebra Its Appl.},
volume = {358},
number = {1},
pages = {71-94},
year = {2003},
issn = {0024-3795},
doi = {https://doi.org/10.1016/S0024-3795(01)00423-2},
author = {Jean-Pierre Dedieu and Françoise Tisseur}
}

@article{Dirac1931,
    author = {Dirac, Paul Adrien Maurice},
    title = {{Quantised singularities in the electromagnetic field}},
    doi = {10.1098/rspa.1931.0130},
    journal = {Proc. R. Soc. Lond. A},
    volume = {133},
    pages = {60-72},
    year = {1931}
}

@article{Dirac1948,
  title = {{The Theory of Magnetic Poles}},
  author = {Dirac, P. A. M.},
  journal = {Phys. Rev.},
  volume = {74},
  issue = {7},
  pages = {817--830},
  numpages = {0},
  year = {1948},
  month = {Oct},
  publisher = {American Physical Society},
  doi = {10.1103/PhysRev.74.817},
  url = {https://link.aps.org/doi/10.1103/PhysRev.74.817}
}

@inbook{FrohlichPedrini2000,
author = {Frohlich,Jurg and Pedrini,Bill},
title = {{New applications of the chiral anomaly}},
booktitle = {Mathematical Physics 2000},
year = {2000},
chapter = {},
pages = {9-47},
doi = {10.1142/9781848160224_0002},
publisher = {World Scientific Publishing}
}

@book{Girvin_Yang_2019, 
place={Cambridge}, 
title={{Modern Condensed Matter Physics}}, 
publisher={Cambridge University Press}, 
author={Girvin, Steven M. and Yang, Kun}, 
year={2019}}

@article{Guttel2017, 
title={{The nonlinear eigenvalue problem}}, 
volume={26}, 
DOI={10.1017/S0962492917000034}, 
journal={Acta Numer.}, 
author={Guttel, Stefan and Tisseur, Françoise}, year={2017}, 
pages={1–94}
}

@article{Grass2025,
  title = {{Colloquium: Synthetic quantum matter in nonstandard geometries}},
  author = {Grass, Tobias and Bercioux, Dario and Bhattacharya, Utso and Lewenstein, Maciej and Nguyen, Hai Son and Weitenberg, Christof},
  journal = {Rev. Mod. Phys.},
  volume = {97},
  issue = {1},
  pages = {011001},
  numpages = {31},
  year = {2025},
  month = {Mar},
  publisher = {American Physical Society},
  doi = {10.1103/RevModPhys.97.011001},
}

@article{He2023,
  title = {{Polaritonic Chern Insulators in Monolayer Semiconductors}},
  author = {He, Li and Wu, Jingda and Jin, Jicheng and Mele, Eugene J. and Zhen, Bo},
  journal = {Phys. Rev. Lett.},
  volume = {130},
  issue = {4},
  pages = {043801},
  numpages = {6},
  year = {2023},
  month = {Jan},
  publisher = {American Physical Society},
  doi = {10.1103/PhysRevLett.130.043801},
}

@article{Johnson2001,
author = {Steven G. Johnson and J. D. Joannopoulos},
journal = {Opt. Express},
number = {3},
pages = {173--190},
publisher = {Optica Publishing Group},
title = {{Block-iterative frequency-domain methods for Maxwell's equations in a planewave basis}},
volume = {8},
month = {Jan},
year = {2001},
doi = {10.1364/OE.8.000173},
}

@article{Jukic2013,
  title = {{Four-dimensional photonic lattices and discrete tesseract solitons}},
  author = {Juki\ifmmode \acute{c}\else \'{c}\fi{}, D. and Buljan, H.},
  journal = {Phys. Rev. A},
  volume = {87},
  issue = {1},
  pages = {013814},
  numpages = {4},
  year = {2013},
  month = {Jan},
  publisher = {American Physical Society},
  doi = {10.1103/PhysRevA.87.013814},
}

@article{Konig2026,
  title = {Exceptional topology on nonorientable manifolds},
  author = {K\"onig, J. Lukas K. and Yang, Kang and Fonseca, Andr\'e Grossi and Vaidya, Sachin and Solja\ifmmode \check{c}\else \v{c}\fi{}i\ifmmode \acute{c}\else \'{c}\fi{}, Marin and Bergholtz, Emil J.},
  journal = {Phys. Rev. Res.},
  volume = {8},
  issue = {1},
  pages = {013233},
  numpages = {10},
  year = {2026},
  month = {Mar},
  publisher = {American Physical Society},
  doi = {10.1103/g5cr-dwxz},
  url = {https://link.aps.org/doi/10.1103/g5cr-dwxz}
}

@article{Kraus2013,
  title = {{Four-Dimensional Quantum Hall Effect in a Two-Dimensional Quasicrystal}},
  author = {Kraus, Yaacov E. and Ringel, Zohar and Zilberberg, Oded},
  journal = {Phys. Rev. Lett.},
  volume = {111},
  issue = {22},
  pages = {226401},
  numpages = {5},
  year = {2013},
  month = {Nov},
  publisher = {American Physical Society},
  doi = {10.1103/PhysRevLett.111.226401},
}

@article{Lee2022,
  title = {{Synthetic Topological Nodal Phase in Bilayer Resonant Gratings}},
  author = {Lee, Ki Young and Yoo, Kwang Wook and Cheon, Sangmo and Joo, Won-Jae and Yoon, Jae Woong and Song, Seok Ho},
  journal = {Phys. Rev. Lett.},
  volume = {128},
  issue = {5},
  pages = {053002},
  numpages = {6},
  year = {2022},
  month = {Feb},
  publisher = {American Physical Society},
  doi = {10.1103/PhysRevLett.128.053002},
}

@article{Liang2011,
  title = {{Three-dimensional coupled-wave model for square-lattice photonic crystal lasers with transverse electric polarization: A general approach}},
  author = {Liang, Yong and Peng, Chao and Sakai, Kyosuke and Iwahashi, Seita and Noda, Susumu},
  journal = {Phys. Rev. B},
  volume = {84},
  issue = {19},
  pages = {195119},
  numpages = {11},
  year = {2011},
  month = {Nov},
  publisher = {American Physical Society},
  doi = {10.1103/PhysRevB.84.195119},
}

@article{Lohse2018,
author={Lohse, Michael
and Schweizer, Christian
and Price, Hannah M.
and Zilberberg, Oded
and Bloch, Immanuel},
title={{Exploring 4D quantum Hall physics with a 2D topological charge pump}},
journal={Nature},
year={2018},
month={Jan},
day={01},
volume={553},
number={7686},
pages={55-58},
issn={1476-4687},
doi={10.1038/nature25000},
}

@article{Ma2021,
author = {Shaojie Ma  and Yangang Bi  and Qinghua Guo  and Biao Yang  and Oubo You  and Jing Feng  and Hong-Bo Sun  and Shuang Zhang },
title = {{Linked Weyl surfaces and Weyl arcs in photonic metamaterials}},
journal = {Science},
volume = {373},
number = {6554},
pages = {572-576},
year = {2021},
doi = {10.1126/science.abi7803}
}

@article{Milde2013,
author = {P. Milde  and D. Köhler  and J. Seidel  and L. M. Eng  and A. Bauer  and A. Chacon  and J. Kindervater  and S. Mühlbauer  and C. Pfleiderer  and S. Buhrandt  and C. Schütte  and A. Rosch },
title = {{Unwinding of a Skyrmion Lattice by Magnetic Monopoles}},
journal = {Science},
volume = {340},
number = {6136},
pages = {1076-1080},
year = {2013},
doi = {10.1126/science.1234657}}

@article{Morris2009,
author = {D. J. P. Morris  and D. A. Tennant  and S. A. Grigera  and B. Klemke  and C. Castelnovo  and R. Moessner  and C. Czternasty  and M. Meissner  and K. C. Rule  and J.-U. Hoffmann  and K. Kiefer  and S. Gerischer  and D. Slobinsky  and R. S. Perry },
title = {{Dirac Strings and Magnetic Monopoles in the Spin Ice Dy$_2$Ti$_2$O$_7$}},
journal = {Science},
volume = {326},
number = {5951},
pages = {411-414},
year = {2009},
doi = {10.1126/science.1178868}}

@article{Nguyen2021,
author={Nguyen, D.-H.-Minh
and Nguyen, Dung Xuan
and Nguyen, H. Chau
and Louvet, Thibaud
and Drouard, Emmanuel
and Letartre, Xavier
and Bercioux, Dario
and Nguyen, Hai Son},
title={Topological Lasing from Thouless Pumping in Bilayer Photonic Crystal},
journal={ACS Photonics},
year={2026},
month={Mar},
day={04},
publisher={American Chemical Society},
volume={13},
number={5},
pages={1399-1406},
doi={10.1021/acsphotonics.5c02664},
url={https://doi.org/10.1021/acsphotonics.5c02664}
}

@article{Nguyen2023,
  title = {{Fermi Arc Reconstruction in Synthetic Photonic Lattice}},
  author = {Nguyen, D.-H.-Minh and Devescovi, Chiara and Nguyen, Dung Xuan and Nguyen, Hai Son and Bercioux, Dario},
  journal = {Phys. Rev. Lett.},
  volume = {131},
  issue = {5},
  pages = {053602},
  numpages = {7},
  year = {2023},
  month = {Aug},
  publisher = {American Physical Society},
  doi = {10.1103/PhysRevLett.131.053602},
}

@article{Oskooi2008,
author = {Ardavan F. Oskooi and Lei Zhang and Yehuda Avniel and Steven G. Johnson},
journal = {Opt. Express},
number = {15},
pages = {11376--11392},
publisher = {Optica Publishing Group},
title = {{The failure of perfectly matched layers, and towards their redemption by adiabatic absorbers}},
volume = {16},
month = {Jul},
year = {2008},
doi = {10.1364/OE.16.011376},
}

@article{Oskooi2010,
title = {{\textsc{Meep}: A flexible free-software package for electromagnetic simulations by the {FDTD} method}},
journal = {Comput. Phys. Commun.},
volume = {181},
number = {3},
pages = {687-702},
year = {2010},
issn = {0010-4655},
author = {Ardavan F. Oskooi and David Roundy and Mihai Ibanescu and Peter Bermel and J.D. Joannopoulos and Steven G. Johnson},
doi = {10.1016/j.cpc.2009.11.008}
}

@article{Oskooi2011,
title = {{Distinguishing correct from incorrect PML proposals and a corrected unsplit PML for anisotropic, dispersive media}},
journal = {J. Comput. Phys.},
volume = {230},
number = {7},
pages = {2369-2377},
year = {2011},
issn = {0021-9991},
doi = {https://doi.org/10.1016/j.jcp.2011.01.006},
author = {Ardavan Oskooi and Steven G. Johnson}
}

@article{Palumbo2018,
  title = {{Revealing Tensor Monopoles through Quantum-Metric Measurements}},
  author = {Palumbo, Giandomenico and Goldman, Nathan},
  journal = {Phys. Rev. Lett.},
  volume = {121},
  issue = {17},
  pages = {170401},
  numpages = {6},
  year = {2018},
  month = {Oct},
  publisher = {American Physical Society},
  doi = {10.1103/PhysRevLett.121.170401}
}

@article{Palumbo2019,
  title = {{Tensor Berry connections and their topological invariants}},
  author = {Palumbo, Giandomenico and Goldman, Nathan},
  journal = {Phys. Rev. B},
  volume = {99},
  issue = {4},
  pages = {045154},
  numpages = {11},
  year = {2019},
  month = {Jan},
  publisher = {American Physical Society},
  doi = {10.1103/PhysRevB.99.045154}
}

@article{Peng2011,
author = {Chao Peng and Yong Liang and Kyosuke Sakai and Seita Iwahashi and Susumu Noda},
journal = {Opt. Express},
number = {24},
pages = {24672--24686},
publisher = {Optica Publishing Group},
title = {{Coupled-wave analysis for photonic-crystal surface-emitting lasers on air holes with arbitrary sidewalls}},
volume = {19},
month = {Nov},
year = {2011},
doi = {10.1364/OE.19.024672},
}

@article{Peng2012,
  title = {{Three-dimensional coupled-wave theory analysis of a centered-rectangular lattice photonic crystal laser with a transverse-electric-like mode}},
  author = {Peng, Chao and Liang, Yong and Sakai, Kyosuke and Iwahashi, Seita and Noda, Susumu},
  journal = {Phys. Rev. B},
  volume = {86},
  issue = {3},
  pages = {035108},
  numpages = {11},
  year = {2012},
  month = {Jul},
  publisher = {American Physical Society},
  doi = {10.1103/PhysRevB.86.035108},
}

@article{Price2015,
  title = {{Four-Dimensional Quantum Hall Effect with Ultracold Atoms}},
  author = {Price, H. M. and Zilberberg, O. and Ozawa, T. and Carusotto, I. and Goldman, N.},
  journal = {Phys. Rev. Lett.},
  volume = {115},
  issue = {19},
  pages = {195303},
  numpages = {6},
  year = {2015},
  month = {Nov},
  publisher = {American Physical Society},
  doi = {10.1103/PhysRevLett.115.195303},
}

@article{Price2020,
  title = {{Four-dimensional topological lattices through connectivity}},
  author = {Price, Hannah M.},
  journal = {Phys. Rev. B},
  volume = {101},
  issue = {20},
  pages = {205141},
  numpages = {8},
  year = {2020},
  month = {May},
  publisher = {American Physical Society},
  doi = {10.1103/PhysRevB.101.205141},
}

@article{Streifer1977,
  author={Streifer, W. and Scifres, D. and Burnham, R.},
  journal={IEEE J. Quantum Electron.}, 
  title={{Coupled wave analysis of DFB and DBR lasers}}, 
  year={1977},
  volume={13},
  number={4},
  pages={134-141},
  doi={10.1109/JQE.1977.1069328}}

@software{SynMo4Topo,
author = {Le, Ngoc Duc},
month = jul,
title = {{SynMo4Topo}},
url = {https://github.com/ngocducle/SynMo4Topo},
version = {1.0.0},
year = {2025}
}

@article{Tisseur2000,
title = {{Backward error and condition of polynomial eigenvalue problems}},
journal = {Linear Algebra Its Appl.},
volume = {309},
number = {1},
pages = {339-361},
year = {2000},
issn = {0024-3795},
doi = {https://doi.org/10.1016/S0024-3795(99)00063-4},
author = {Françoise Tisseur}
}

@article{Tisseur2001,
author = {Tisseur, Fran\c{c}oise and Meerbergen, Karl},
title = {The Quadratic Eigenvalue Problem},
journal = {SIAM Review},
volume = {43},
number = {2},
pages = {235-286},
year = {2001},
doi = {10.1137/S0036144500381988}
}

@article{Volovik1987,
    title = {Zeros in the fermion spectrum in superfluid systems as diabolical points},
    author = {Volovik, G. E.},
    journal = {JETP Lett.},
    volume = {46},
    issue = {2},
    pages = {81},
    year = {1987},
    doi = {},
    url = {http://jetpletters.ru/ps/0/article_18512.shtml},
}

@article{Wang2017,
  title = {{Optical Interface States Protected by Synthetic Weyl Points}},
  author = {Wang, Qiang and Xiao, Meng and Liu, Hui and Zhu, Shining and Chan, C. T.},
  journal = {Phys. Rev. X},
  volume = {7},
  issue = {3},
  pages = {031032},
  numpages = {14},
  year = {2017},
  month = {Aug},
  publisher = {American Physical Society},
  doi = {10.1103/PhysRevX.7.031032},
}

@article{Xiao2010,
  title = {{Berry phase effects on electronic properties}},
  author = {Xiao, Di and Chang, Ming-Che and Niu, Qian},
  journal = {Rev. Mod. Phys.},
  volume = {82},
  issue = {3},
  pages = {1959--2007},
  numpages = {0},
  year = {2010},
  month = {Jul},
  publisher = {American Physical Society},
  doi = {10.1103/RevModPhys.82.1959},
}

@article{Yu2025,
author = {Danying Yu and Wange Song and Luojia Wang and Rohith Srikanth and Sashank Kaushik Sridhar and Tao Chen and Chenxi Huang and Guangzhen Li and Xin Qiao and Xiaoxiong Wu and Zhaohui Dong and Yanyan He and Meng Xiao and Xianfeng Chen and Avik Dutt and Bryce Gadway and Luqi Yuan},
title = {{Comprehensive review on developments of synthetic dimensions}},
volume = {4},
journal = {Photon. Insights},
number = {2},
publisher = {SPIE},
pages = {R06},
year = {2025},
doi = {10.3788/PI.2025.R06},
}

@article{ZhangHu2001,
author = {Shou-Cheng Zhang  and Jiangping Hu },
title = {{A Four-Dimensional Generalization of the Quantum Hall Effect}},
journal = {Science},
volume = {294},
number = {5543},
pages = {823-828},
year = {2001},
doi = {10.1126/science.294.5543.823}}

@article{Zhu2020,
  title = {{Four-dimensional semimetals with tensor monopoles: From surface states to topological responses}},
  author = {Zhu, Yan-Qing and Goldman, Nathan and Palumbo, Giandomenico},
  journal = {Phys. Rev. B},
  volume = {102},
  issue = {8},
  pages = {081109},
  numpages = {5},
  year = {2020},
  month = {Aug},
  publisher = {American Physical Society},
  doi = {10.1103/PhysRevB.102.081109},
}

@article{Zilberberg2018,
author={Zilberberg, Oded
and Huang, Sheng
and Guglielmon, Jonathan
and Wang, Mohan
and Chen, Kevin P.
and Kraus, Yaacov E.
and Rechtsman, Mikael C.},
title={{Photonic topological boundary pumping as a probe of 4D quantum Hall physics}},
journal={Nature},
year={2018},
month={Jan},
day={01},
volume={553},
number={7686},
pages={59-62},
issn={1476-4687},
doi={10.1038/nature25011},
}

@article{Cayssol_2021,
doi = {10.1088/2515-7639/abf0b5},
year = {2021},
month = {apr},
publisher = {IOP Publishing},
volume = {4},
number = {3},
pages = {034007},
author = {Cayssol, J and Fuchs, J N},
title = {{Topological and geometrical aspects of band theory}},
journal = {J. Phys. Mater.}
}

@article{Skjrvo_2019, 
    title={{Advances in artificial spin ice}}, 
    volume={2}, 
    ISSN={2522-5820}, 
    url={http://dx.doi.org/10.1038/s42254-019-0118-3}, 
    DOI={10.1038/s42254-019-0118-3}, 
    number={1}, journal={Nat. Rev. Phys.}, 
    publisher={Springer Science and Business Media LLC}, 
    author={{Skjærvø, S. H. and Marrows, C. H. and Stamps, R. L. and Heyderman, L. J.}}, 
    year={2019}, 
    month=nov, 
    pages={13} 
}

@article{Hoffmann_2017,
    title={{Antiskyrmions stabilized at interfaces by anisotropic Dzyaloshinskii-Moriya interactions}}, 
    volume={8}, 
    pages={308},
    ISSN={2041-1723}, 
    DOI={10.1038/s41467-017-00313-0}, 
    number={1}, 
    journal={Nat. Commun.}, 
    publisher={Springer Science and Business Media LLC}, 
    author={Hoffmann, Markus and Zimmermann, Bernd and Müller, Gideon P. and Schürhoff, Daniel and Kiselev, Nikolai S. and Melcher, Christof and Blügel, Stefan}, 
    year={2017}, 
    month=aug 
}
